\documentclass[3p,12pt]{elsarticle}
\usepackage{threeparttable,booktabs,tabularx}
\usepackage[fleqn]{amsmath}
\usepackage{cases}
\usepackage{multirow}
\usepackage{xcolor}
\usepackage[normalem]{ulem}
\usepackage{tabularx}
\usepackage{siunitx}
\usepackage{comment}
\usepackage{algorithm}
\usepackage{algpseudocode}
\usepackage{algorithmicx}
\usepackage{grffile}
\usepackage{rotating}
\usepackage{array}
\usepackage{lineno}
\usepackage{hyperref}
\usepackage{makecell}
\usepackage{graphicx,enumerate}
\usepackage{amssymb}
\usepackage{bm,amsmath}
\allowdisplaybreaks  
\usepackage{subfigure}
\usepackage{caption,color}
\usepackage{threeparttable}
\usepackage{subeqnarray}
\usepackage{multirow}
\usepackage{lscape}
\usepackage{rotating}
\usepackage{graphics}
\floatname{algorithm}{Algorithm}
\usepackage{epstopdf}
\journal{}

\biboptions{numbers,sort&compress} 

\begin{document}
	
\begin{frontmatter}

\title{Enhancing DSMC simulations of rarefied gas mixtures using a fast-converging and asymptotic-preserving scheme}
\author[sustech]{Liyan Luo}
\author[sustech]{Jianan Zeng}
\author[sustech]{Yanbin Zhang}
\author[sustech]{Wei Li}
\author[sysu]{Qi Li} 
\author[sustech]{Lei Wu\corref{mycorrespondingauthor1}}
\ead{wul@sustech.edu.cn}
\cortext[mycorrespondingauthor1]{Corresponding author}
\address[sustech]{Department of Mechanics and Aerospace Engineering, Southern University of Science and Technology, 518055 Shenzhen, China}
\address[sysu]{School of Aeronautics and Astronautics, Sun Yat-sen University, Shenzhen, 518107, China}


\begin{abstract}
The numerical simulation of rarefied gas mixture dynamics with disparate masses using the direct simulation Monte Carlo (DSMC) method is slow, primarily because the time step is constrained by that of the lighter species, necessitating an enormous number of evolution steps to reach a steady state. Here, we address this issue by developing a general synthetic iterative scheme, in which the traditional DSMC simulation is intermittently enhanced using a macroscopic synthetic equation. Specifically, after running the DSMC for a certain number of time steps, the high-order constitutive relations for stress and heat flux, as well as the momentum and energy exchange terms from inter-species collisions, are extracted from the DSMC and incorporated into the macroscopic synthetic equations. These equations are solved to obtain the steady state, and the solution is then used to update the particle distribution in DSMC, thereby skipping unnecessary intermediate evolutions. This two-way coupling not only accelerates convergence to the steady state but also asymptotically preserves the  Navier-Stokes limit in the continuum flow regime, allowing the spatial cell size to be much larger than the molecular mean free path. The accuracy of our method is validated for one-dimensional force-driven Poiseuille flow and two-dimensional hypersonic flow past cylinders, considering Maxwell and hard-sphere gases with mass ratios of 10 and 100. Although our in-house DSMC method is approximately an order of magnitude slower than the open-source DSMC code SPARTA, intermittently augmenting it with the synthetic equation makes it roughly 30 times faster at a Knudsen number of 0.01, with even greater computational gains anticipated at smaller Knudsen numbers. This work represents a critical step toward developing fast-converging and asymptotic-preserving schemes for hypersonic chemical reactions.
\end{abstract}

\begin{keyword}
Rarefied gas dynamics;  gas mixtures; direct simulation Monte Carlo; fast convergence; asymptotic preserving; 
\end{keyword}

\end{frontmatter}

\nolinenumbers

\section{Introduction}\label{sec:1}

Multiscale simulation of rarefied gas mixture flows has become increasingly important in a wide range of engineering applications, including space exploration~\cite{votta-2013}, plasma physics~\cite{chen-1984}, and extreme ultraviolet lithography~\cite{fu2019euv}.
In particular, gas mixtures with disparate molecular masses tend to exhibit pronounced non-equilibrium behavior, making their simulation especially demanding in multiscale frameworks. For instances, during the atmospheric re-entry of a space vehicle, the extremely high temperatures can cause gas dissociation, leading to the formation of plasma—comprising ions, electrons, and neutral species—with mass ratios between different species reaching up to $10^{5}$~\cite{brun-2012}. In extreme ultraviolet lithography, dynamic gas locks are employed to continuously inject the clean gas to reduce the partial pressure of hydrocarbon contaminants in the vacuum optical cavity. The molecular masses of hydrogen and hydrocarbon can differ by up to two orders of magnitude~\cite{teng2023pollutant}. While the  Navier–Stokes (NS) equations works in the continuum regime where the Knudsen number ($\text{Kn}$, defined as the ratio of the molecular mean free path $\lambda$ to the characteristic length scale $L$) is small, the Boltzmann equation provides fundamental description in all flow regimes. Due to the disparate time scales in various species, the non-equilibrium dynamics become far more significant than in a single-species gas, necessitating the use of the Boltzmann equation.


Since the Boltzmann equation is defined in the seven-dimensional phase space, and its collision operator is a five-fold nonlinear integral, its numerical simulation has long posed a significant research challenge.
The direct simulation Monte Carlo (DSMC) method is widely used for simulating rarefied gas dynamics~\cite{bird-1994}. In DSMC, each simulation particle represents a group of real gas molecules, and their motion and binary collisions mimic those of real molecules, provided that the collision probability is chosen appropriately. It has been shown to recover the Boltzmann solution for monatomic gases in the limit of an infinite number of simulation particles~\cite{wagner1992convergence}. Moreover, DSMC is well-suited to incorporating complex physical and chemical processes, making it a practical choice for simulating complex chemical reactions~\cite{bird-2011}.
Nevertheless, when the Knudsen number is small, to reduce the numerical dissipation, the grid size and time step in DSMC should be smaller than one-third of mean free path and mean collision time, respectively. This results in a prohibitively large number of evolution steps to reach statistical convergence for steady-state solutions. This issue becomes even more pronounced in mixtures with disparate molecular masses, where the time step is constrained by that of the lighter gas, causing the heavier molecules to evolve slowly and significantly increasing the time required to reach a steady-state solution. Moreover, to reduce statistical noise, the simulation time for gas mixtures is expected to scale approximately linearly with the mass ratio, as estimated from the Mach number~\cite{Hadjiconstantinou2003JCP}.

Over the past decades, numerous strategies have been proposed to improve the efficiency of DSMC simulation in the near-continuum regime. One such approach is the hybrid NS–DSMC method, which partitions the computational domain into continuum and rarefied regions, where the NS equations and DSMC are applied, respectively~\cite{schwartzentruber-2006,schwartzentruber-2007}. However, in practical engineering applications, accurately identifying the interface between these two regimes remains a significant challenge. To circumvent the need for domain decomposition, various asymptotic-preserving schemes have been developed, enabling the use of large time steps and/or coarse spatial cells. These include time-relaxed~\cite{pareschi-2001}, exponential Runge–Kutta~\cite{dimarco-2011}, and asymptotic-preserving Monte Carlo methods~\cite{ren-2014}, many of which have been extended to multi-species systems~\cite{Binzhang-2024,Li_Yang_2014}. The key idea of these asymptotic-preserving methods is to decompose the collision operator into a stiff linear component and a less stiff nonlinear component. The stiff component is typically approximated using a relaxation-time model, ensuring the correct asymptotic behavior and consistency with the Euler equations in the continuum limit.
However, in the continuum regime, they often fail to accurately recover transport coefficients-such as viscosity, thermal conductivity, and diffusion in gas mixtures—due to their asymptotic preservation of the Euler equations~\cite{JIN199551}.
To address these limitations, an asymptotic accurate and preserving Monte Carlo method has been developed for single-species~\cite{fei-2023} and multi-species~\cite{FEI2025114196} gas flows, where the micro–macro decomposition of the collision operator is constructed based on the Chapman–Enskog expansion~\cite{chapman-1990}, enabling both asymptotic preservation of the NS equations and second-order accuracy in the fluid regime.
However, although the time step in these methods can be significantly larger than the mean collision time, their explicit time-stepping nature ties the evolution of the velocity distribution functions to particle velocities, leading to computational inefficiency in many multiscale simulations. 
Since most problems in rarefied gas dynamics focus on steady-state solutions, tracking time evolution is generally unnecessary. 

Recently, the general synthetic iterative scheme (GSIS) has been developed as an accurate and efficient method for obtaining steady-state solutions of the Boltzmann equation and its simplified kinetic models~\cite{su-2020-can,zhu-2021,Su2021CMAME}. The core idea is to deterministically and iteratively solve both the mesoscopic Boltzmann equation and the corresponding macroscopic synthetic equations. As a result, steady-state solutions can be obtained within hundreds of iterations across all flow regimes, and the spatial cell size can be much larger than the mean free path in the near-continuum regime~\cite{zhang-2024}.
Building on these fast convergence and asymptotic preserving properties, the direct intermittent GSIS–DSMC coupling (DIG) method has been proposed for the enhance the DSMC simulation~\cite{luo-2024}. By intermittently coupling a standard DSMC solver with steady-state macroscopic synthetic equations derived from the Boltzmann equation, DIG exhibits the asymptotic-preserving property: it recovers the standard DSMC method in the rarefied regime and the NS equations in the continuum regime. 
As a result, the cell size can be significantly enlarged, allowing for a larger time step. Furthermore, the synthetic equations, when solved to steady state, enhance global fluid information exchange across the entire computational domain—clearly offering greater efficiency than the NS–DSMC coupling. Together, these two factors enable DIG to reduce computational time by orders of magnitude compared to conventional DSMC, in the near-continuum regime.
For example, in a hypersonic argon gas flow past a cylinder in the near-continuum regime, DIG requires only 40,000 computational cells, compared with over 2,000,000 in standard DSMC simulations; moreover, its computational time is reduced by nearly two orders of magnitude owing to the additional reduction in transitional evolutions~\cite{luo-2024}.

While DSMC has been successfully extended to multi-species flows, gas mixtures with disparate mass present additional challenges. In such systems, the mean collision times for different collision pairs can vary by several orders of magnitude, forcing the DSMC time step to be dictated by the shortest relaxation time and thus significantly decelerating the numerical evolution of the system. In this work, we propose the DIG method~\cite{luo-2024} to simulate the multi-species multiscale problems, where the convergence of both light and heavy species is accelerated simultaneously. Moreover, owing to the implicit treatment of macroscopic synthetic equations, the proposed multi-species DIG achieves rapid convergence to steady state, offering a substantial acceleration over other explicit schemes.

Our method will be applied directly to accelerate DSMC simulations, rather than to simplified kinetic models. This choice reflects our ultimate objective: developing efficient and accurate numerical methods for chemical reactions—a longstanding challenge that has remained unresolved for decades, despite numerous multiscale methods proposed for single- and multi-species Boltzmann equations without chemical reactions\footnote{We refer to recent progress in multiscale methods based on the simplified kinetic models~\cite{Xu2010JCP,xin2023discrete,Guo2013PRE,Gorji2014JCP,Gorji_2012,Eunji2025, MPfeiffer2024JCP,fei2020JCP}.} If the present method proves effective for gas mixtures, its extension to DSMC with chemical reactions can be pursued in the near future. By contrast, simplified kinetic models for chemical reactions still have a long way to go, as kinetic modeling is a complex task, in contrast to DSMC’s straightforward incorporation of multi-physical–chemical processes. 

The remainder of the paper is organized as follows. Section~\ref{sec:2} introduces the Boltzmann equation for monatomic gas mixtures and the DSMC method. Section~\ref{sec:macroscopic} derives the macroscopic synthetic equation from the Boltzmann equation and presents high-order closures for the constitutive relations, as well as the momentum and energy exchange terms among different gas species. Section~\ref{sec:DIG} details the implementation of the DIG method to enhance DSMC simulations. Section~\ref{sec:num_shock} validates the DIG method through simulations of the one dimensional force driven Poiseuille flow, and a two-dimensional shock wave passing a cylinder for both Maxwell and hard-sphere gases with mass ratios of 10 and 100. Finally, Section~\ref{sec:conclusion} provides concluding remarks and an outlook.

\section{Mesoscopic description of monatomic gas mixture}\label{sec:2}

In this section, we present an overview of the full collision operator in the Boltzmann equation, and the DSMC for simulating multi-species gas flows. 



\subsection{The Boltzmann equation for multi-species gas flows}

Consider a mixture of $N$-species monatomic gas with the velocity distribution functions $f_s(t,\bm{x},\bm{v})$ of each species describing their mesoscopic states, where $s$ represents the species, $t$ is the time, $\bm{x}$ is the spatial coordinate and $\bm{v}$ is the molecular velocity for $s$ species. The evolution of the distribution function $f_s(t,\bm{x},\bm{v})$ is governed by the Boltzmann equation~\cite{chapman-1990}:
\begin{equation}
    \underbrace{\frac{\partial f_s}{\partial t}+\boldsymbol{v} \cdot \frac{\partial f_s}{\partial \boldsymbol{x}}+\bm{a}_s\cdot\frac{\partial f_s}{\partial \boldsymbol{v}}}_{\mathcal{D} f_s}=\frac{1}{\varepsilon} \underbrace{\sum_{r=1}^N Q_{s r}\left(f_s, f_r\right)}_{Q_s}, \quad s=1,2, \ldots, N,
\label{eq:boltamanneq_full}
\end{equation}
where $\varepsilon$ denotes the Knudsen number, $\bm{a}_s$ represents the external acceleration (suppose it is independent of the molecular velocity),  $\mathcal{D}f_s$ represents the streaming (or transport) of gas molecules, while $Q_s$ is the full Boltzmann collision operator, which can be decomposed into gain and loss terms:
\begin{equation}
\begin{aligned}
Q_{s r}\left(f_s, f_r\right)&=\underbrace{\int_{-\infty}^{+\infty} \int_{4 \pi}\left|\boldsymbol{v}_{*}\right| \sigma_{s r}f_s\left(\boldsymbol{v}^{\prime}\right) f_r\left(\boldsymbol{w}^{\prime}\right) \mathrm{d} \boldsymbol{\Omega} \mathrm{~d} \boldsymbol{w}}_{Q_{sr}^+}
-\underbrace{\nu_{sr}f_s(\boldsymbol{v})}_{Q_{sr}^-},
\end{aligned}
\label{eq:collision_operator}
\end{equation}
with the collision frequency
$\nu_{sr}^-= \int_{-\infty}^{+\infty} \int_{4 \pi}\left|\boldsymbol{v}_{*}\right| \sigma_{s r} f_r(\boldsymbol{w}) \mathrm{d} \boldsymbol{\Omega} \mathrm{~d} \boldsymbol{w}$.  
$Q_{s r}\left(f_s, f_r\right)$ denotes an intra-species collision when $s = r$, and inter-species collision when $s \neq r$. $\sigma_{sr}$ represents the differential cross-section of the binary collision, which depends on the intermolecular potential between the two species. It is normalized by $\pi d^2$, where $d$ is the effective molecular diameter. $\bm{\Omega}$ is the solid angle, while $\bm{v}$ and $\bm{w}$ refer to the pre-collision velocities of species $s$ and $r$, respectively. The relative molecular velocity is $\bm{v}_{*} = \bm{v} - \bm{w}$, which determines the post-collision velocities $\bm{v}$ and $\bm{w}$ for the two species:
\begin{equation}
\begin{aligned}
\boldsymbol{v}^{\prime}=&\boldsymbol{v}-\frac{2 m_s}{m_s+m_r}\left(\boldsymbol{v}_{*} \cdot \boldsymbol{\Omega}\right) \boldsymbol{\Omega}, \quad 
\boldsymbol{w}^{\prime}=\boldsymbol{w}+\frac{2 m_r}{m_s+m_r}\left(\boldsymbol{v}_{*} \cdot \boldsymbol{\Omega}\right) \boldsymbol{\Omega}.
\end{aligned}
\end{equation}

By taking moments of the distribution function, the macroscopic properties of species $s$, i.e., mass density $\rho_s$, velocity $\bm{u}_s$, temperature $T_s$, deviatoric stress $\bm{\sigma}_s$ and heat flux $\bm{q}_s$, can be obtained as:
\begin{equation}
\begin{aligned}
&n_s=\left<1,f_s\right>\quad\rho_s=\left<m_s,f_s\right>,\quad
\rho_s\bm{u}_s=\left<m_s\bm{v},f_s\right>,\quad
\frac{3}{2}\rho_sT_s=\left<m_s\frac{c_s^2}{2},f_s\right>,\\
&\bm{\sigma}_s=\left<m_s\left(\bm{c}_s\bm{c}_s-\frac{c_s^2}{3}\text{\textbf{I}}\right),f_s\right>,\quad\bm{q}_s=\left<m_s\bm{c}_s\frac{c_s^2}{2},f_s\right>,
\end{aligned}
\label{eq:macroscopic_properties_int}
\end{equation}
where the operator $\left<h,\psi\right>=\int h\psi\mathrm{d}\bm{v}$ is defined as the integral of $ h\psi$ over the whole velocity space, \textbf{I} is the $3\times3$ identity matrix, and $\bm{c}_s=\bm{v}-\bm{u}_s$ is the peculiar velocity for species $s$, whose magnitude is $c_s$. The macroscopic properties for the gas mixture, i.e., mass density $\rho$, velocity $\bm{u}$, temperature $T$, pressure tensor $\bm{P}$ and heat flux $\bm{q}$, can be obtained accordingly,
\begin{equation}\label{macro_mixture}
\begin{aligned}
n_s&=\sum_s\left\langle 1, f_s\right\rangle=\sum_s n_s,\quad
\rho=\sum_s\left\langle m_s, f_s\right\rangle=\sum_s \rho_s, \\
\rho \boldsymbol{u}&=\sum_s\left\langle  m_s\boldsymbol{v}, f_s\right\rangle=\sum_s \rho_s \boldsymbol{u}_s, \\
\frac{3}{2}\rho T&=\sum_s\left\langle\frac{1}{2} m_sc^2, f_s\right\rangle=\sum_s \frac{3}{2}\rho_sT_s+\frac{1}{2} \sum_s \rho_s\left|\boldsymbol{u}_s-\boldsymbol{u}\right|^2, \\
\boldsymbol{P}&=\sum_s\left\langle m_s\boldsymbol{c} \boldsymbol{c}, f_s\right\rangle=\sum_s \boldsymbol{P}_s+\sum_s \rho_s\left(\boldsymbol{u}_s-\boldsymbol{u}\right)\left(\boldsymbol{u}_s-\boldsymbol{u}\right), \\
\boldsymbol{q}&=\sum_s\left\langle\frac{1}{2} m_sc^2 \boldsymbol{c}, f_s\right\rangle=\sum_s \boldsymbol{q}_s+\sum_s \frac{3}{2} \rho_s \boldsymbol{T}_s\left(\boldsymbol{u}_s-\boldsymbol{u}\right) \\
&+\frac{1}{2} \sum_s \rho_s\left|\boldsymbol{u}_s-\boldsymbol{u}\right|^2\left(\boldsymbol{u}_s-\boldsymbol{u}\right)+\sum_s \boldsymbol{P}_s \cdot\left(\boldsymbol{u}_s-\boldsymbol{u}\right),
\end{aligned}
\end{equation}
where $\bm{c}=\bm{v}-\bm{u}$ is the peculiar velocity with respect to the velocity for the gas mixture $\bm{u}$. 

Note that all macroscopic properties are normalized by reference quantities: the characteristic number density $n_0$, reference temperature $T_0$, characteristic length $L_0$, and the thermal speed $c_0=\sqrt{k_BT_0/m_0}$, where $k_B$ is the Boltzmann constant and $m_0$ denotes the reference molecular mass. We also define the Knudsen number for species $s$ to characterize the rarefaction level of the gas mixture,
\begin{equation}
    \text{Kn}_s=\frac{\mu_s(T_0)}{n_0L_0}\sqrt{\frac{\pi}{2m_sk_BT_0}},
\label{eq:Knudsen_definition}
\end{equation}
where $m_s$ denotes the molecular mass of species $s$, normalized by the reference mass $m_0$. The quantity $\mu_s$ represents the shear viscosity of species $s$. Under the assumption of an inverse power-law intermolecular potential, the shear viscosity takes the form 
\begin{equation}
    \mu_s(T_s) = \mu_s(T_0)\left(\frac{T_s}{T_0}\right)^\omega,
\end{equation}
where $\omega$ is the viscosity index. In particular, for Maxwell molecules and hard-sphere gases, $\omega$ is taken as 1 and 0.5, respectively. For other gases, $\omega$ usually varies between 0.5 and 1.

\subsection{The DSMC}

The Boltzmann equation is difficult to solve because it is a differential–integro equation defined in a seven-dimensional phase space, with a collision operator given by a five-fold nonlinear integral. 
In the prevailing DSMC method, gas molecules are modeled as a collection of simulation particles, each representing $N_{\text{eff}}$ real molecules, thereby avoiding discretization of the three-dimensional molecular velocity space. The velocity distribution function $f_s$ in a single cell is represented by these simulation particles:
\begin{equation}
    f_s(\bm{v},\bm{x})=\frac{N_{\text{eff}}}{V_{\text{cell}}}\sum_{p=1}^{N_{s,p}}\delta\left(\bm{v}-\bm{v}^{(p)}\right)\delta\left(\bm{x}-\bm{x}^{(p)}\right),
    \label{eq:fs_expression}
\end{equation}
where $\bm{v}^{(p)}$ and $\bm{x}^{(p)}$ denote the velocity and position of the simulation particle $p$ within the cell, and $\delta$ is the Dirac delta function. The cell volume is $V_{\text{cell}}$, and the total number of simulated particles of species $s$ in this cell is $N_{s,p}$. According to Eqs.~\eqref{eq:macroscopic_properties_int} and \eqref{eq:fs_expression}, the macroscopic variables at the $n$-th time step are obtained by taking moments of the distribution function $f_s$:
\begin{equation}
\begin{aligned}
& n_s=\frac{N_{\text {eff }}}{V_{\text {cell }}} N_{s,p},\,\,\rho_s=\frac{m_sN_{\text {eff }}}{V_{\text {cell }}} N_{s,p}, \,\,\bm{u}_s=\frac{1}{N_{s,p}} \sum_{p=1}^{N_{s,p}} \bm{v},\,\, T=\frac{m_s}{3 N_{s,p}} \sum_{p=1}^{N_{s,p}}|\boldsymbol{v}-\boldsymbol{u}_s|^2, \\
& \sigma_{s,i j}=\frac{m_sN_{\text {eff }}}{V_{\text {cell }}} \sum_{p=1}^{N_{s,p}}\left[\left(v_i-u_{s,i}\right)\left(v_j-u_{s,j}\right)-\frac{\delta_{i j}}{3}|\boldsymbol{v}-\boldsymbol{u}_s|^2\right], \\
& q_{s,i}=\frac{m_sN_{\text {eff }}}{2 V_{\text {cell }}} \sum_{p=1}^{N_{s,p}}\left(v_{i}-u_{s,i}\right)|\boldsymbol{v}-\boldsymbol{u}_s|^2,
\end{aligned}
\label{eq:statistic_macroscopic_variable}
\end{equation}
where $\delta_{ij}$ is the Kronecker delta function.

The DSMC employs a splitting scheme that decouples the governing equation into two distinct physical processes~\cite{bird-1994}: the molecular convection part and intermolecular collisions, i.e.,
\begin{equation}
\begin{aligned}
\text{Convection: }\quad& \frac{\partial f_s}{\partial t}+\bm{v} \cdot \frac{\partial f_s}{\partial \bm{x}} + \bm{a}_s\cdot\frac{\partial f_s}{\partial \bm{v}}=0 , \\
\text{Collision: }\quad& \left[\frac{\partial f_s}{\partial t}\right]_{\text {coll }}=\frac{1}{\varepsilon}Q_s.
\end{aligned}
\end{equation}
For the convection part, the velocities of simulation particles remain unchanged, while their positions are modified according as $\bm{x}^p(t+\Delta{t})=\bm{x}^p(t)+\bm{v}^p(t)\Delta{t}+\frac{1}{2}\bm{a}_s\Delta t^2$.
After the convection step, the DSMC computes binary collisions between particles, accounting for both inter-species and intra-species interactions. Follow the Nanbu-Babovsky scheme~\cite{nanbu-1980}, the collision term in the Boltzmann equation for multi-species systems can be reformulated as:
\begin{equation}
\left[\frac{\partial f_s}{\partial t}\right]_{\text{coll}}=\frac{1}{\varepsilon}\left[P_s(f, f)-\beta f_s\right], \text { with } P_s(f, f)=Q_s+\beta f_s ,
\label{eq:nanbu-babovsky1}
\end{equation}
where $\beta$ denotes the upper bound of the total collision frequency, i.e., the coefficient of the loss term $Q_s^-\equiv \beta f_s$. This parameter can be estimated as $n (\left|\bm{v}_*\right|\sigma_{T,sr})_{\text{max}}$, where the maximum value is determined by evaluating all collisional particles within a single computational cell in the DSMC. 
Here, $\sigma_{T,sr}$ refers to the total collision cross-section for the $s$-th and $r$-th species, defined as $\sigma_{T,sr}=\int_{4\pi} \sigma_{sr}d\Omega$.
By applying the forward Euler scheme, the collision operator in Eq.~\eqref{eq:nanbu-babovsky1} can be rewritten as:
\begin{equation}
f_s^{n+1}(\boldsymbol{v} , \boldsymbol{x})=\left(1-\frac{\beta \Delta t}{\varepsilon}\right) f_s^*(\boldsymbol{v} , \boldsymbol{x})+\frac{\beta \Delta t}{\varepsilon} \frac{P_s\left(f^*, f^*\right)}{\beta} .
\label{eq:nanbu_babovsky2}
\end{equation}
That is, the collision process within each time step $\Delta t$ in the DSMC divides the simulated particles in the single cell into two groups: collisionless particles, with a proportion of $(1 - \beta\Delta t/\varepsilon)$, and collisional particles, with a proportion of $\beta\Delta t/\varepsilon$. To ensure the non-negativity of Eq.~\eqref{eq:nanbu_babovsky2}, the time step must satisfy
$\Delta t < {\varepsilon}/{\beta}$.
Consequently, as the Knudsen number approaches zero, the time step $\Delta t$ must be chosen much smaller than the characteristic time scale of macroscopic flows, which significantly increases the computational cost. Additionally, the DSMC requires the cell size to be less than approximately one-third of the local mean free path, imposing a further constraint in the near-continuum regime and exacerbating the computational burden.

To overcome these defects, several stochastic methods, such as the asymptotic accurate and preserving Monte Carlo method~\cite{fei-2023,FEI2025114196}, have been developed. These approaches not only allow for significantly larger time steps, but also enable the use of grid sizes much greater than the local mean free path in the near-continuum regime, thereby improving computational efficiency.  However, due to their explicit time-stepping nature, the evolution of the particle distribution remains slow in practice. In the following sections, we develop the DIG for monatomic gas mixtures, which intermittently couples the explicit DSMC with the implicit solution of macroscopic synthetic equations, thereby significantly accelerating convergence to the steady state~\cite{luo-2024}.

\section{Macroscopic description of monatomic gas mixture}\label{sec:macroscopic}



By taking moments of Eq.~\eqref{eq:boltamanneq_full} with respect to the velocity space, the macroscopic equations for species $s$ can be obtained as,
\begin{equation}
\begin{aligned}
\frac{\partial\rho_s}{\partial t}+\nabla\cdot(\rho_su_s)&=0,\\\frac{\partial}{\partial t}\left(\rho_s\boldsymbol{u}_s\right)+\nabla\cdot\left(\rho_s\boldsymbol{u}_s\boldsymbol{u}_s\right)+\nabla\cdot\left(n_sT_s\mathbf{I}+\boldsymbol{\sigma}_s\right)&=\rho_s\bm{a}_s+\bm{Q}_{s}^{M},\\\frac{\partial}{\partial t}\left(E_s\right)+\nabla\cdot\left(E_s\boldsymbol{u}_s\right)+\nabla\cdot\left[\left(n_sT_s\mathbf{I}+\bm{\sigma}_s\right)\cdot\boldsymbol{u}_s+\boldsymbol{q}_s\right]&=\rho_s\bm{a}_s\cdot \bm{u}_s + Q_{s}^{E},
\end{aligned}
\label{eq:macroscopicequation}
\end{equation}
where $\bm{Q}_s^M=\int m_s\bm{v}Q_s(f_s)d\bm{v}$ and $Q_s^E=\int \frac{1}{2}m_sv^2Q_s(f_s)d\bm{v}$ represents the momentum and energy exchange between species, respectively. Owing to the conservation of total momentum and energy, these exchange terms satisfy the relations $\sum_s \bm{Q}_{s}^{M} = 0$ and $\sum_s Q_{s}^{E} = 0$. Moreover, the kinetic energies $E_{s}$ is defined as $E_s = \frac{3}{2} n_s T_s + \frac{1}{2} \rho_s \bm{u}_s^2$.

The macroscopic equations~\eqref{eq:macroscopicequation} remain unclosed because the shear stress $\bm{\sigma}_s$, heat flux $\bm{q}_s$, and exchange terms are not expressed in terms of the lower-order moments. Their proper modeling is critical to the development of efficient and accurate numerical methods.

\subsection{General constitutive relations for stress and heat flux}

In the continuum flow regime, the stress and heat flux are obtained from the Chapman-Enskog expansion of the Boltzmann equation as~\cite{chapman-1990}:
\begin{equation}
\begin{aligned}
    \boldsymbol{\sigma}_{s}^{\text{NS}} = &
        -\mu_{s} \left[
            \nabla\boldsymbol{u}_{s} + \nabla\boldsymbol{u}_{s}^{\top} - 
            \frac{2}{3}(\nabla \cdot \boldsymbol{u}_{s}) \mathbf{I}
        \right], \\
    \boldsymbol{q}_{s}^{\text{NS}} = &
        -\kappa_{s} \nabla T_{s},  
\end{aligned}
\label{eq:constitutive_relationsNS}
\end{equation}
where $\mu_s$ and $\kappa_s$ are the  shear viscosity and heat conductivity for species $s$, respectively.

Beyond the continuum flow regime, although numerous macroscopic models have been proposed historically, none has successfully provided accurate constitutive relations across the entire range of gas rarefaction~\cite{LeiAiA}. They can only be closed with the aid of numerical simulations. For example,
in the GSIS framework, $\bm{\sigma}_s$ and $\bm{q}_s$ are decomposed into two parts~\cite{zeng-2024}: a linear contribution given by the NS constitutive relations, and the higher-order terms (HoTs) explicitly determined from the velocity distribution function, given by
\begin{equation}
\begin{aligned}
    \boldsymbol{\sigma}_{s} =& \boldsymbol{\sigma}_{s}^{\mathrm{NS}} + 
    \underbrace{
        \int \left( \boldsymbol{c}_{s} \boldsymbol{c}_{s} - \frac{c_{s}^{2}}{3} \mathbf{I} \right) f_{s}^{*} \mathrm{d} \boldsymbol{v} - \boldsymbol{\sigma}_{s}^{\mathrm{NS}*}
    }_{\mathrm{HoT}_{\boldsymbol{\sigma}_{s}}},\\  
    \boldsymbol{q}_{s} =& \boldsymbol{q}_{s}^{\mathrm{NS}} + 
    \underbrace{
        \frac{1}{2} \int \boldsymbol{c}_{s} c_{s}^{2} f_{s}^{*} \mathrm{d} \boldsymbol{v} - \boldsymbol{q}_{s}^{\mathrm{NS}*}
    }_{\mathrm{HoT}_{\boldsymbol{q}_{s}}},
\end{aligned}
\label{eq:constitutive_relations}
\end{equation}
where $f_s^*$ represents the temporal VDF, while $\bm{\sigma}_{s}^{\text{NS}*}$ and $\bm{q}_{s}^{\text{NS}*}$ are computed using the NS constitutive relations based on the macroscopic properties derived from $f_s^*$.

It should be noted that, the HoTs vanish in the continuum limit and become significant in the rarefied regime~\cite{su-2020}. Consequently, the solution of the macroscopic synthetic equations asymptotically recovers the NS equations for continuum flows and approaches the mesoscopic results under highly rarefied conditions, thereby providing an adaptive description across all flow regimes. 
It should also be emphasized that $\bm{\sigma}_{s}^{\text{NS}*}$ and $\bm{\sigma}_{s}^{\text{NS}}$ do not cancel each other out until the steady state is reached. Specifically, $\bm{\sigma}_{s}^{\text{NS}*}$ is extracted from the DSMC simulation in previous time steps, while $\bm{\sigma}_{s}^{\text{NS}}$ is obtained by solving Eq.~\eqref{eq:macroscopicequation} in the next time step; the same applies to the heat flux.
This property enhances numerical stability when solving the synthetic equation with large, or even infinite, time steps, thereby ensuring fast convergence. 


\subsection{Momentum and energy exchange terms }

The inter-species momentum and energy exchange terms, $\bm{Q}_s^M$ and $Q_s^E$ in Eq.~\eqref{eq:macroscopicequation}, can be formally obtained from the DSMC simulation as:
\begin{equation} \label{eq:DSMCsourceterm}
\begin{aligned}
    \bm{Q}_{s,\text{DSMC}}^{M}=& \frac{m_sN_{\text{eff}}}{V_{\text{cell}}}\sum_{p=1}^{N_{s,p}}\frac{\bm{v}'-\bm{v}}{\Delta t},
    \\
    Q_{s,\text{DSMC}}^{E}=& \frac{m_sN_{\text{eff}}}{V_{\text{cell}}}\sum_{p=1}^{N_{s,p}}\frac{v'^{2}-v^2}{2\Delta t}.
\end{aligned}
\end{equation}
For a Maxwell gas, these exchange terms can be written explicitly in terms of lower-order moments, see Eq.~\eqref{eq:modelsourceterm}; for example, the momentum exchange term takes a velocity-relaxation form.  In contrast, for general intermolecular potentials, no exact relation analogous to Eq.~\eqref{eq:modelsourceterm} exists.


Notably, Eq.~\eqref{macro_mixture} indicates that the diffusion velocity, $\bm{u}_s - \bm{u}$, plays a crucial role in determining macroscopic transport properties, including the mixture temperature, shear stress tensor, and heat flux. Consequently, accurately modeling momentum and energy exchanges—expressed explicitly in terms of macroscopic variables—is essential for achieving both rapid convergence and asymptotic-preserving behavior.
Inspired by the construction of constitutive relations for stress and heat flux, we divide the exchange terms into two parts: (i) the exchange terms derived from the model equation~\cite{li2024jfm}, expressed explicitly in terms of macroscopic variables and updated at each iteration when solving the macroscopic equations;
(ii) the difference between the model-derived exchange terms and those directly extracted from the DSMC, which remains constant when solving the macroscopic equations. Specifically, the expression of the exchange terms can be expressed as,
\begin{equation}
\begin{aligned}
    \bm{Q}_s^M =& \bm{Q}_{s,\text{model}}^{M}+\Delta \bm{Q}_{s}^{M*},
    \\
    Q_s^E =& Q_{s,\text{model}}^{E}+\Delta Q_{s}^{E*},
\end{aligned}
\label{eq:sourcetermexpression1}
\end{equation}
where the modeled momentum and energy exchange terms are given in~\ref{Appendix_model}, and the correction terms are given by
\begin{equation}
\begin{aligned}
    \Delta \bm{Q}_{s}^{M*}=& \bm{Q}_{s,\text{DSMC}}^{M}-\bm{Q}_{s,\text{model}}^{M*},
    \\
    \Delta Q_{s}^{E*}=& Q_{s,\text{DSMC}}^{E}-Q_{s,\text{model}}^{E*}.
\end{aligned}
\label{eq:sourcetermexpression1}
\end{equation}
Here, the superscript * indicates that the macroscopic properties are directly obtained from the DSMC.

\section{The DIG for gas mixture}\label{sec:DIG}





\begin{figure}[!t]
\centering
\includegraphics[width=0.85\textwidth]{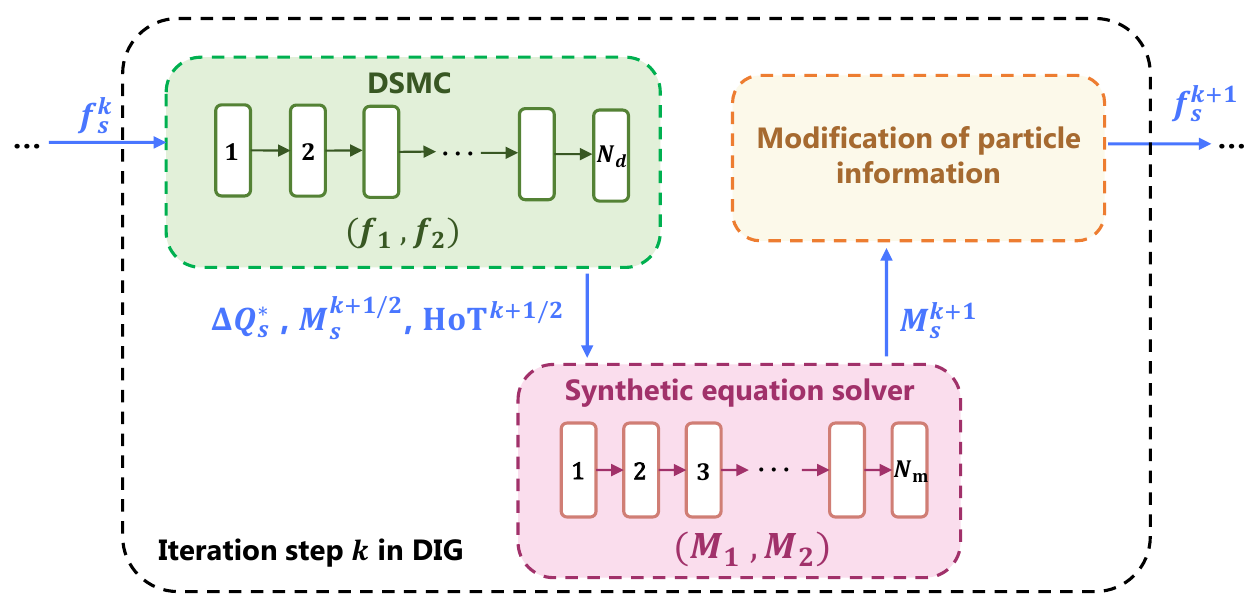}
\caption{Flowchart of the DIG algorithm for two-species monatomic gas mixture. In each iteration step, the DSMC is run for $N_d=100$ steps and the synthetic equations are solved for $N_\text{m}=200\sim1000$ iterations, or until the maximum relative error of the conservative variables among the two species between two successive steps drops below $10^{-5}$. 
}
\label{fig:flowchart}
\end{figure}

The general flowchart of the DIG for simulating gas mixture flows is shown in Fig.~\ref{fig:flowchart}.
Within each unit cycle, the traditional DSMC is first executed for a prescribed number of time steps $N_d$\footnote{The selection of an appropriate $N_d$ is a trade-off between numerical stability and convergence acceleration. 
When $N_d$ is excessively large, the macroscopic solver is employed infrequently, such that the simulation is effectively governed by the DSMC, thereby forfeiting the advantages of rapid convergence and asymptotic-preserving. In contrast, an overly small $N_d$ yields insufficient statistical samples to adequately suppress stochastic fluctuations, which in turn undermines the stability of the macroscopic synthetic equations.
In our previous work~\cite{luo-2024}, the value of $N_d$ is empirically determined to be 50$\sim$100. 
}, during which the time-averaged source terms $\Delta\bm{Q}^*$ in Eq.~\eqref{eq:sourcetermexpression1}, macroscopic properties $\bm{M}_s^{k+1/2}$, and higher-order terms $\textbf{HoT}^{k+1/2}$ in Eq.~\eqref{eq:constitutive_relations} are sampled.
These quantities are then used to solve the macroscopic synthetic equations~\eqref{eq:macroscopicequation} via a time-implicit finite-volume scheme, with boundary conditions imposed following the Grad 13-moment method~\cite{liuw-2024}.
The updated macroscopic properties $\bm{M}_s^{k+1}$ are subsequently employed to adjust the particle information for each species, thereby driving the particle distribution in the flow field closer to the steady state.

The DIG procedure is summarized in Algorithm~\ref{algo:DIG_mixture}, while the detailed sampling strategy and particle modification method, which are respectively critical before and after solving the macroscopic synthetic equation, are described in the following sections.


\begin{algorithm}[!t]
    \caption{Overall algorithm of DIG for monatomic gas mixture} 
    \label{algo:DIG_mixture}
    \begin{algorithmic}[1]
        \Require
            Initial distribution of macroscopic properties $\bm{M}_s$ for species $s$;
        \Ensure
            Time-averaged macroscopic properties $\bm{M}_s$ after steady state;
        \State Draw $N_{s,p}$ simulation particles for species $s$
        within each computational cell according to the initial $\bm{M}_s$ and Maxwellian distribution;
        \State Run standard DSMC for $n_a$ time steps to obtain sufficient samples;
        \State Set iteration step $k = 1$;
        \While {$k \le \text{MaxSteps}$}
            \State Run standard DSMC for $N_d$ time steps;
            \State Extract $\bm{Q}_{s,\text{DSMC}}^{M}$ and $Q_{s,\text{DSMC}}^{E}$ from the DSMC according to Eq.~\eqref{eq:DSMCsourceterm};
            \State Calculate the time-averaged macroscopic properties $\bm{M}^{k+1/2}_s$ based on Eq.~\eqref{eq:EWMTA};
            \State Obtain the higher-order terms $\text{HoT}^{k+1/2}$ based on Eq.~\eqref{eq:highorderterms_DSMC};
            \State Calculate the correction of the exchange terms $\Delta \bm{Q}_s^{M*}$ and $\Delta Q_{s}^{E*}$ in Eq.~\eqref{eq:sourcetermexpression1}.
            \State Solve Eqs.~\eqref{eq:macroscopicequation}, \eqref{eq:constitutive_relations} and \eqref{eq:sourcetermexpression1} by $N_\text{m}$ steps to obtain $\bm{M}^{k+1}_s$;
            \State Replicating and discarding particles to match species mass density $\rho_s^{k+1}$;
            \State Particle velocity scaling to match species velocity $\bm{u}_s^{k+1}$ and temperature $T_s^{k+1}$;
            \State $k ++$;
        \EndWhile
    \end{algorithmic}
\end{algorithm}

\subsection{Time-averaged sampling method}

Due to the significant fluctuations in DSMC,  macroscopic properties sampled from a single step by Eq.~\eqref{eq:statistic_macroscopic_variable} cannot be directly used in solving the macroscopic synthetic equations. Therefore, a time-averaged sampling technique is required. Here we adopt the exponentially weighted moving time average method to reduce thermal fluctuations~\cite{Jenny2010JCP}. That is, the summation variables for macroscopic properties of $s$ species $\bm{\Xi}_s=\{n_s,\rho_s,\rho_s\bm{u}_s,\rho T_s,\bm{\sigma}_{s},\bm{q}_{s}\}$ in Eq.~\eqref{eq:statistic_macroscopic_variable} is written as,
\begin{equation}
\begin{aligned}
    &\bm{\Xi}_s(t)=\frac{n_a-1}{n_a}\bm{\Xi}_s(t-\Delta t)+\frac{1}{n_a}\frac{m_sN_{\text{eff}}}{V_{\text{cell}}}\sum_{p=1}^{N_{s,p}}\bm{\zeta}_s(t),\\
    &\bm{\zeta}_s=\left\{\frac{1}{m_s},1,\bm{v},\frac{c_s^2}{3},\bm{c}_s\bm{c}_s-\frac{c_s^2}{3}\text{\textbf{I}},\bm{c}_sc_s^2\right\},
\end{aligned}
    \label{eq:EWMTA}
\end{equation}
where $n_a$ is the number of sampling steps applied in the time-averaging process. Meanwhile, HoTs in Eq.~\eqref{eq:constitutive_relations} are directly constructed according to the definition of shear stress and heat flux:
\begin{equation}
\begin{aligned}
    \mathrm{HoT}_{\boldsymbol{\sigma}_s} = &\boldsymbol{\sigma}^{\mathrm{DSMC}}_s - \boldsymbol{\sigma}^{\mathrm{NS}*}_s, \\
    \mathrm{HoT}_{\boldsymbol{q}_s} = &\boldsymbol{q}_{s}^{\mathrm{DSMC}} - \boldsymbol{q}_{s}^{\mathrm{NS}*},
    \end{aligned}
    \label{eq:highorderterms_DSMC}
\end{equation}
where $\boldsymbol{\sigma}^{\mathrm{DSMC}}_s$ and $\boldsymbol{q}_{s}^{\mathrm{DSMC}}$ are the time-averaged results sampled from the DSMC via Eq.~\eqref{eq:EWMTA}, whereas $\boldsymbol{\sigma}^{\mathrm{NS}}_s$ and $\boldsymbol{q}_{s}^{\mathrm{NS}}$ are determined by Eq.~\eqref{eq:constitutive_relationsNS} using the time-averaged macroscopic variables $\rho_s$, $\bm{u}_s$, and $T_s$, likewise evaluated through Eq.~\eqref{eq:EWMTA}.

Moreover, since the exchange terms in Eq.~\eqref{eq:DSMCsourceterm} are calculated based on the time step $\Delta t$, excessively small $\Delta t$ results in significant fluctuations, which may cause numerical instabilities when solving the macroscopic synthetic equations. Therefore, in practice, the exchange terms $\bm{Q}_{s,\text{DSMC}}^M$ and $Q_{s,\text{DSMC}}^E$ are taken as time-averaged values sampled from the beginning of the simulation. 

\subsection{Modification of particle distribution for mixture gas}

The macroscopic synthetic equation is solved by the finite-volume method in~\ref{Finite_volume_synthetic}. After that, the updated macroscopic properties of species $s$, $\bm{M}_s^{n+1} = \left\{\rho_s^{n+1}, \bm{u}_s^{n+1}, T_s^{n+1}\right\}$, are fed back to DSMC to skip unnecessary intermediate time evolutions.
This is realized in the following two steps.

Firstly, we determine the target number of $s$ species particles, based on the updated mass density $\rho_s^{n+1}$:
\begin{equation}    N_{s,p}^{n+1}=\text{Iround}\left(\frac{\rho_s^{n+1}V_{\text{cell}}}{m_sN_{\text{eff}}}\right),\,\,\text{Iround}(x)=\begin{cases} \left\lfloor x \right\rfloor+1, & \text{with probability } x-\left\lfloor x \right\rfloor, \\ \left\lfloor x \right\rfloor, &  \text{with probability } 1-x+\left\lfloor x \right\rfloor ,\end{cases}
\label{eq:particlenumber_modification}
\end{equation}
where $\lfloor x\rfloor$ is the integer part of $x$. If $N_{s,p}^{n+1} > N_{s,p}^n$, the difference $N_{s,p}^{n+1} - N_{s,p}^n$ is compensated by creating new particles. Their velocities are assigned by randomly replicating existing $s$ species particles in the cell, while their positions are uniformly distributed within the cell. Conversely, if $N_{s,p}^{n+1} < N_{s,p}^n$, $N_{s,p}^{n} - N_{s,p}^{n+1}$ particles are randomly selected and deleted from the cell.

Secondly, we calculate the temporary velocity $\bm{u}^*_s$ and temperature $T_s^*$ for $s$ species in the cell by sampling the current particle distribution according to Eq.~\eqref{eq:statistic_macroscopic_variable}. Then, the velocities of simulated particles for $s$ species are scaled via a linear transformation,
\begin{equation}
    \bm{v}^{n+1}=\sqrt{\frac{T^{n+1}_s}{T_s^*}} \left( \bm{v}^n -\bm{u}^{*}_s \right) 
    +\bm{u}^{n+1}_s.
\end{equation}
Overall, the replicating and deleting procedures serve to adjust the particle count, ensuring consistency with the updated mass density obtained from the macroscopic synthetic equations. Subsequently, the velocity scaling operation corrects the particle velocities so that the mean flow velocity and temperature align with the updated macroscopic properties. Since the macroscopic synthetic equations are formulated to preserve mass, momentum, and energy, these conservation laws remain satisfied when the particle information is fully adjusted.

\section{Numerical results}\label{sec:num_shock}

To assess the performance of the proposed DIG for monatomic gas mixtures, numerical simulations are carried out for hypersonic flows over a cylinder involving different types of gas mixtures summarized in Table~\ref{tab:mixture}. Molecular collisions in DSMC are controlled by the parameter $\alpha$: for Maxwell gas mixture, $\alpha=2.14$, while for hard-sphere gas mixture, $\alpha=1.0$. The light species in each mixture (denoted as species 1) are used as the reference, with characteristic parameters given as: reference molecular mass $m_0 = 1.66 \times 10^{-27}$ kg, reference temperature $T_0 = 273.15$ K, and reference collision diameter $d_0 = 3 \times 10^{-10}$ m.

As discussed in previous sections, the DIG for monatomic gas mixtures requires sufficient statistical samples to mitigate fluctuations in the macroscopic properties. In the present study, $n_a = 1000$ samples are used, and the two-fluid macroscopic synthetic equations are solved every $N_d = 100$ DSMC steps. The DSMC solutions performed by Stochastic PArallel Rarefied-gas Time-accurate Analyzer (SPARTA, \url{https://sparta.sandia.gov/}) will be used for benchmarking. All simulations are done by Intel(R) Xeon(R) Gold 6148 CPU @ 2.40GHz processor.

In order to increase the numerical instability, the shear viscosity and heat conductivity in Eq.~\eqref{eq:constitutive_relationsNS} are chosen as~\cite{zeng-2024},
\begin{equation}
\begin{aligned}
\mu_{s}^{*}&=\Theta_{s,\mu}n_{s}T_{s}\tau_{ss}\frac{n_{s}}{\sum_{r}n_{s}\phi_{sr}},\quad\Theta_{s,\mu}=\frac{1}{\mathrm{Kn}_{s}}\mathrm{min}\left(\mathrm{Kn}_{s},\frac{1}{n}\sum_{r}n_{s}\phi_{sr}\right),\\
\kappa_{s}^{*}&=\Theta_{s,\kappa}\frac{5n_{s}T_{s}\tau_{ss}}{2m_{s}\Pr}\frac{n_{s}}{\sum_{r}n_{s}\phi_{sr}\varphi_{sr}},\quad\Theta_{s,\kappa}=\frac{1}{\mathrm{Kn}_{s}}\mathrm{min}\left(\mathrm{Kn}_{s},\frac{1}{n}\sum_{r}n_{s}\phi_{sr}\varphi_{sr}\right),
\end{aligned}
\label{eq:viscosity_kappa}
\end{equation}
where the parameter $\varphi_{sr}$ is the ratio of thermal relaxation rates between the inter- and intra-species collisions, and is varied for different types of gas mixture.
Since the model parameters are inherently linked to the transport properties, they are specified for different gas mixture accordingly, see Table~\ref{tab:mixture}. 

\subsection{Force-driven Poiseuille flow}

We simulate the one-dimensional force-driven Poiseuille flow to assess the accuracy and efficiency of the DIG method. A type-1 gas mixture is confined between two parallel plates located at $x_L=0$ and $x_R=L_0$, both maintained at a uniform wall temperature $T_0$. Initially, the two species have equal number densities, i.e., $n_1=n_2=0.5n_0$. Three Knudsen numbers are considered: $\text{Kn}_1=0.1$, $0.01$, and 0.001, corresponding to plate separations of $L_0=10\lambda_0$,  $100\lambda_0$, and $1000\lambda_0$, respectively. Uniform external forces in the $y$-direction are applied to drive the flow,  and the acceleration $a_{s,y}$ of each species is given by
\begin{equation}
    a_{1,y}=a_{2,y}=\frac{1}{m_1n_1+m_2n_2}\frac{\partial P_e}{\partial y},
\end{equation}
where the effective pressure gradient $\partial P_e/\partial y$ is $4\times 10^{10}$, $4\times 10^{8}$ and $4\times 10^{6}$ Pa·m$^{-1}$ for $\text{Kn}_1=0.1$, $0.01$ and $0.001$, respectively.

The DSMC time step is chosen as the ratio of the smallest mean free path to the most probable molecular speed. The initial number of simulated particles is set to $N_{s,p}=500$ per cell for each species, and their velocities are sampled from a Maxwellian distribution corresponding to a number density of $0.5n_0$, unit temperature, and zero velocity. Moreover, the non-uniform grid in the x-direction is generated by a symmetric hyperbolic–tangent stretching,
\begin{equation}\label{grid_nonuniform}
\frac{x_i}{L_0} = \frac{1}{2} + \frac{\tanh\!\left[ \vartheta \left( \tfrac{2i}{N_x-1} - 1 \right) \right]}{2\tanh(\vartheta)},
\quad i=0,1,\ldots,N_x-1 ,
\end{equation}
where $N_x=41$ grid points (40 cells) and $\vartheta=2.47$. This yields a minimum cell size of $\Delta x_{\min}=0.002$ adjacent to the walls, in order to capture the Knudsen layers. The maximum cell size is 0.0623, located at the center of the computational domain.

\begin{figure}[!t]
\centering
\vspace{-1.5mm}
\includegraphics[width=0.4\textwidth,trim=10pt 10pt 10pt 10pt,clip]{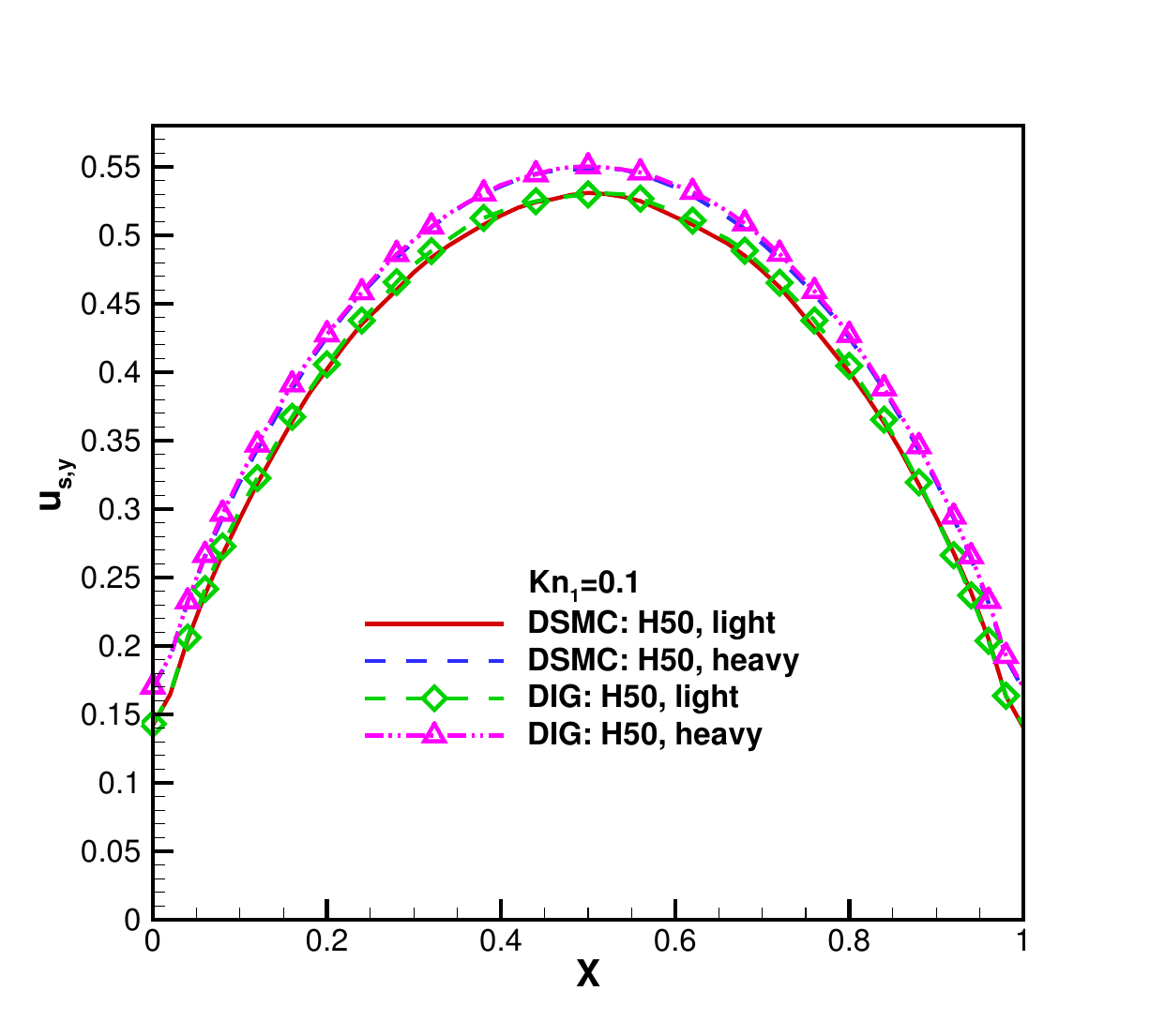}
\\
\vspace{-0.2cm}
\includegraphics[width=0.4\textwidth,trim=10pt 10pt 10pt 10pt,clip]{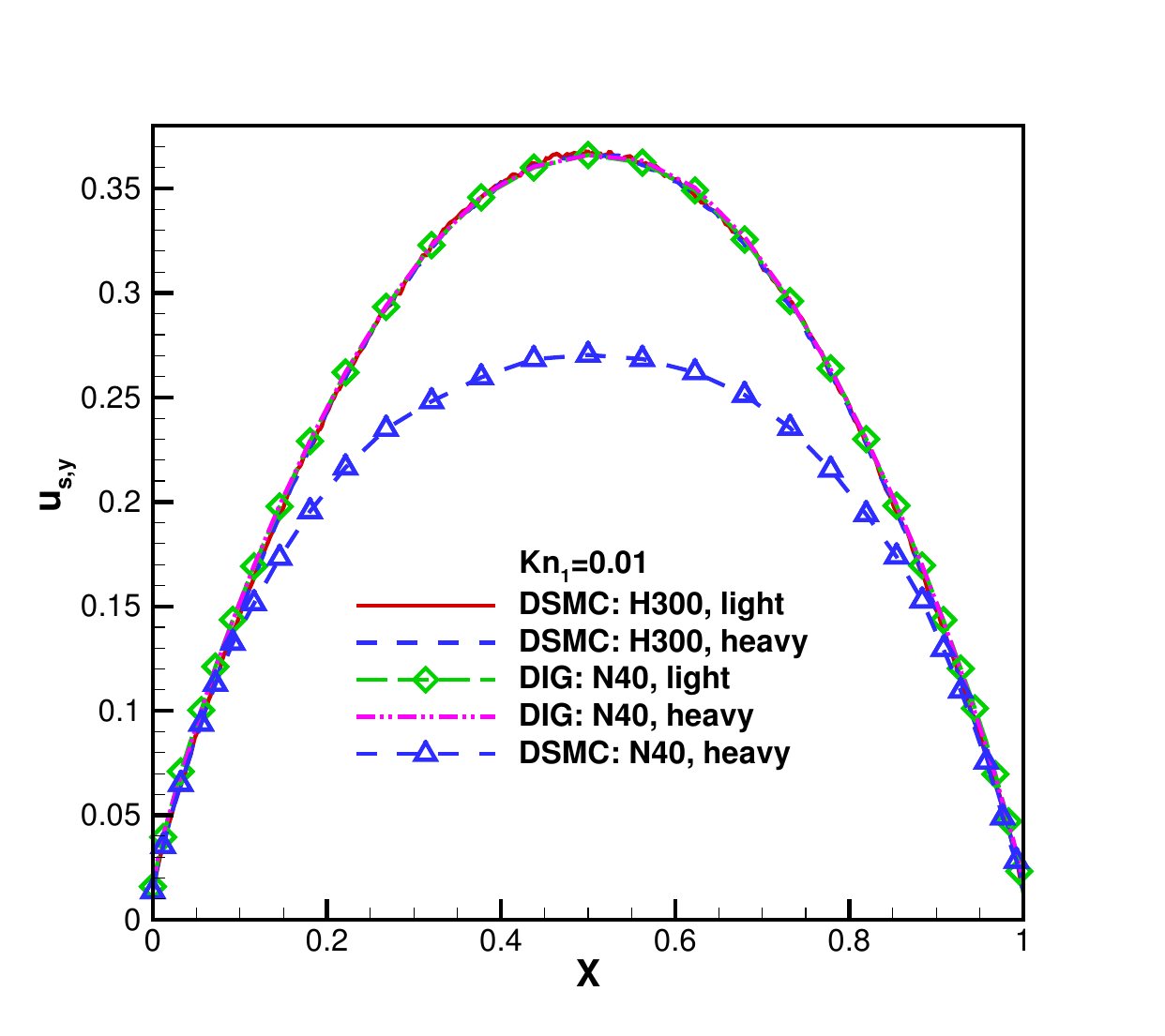}
\includegraphics[width=0.4\textwidth,trim=10pt 10pt 10pt 10pt,clip]{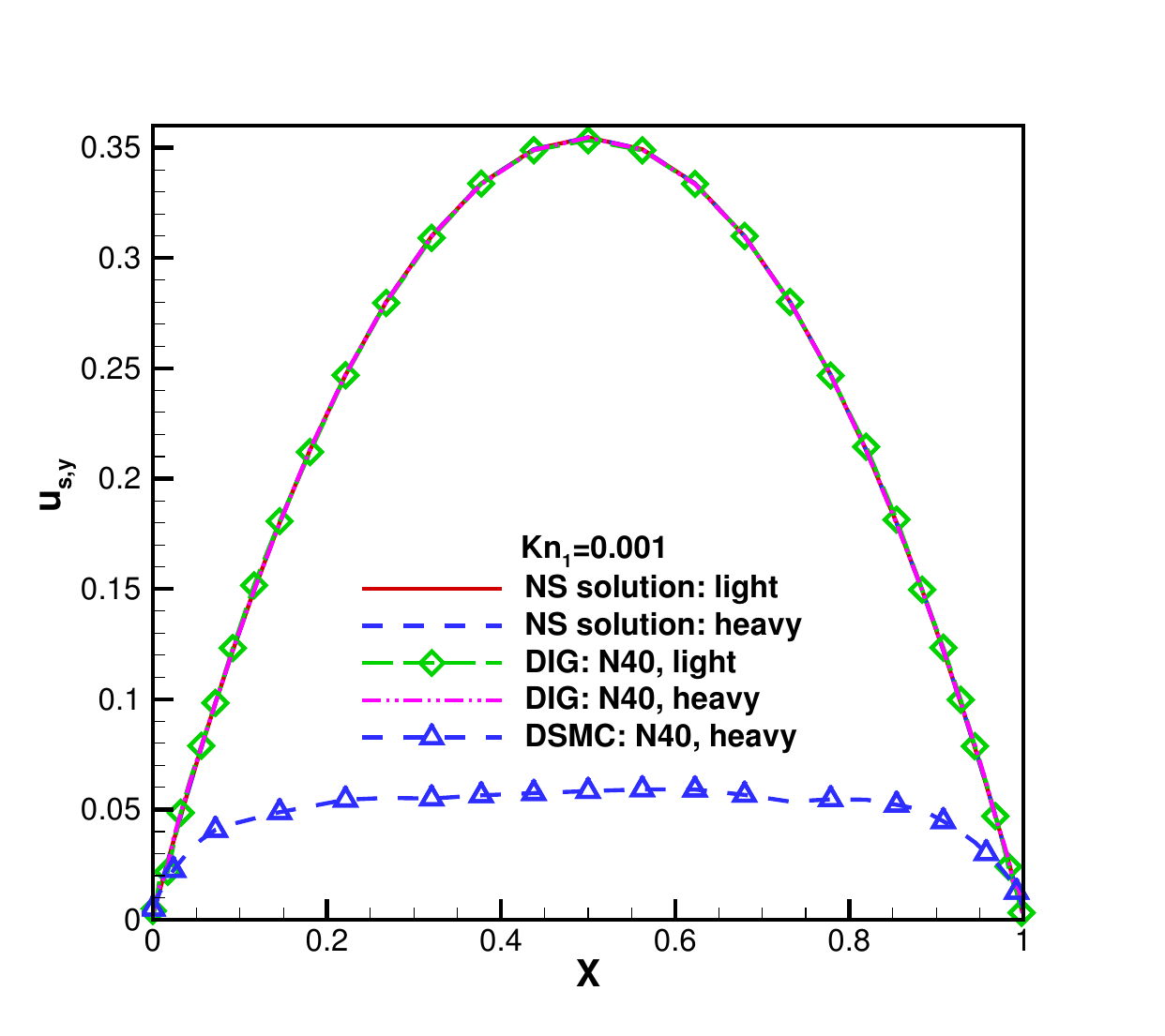}
\caption{Velocity profiles for 1D force-driven Poiseuille flow of type-1 mixtures with a mass ratio of 10, when $\text{Kn}_1=0.1$, 0.01, and 0.001. H300 represents the 300 homogeneous grids, while N40 represents the nonuniform grids generated by Eq.~\eqref{grid_nonuniform}. Note that when $\text{Kn}_1=0.001$, the traditional DSMC is very slow, so that the NS solutions are used to validate the DIG simulations.
}
\label{fig:Poiseuille1D}
\end{figure}

Figure~\ref{fig:Poiseuille1D} presents the velocity profiles of the Poiseuille flow predicted by the DSMC and DIG. For $\text{Kn}_1=0.1$, although the external acceleration applied to both species is identical, noticeable differences appear between the velocity profiles of the light and heavy species, due to the rarefaction effects which go beyond the NS descriptions. 
In this case, 50 uniform spatial grids are employed, with a grid size smaller than the mean free path; therefore, the DIG and DSMC produce identical velocity profiles. When $\text{Kn}_1=0.01$ and 0.001, the flow enters the near-continuum regime, where frequent inter-species collisions drive the system toward global equilibrium, resulting in identical flow velocities for both species. For the traditional DSMC, obtaining the velocity profile in this regime is challenging because: i) the spatial cell size must be about one third of the mean free path, requiring 300 and 3000 uniform cells, respectively; ii) the time step is about one third of the mean collision time, so an enormous number of iterations is needed both to reach the steady state and to sample the velocity profiles thereafter. Therefore, using only 40 non-uniform spatial grids leads to an underestimation of the flow velocity, and a comparison of velocity profiles at $\text{Kn}_1 = 0.01$ and $0.001$ shows that smaller Knudsen numbers correspond to stronger numerical dissipation in the traditional DSMC.
However, owing to the macroscopic guidance incorporated in DIG, the method possesses the asymptotic-preserving property. As a result, correct velocity profiles can be obtained with only 40 non-uniform grids when $\text{Kn}_1=0.01$ and $0.001$, where the maximum cell size is approximately 6 and 60 times larger than the mean free path.

In addition to its asymptotic-preserving property, which allows the use of far fewer spatial grids in near-continuum flow regimes, the DIG also exhibits fast convergence, significantly reducing the required evolution steps.  For example, at $\text{Kn}_1 = 0.01$, the reference DSMC simulation required nearly 300,000 iterations and 4.8 core hours to achieve steady state, whereas the DIG method achieved convergence in only 3,000 steps and 0.025 core hours (all DSMC and DIG results are obtained using our in-house solver running on 10 CPU cores); this corresponding to approximately 100- and 200-fold reductions in iteration step and computational time, respectively. The smaller the Knudsen number, the greater the reduction in the computational time of DIG compared to the traditional DSMC.



\subsection{Hypersonic flows passing over a cylinder}


We further evaluate the performance of DIG in two-dimensional hypersonic flow over a cylinder. Both Maxwell and hard-sphere gases are considered, with mass ratios of 10 and 100.

\subsubsection{Maxwell gas}

\begin{figure}[!t]
\centering
\includegraphics[width=0.4\textwidth,trim=20pt 20pt 50pt 50pt,clip]{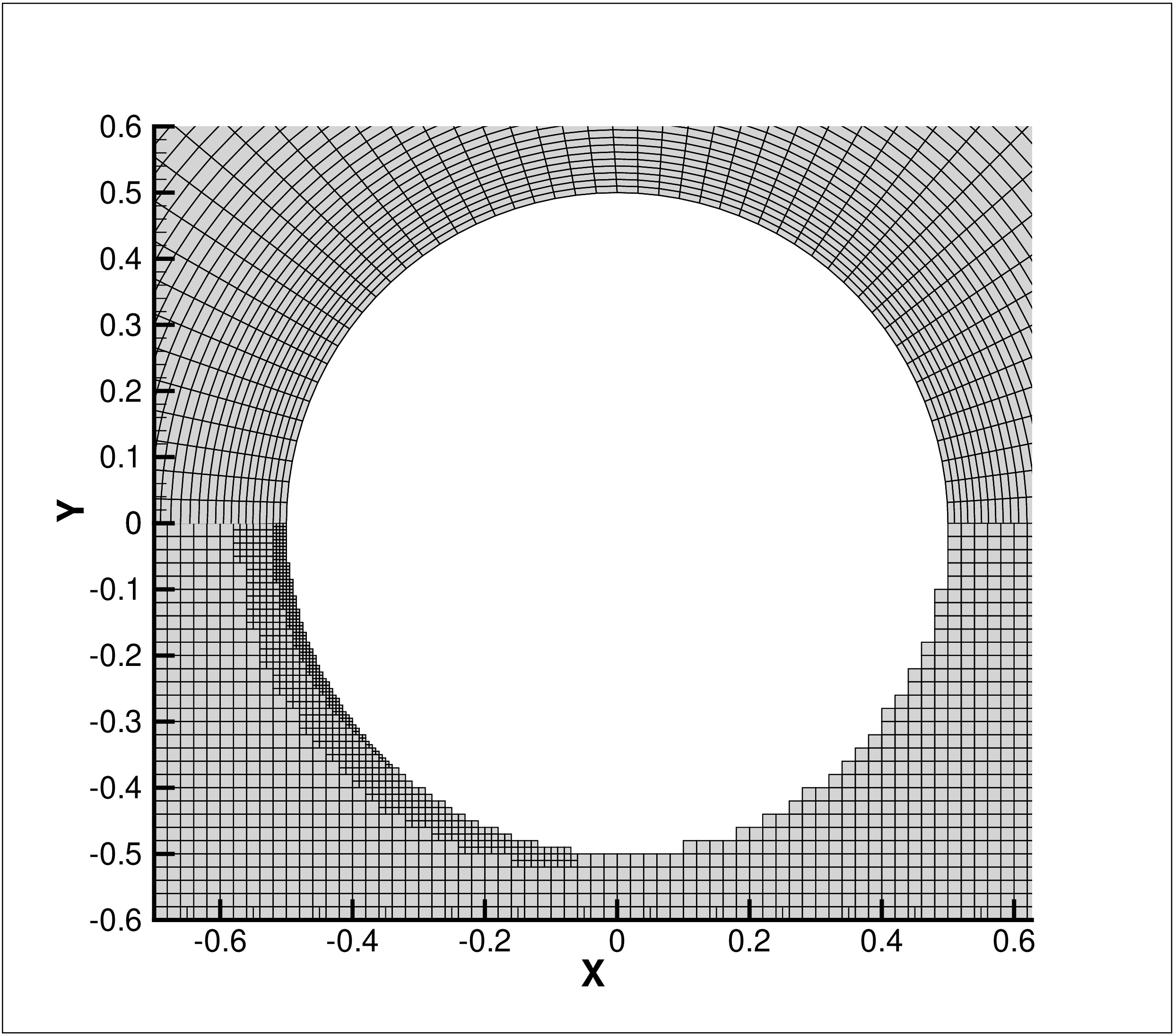}
\includegraphics[width=0.4\textwidth,trim=20pt 20pt 50pt 50pt,clip]{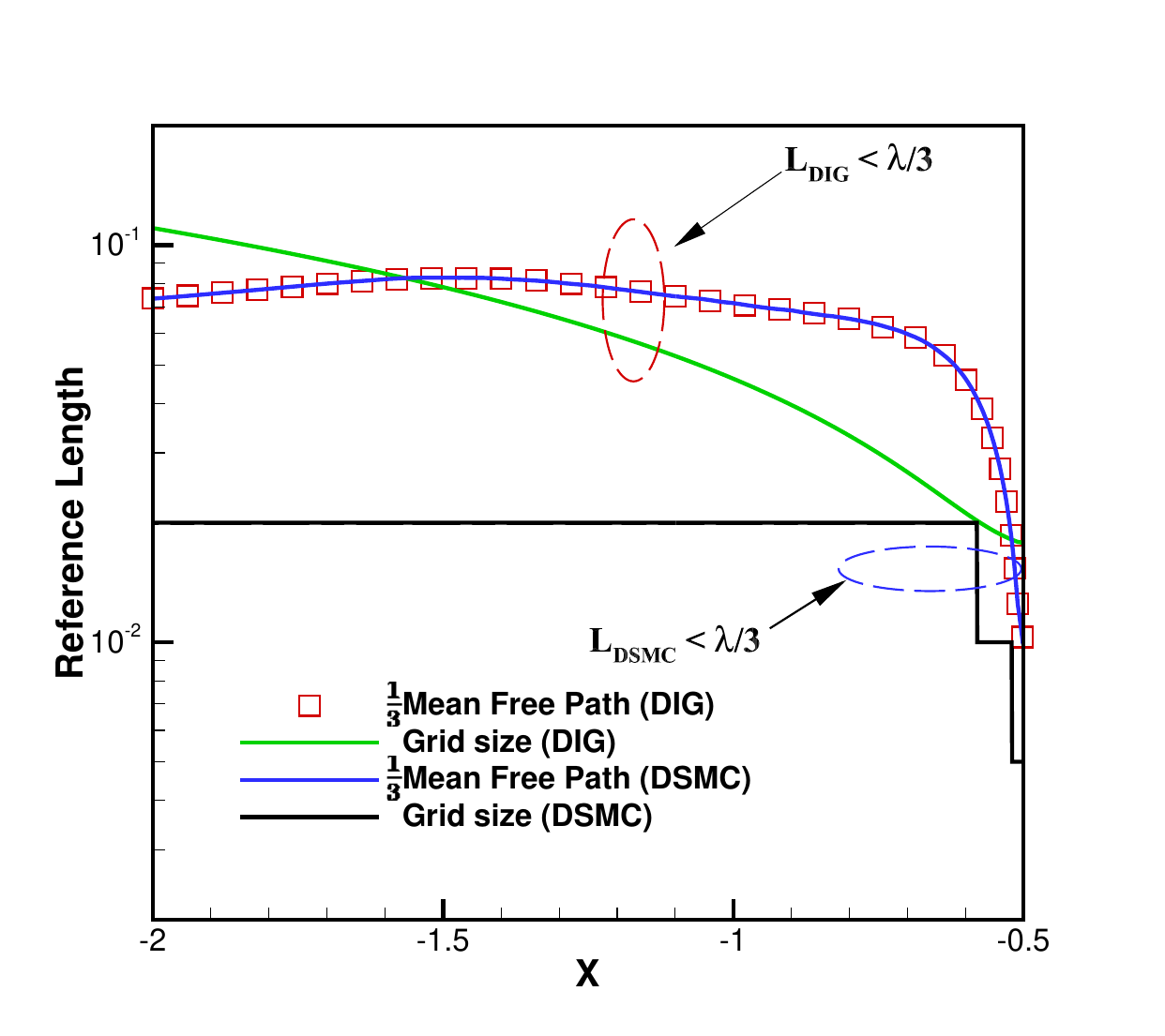}\\
\vspace{0.4cm}
\includegraphics[width=0.4\textwidth,trim=20pt 20pt 50pt 50pt,clip]{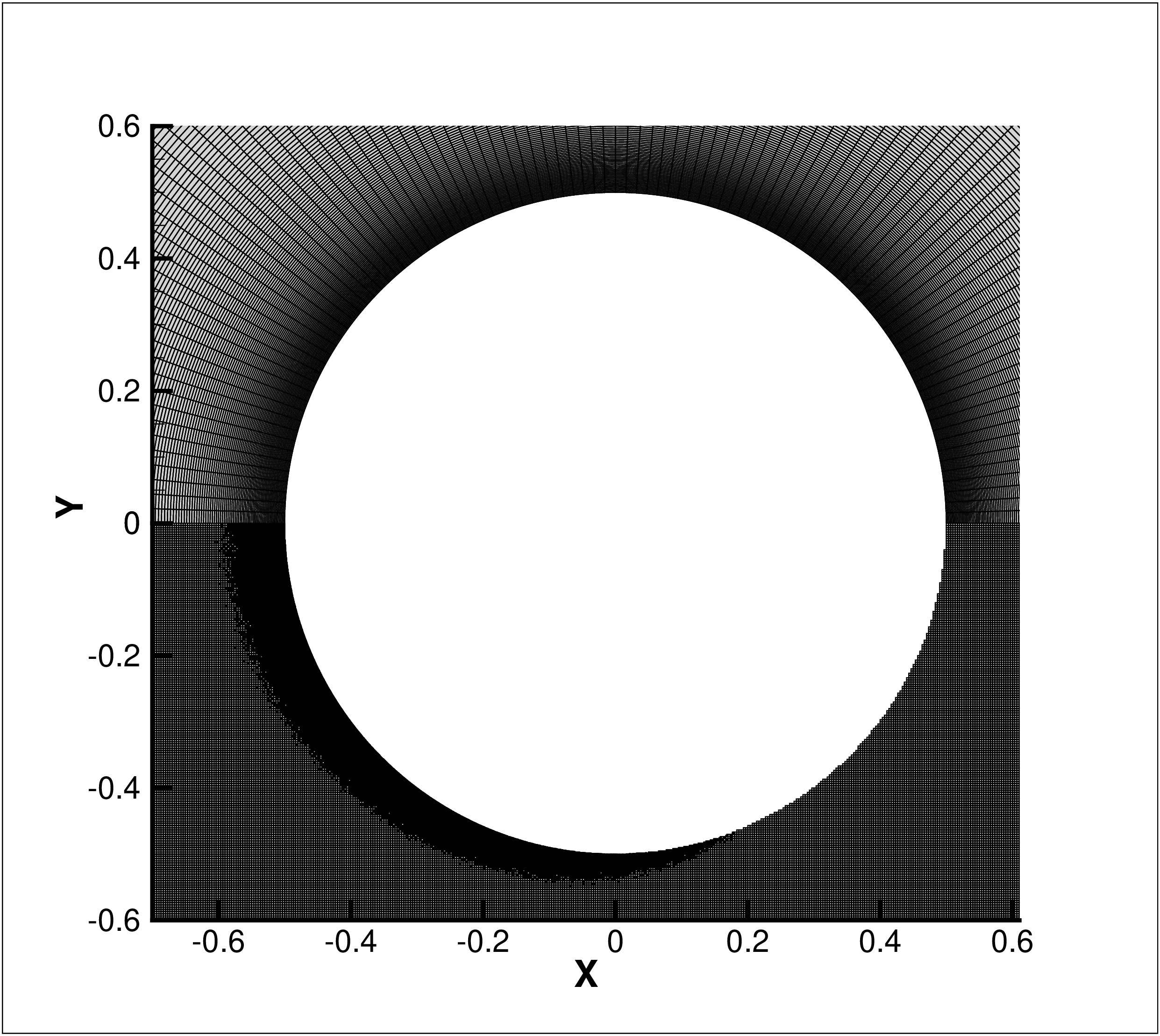}
\includegraphics[width=0.4\textwidth,trim=20pt 20pt 50pt 50pt,clip]{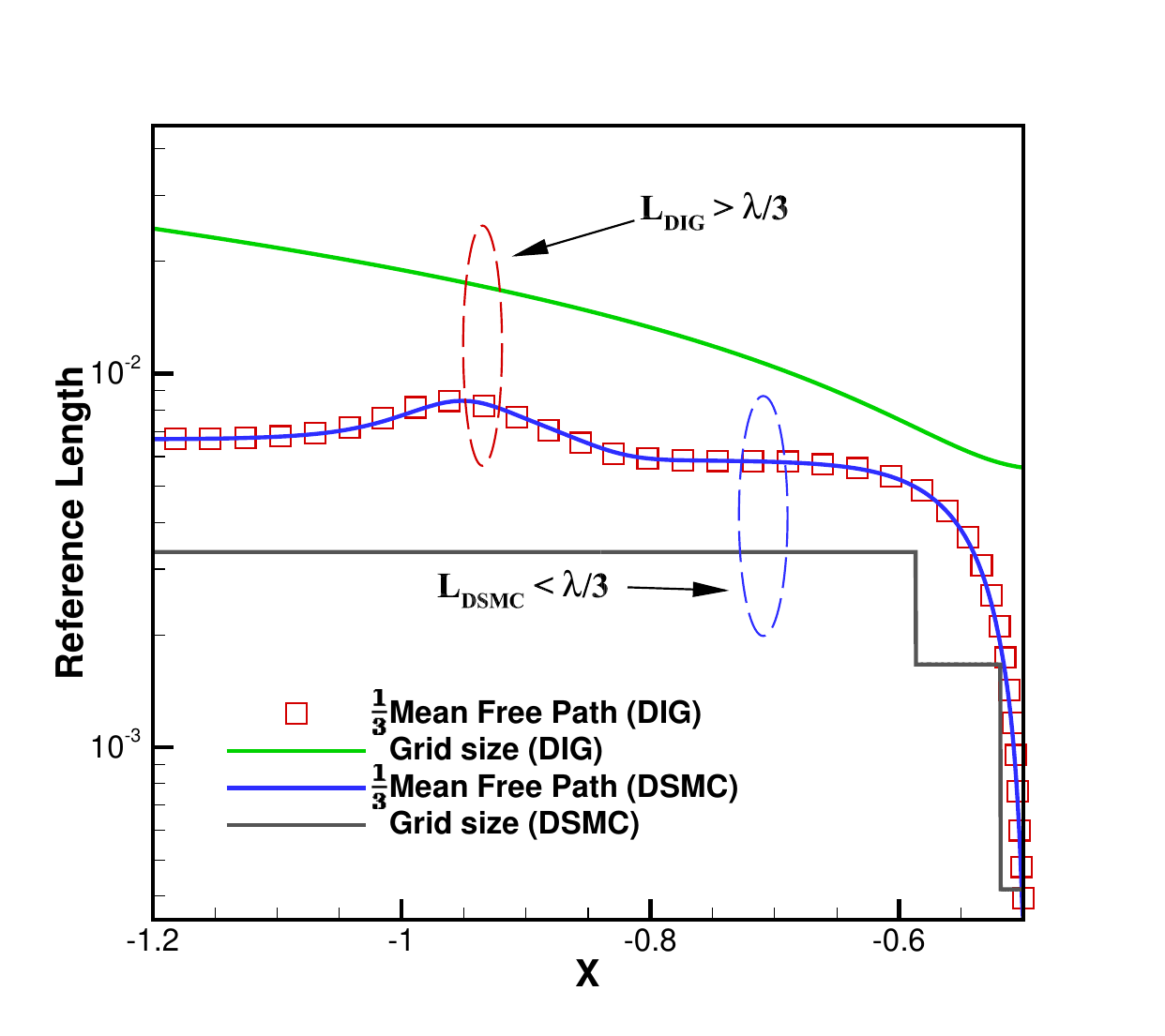}
\caption{Left column: the computational grids  in DSMC (lower half) and DIG (upper half). Right column: the comparison of mean free path and grid size in DIG and DSMC. The global Knudsen number is 0.1 and 0.01 for the results in top and bottom rows, respectively. Note that the mean free path and grid size are normalized by the characteristic length $L_0$. The grid size in DSMC is adaptively refined and strictly lower than one-third of the local mean free path for the heavy species. The time step $\Delta t$, multiplied by the most probable speed, is carefully constrained to remain smaller than the length of the smallest grid.}
\label{fig:meshcompare_Mr10_Maxwell}
\end{figure}

The two-dimensional supersonic flows with the Mach number of $\text{Ma} = 5$ and 10 passing over a cylinder are simulated for mixture types 1 and 2, under Knudsen numbers $\text{Kn}_1 = 0.1$ and $0.01$. The freestream and wall temperatures, $T_\infty$ and $T_w$, are both set to the reference temperature $T_0$. The freestream Mach number is determined based on the mixture's speed of sound, given by $a_\infty = \sqrt{\gamma k_B T_\infty / m_{\text{mix}}}$, where the mixture molecular mass is defined as $m_{\text{mix}} = n_{1,\infty} m_1 + n_{2,\infty} m_2$ with the equimolar composition $n_{1,\infty} = n_{2,\infty}=0.5$. Additionally, the initial number of simulated particles in each cell is set to $N_{s,p} = 100$ per species to ensure a balance between computational efficiency and numerical stability.

The computational domain is an annular region with the inner boundary representing the cylinder surface. The reference length $L_0$ is taken as the cylinder diameter, i.e., $L_0 = D$. For DIG, when $\text{Kn}=0.1$, the outer radius of the domain is set to $5.5L_0$ for mixture type 1 and $10L_0$ for mixture type 2. For DSMC, when $\text{Kn}=0.1$, the domain is defined as a square region $[-L, L] \times [-L, L]$, with $L / L_0 = 5.5$ and 10 for mixture types 1 and 2, respectively. For $\text{Kn}=0.01$, the square domain is set to $L/L_0 = 3$ for both mixture types. The fully diffuse reflection boundary condition is applied at the isothermal cylinder surface. The entire simulation domain is initially set to an equilibrium state consistent with the freestream properties.

\begin{figure}[!t]
\centering
\vspace{-1.5mm}
\hspace{-11mm}
\includegraphics[width=0.38\textwidth,trim=10pt 10pt 10pt 10pt,clip]{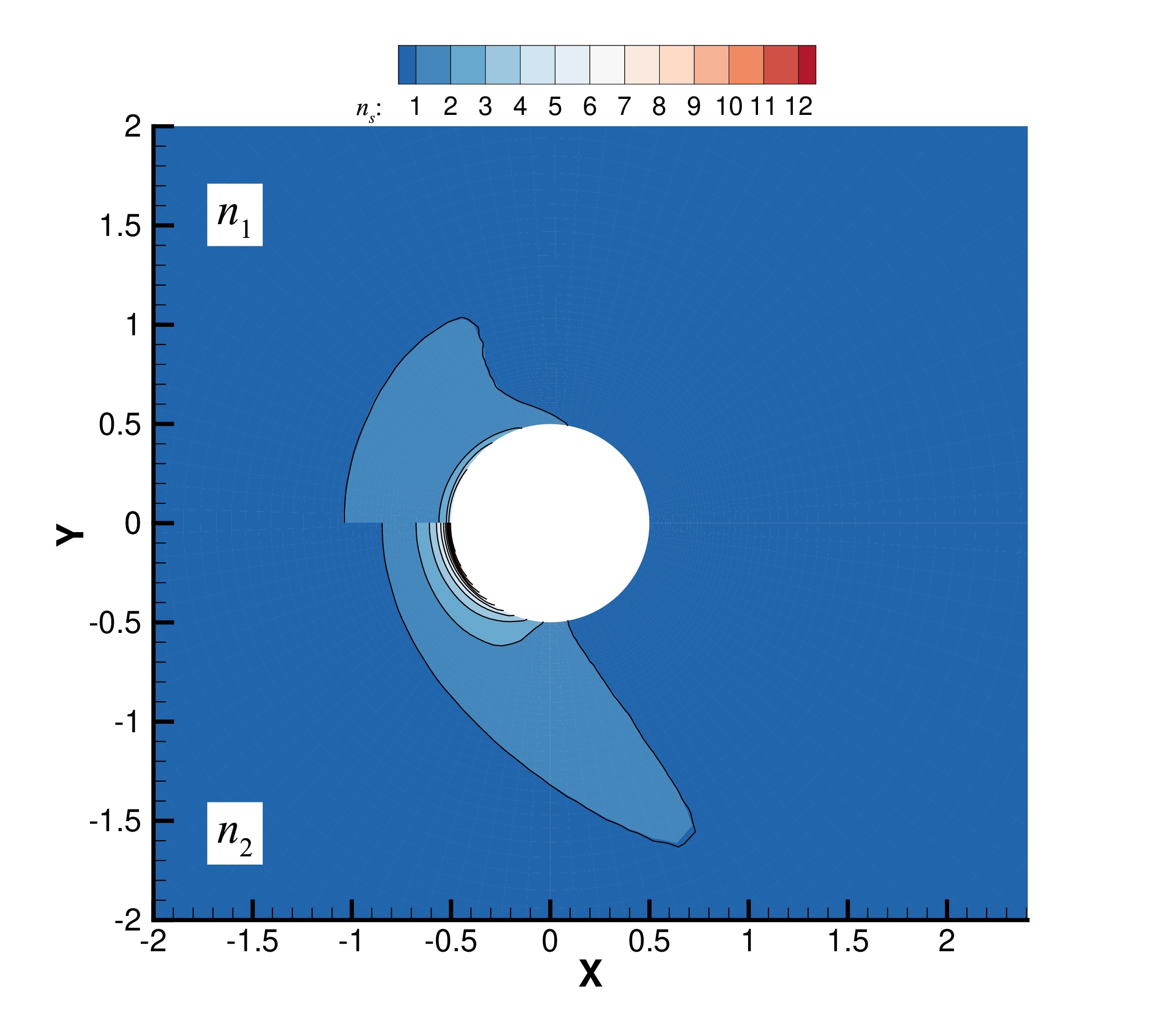}
\hspace{-11mm}
\includegraphics[width=0.38\textwidth,trim=10pt 10pt 10pt 10pt,clip]{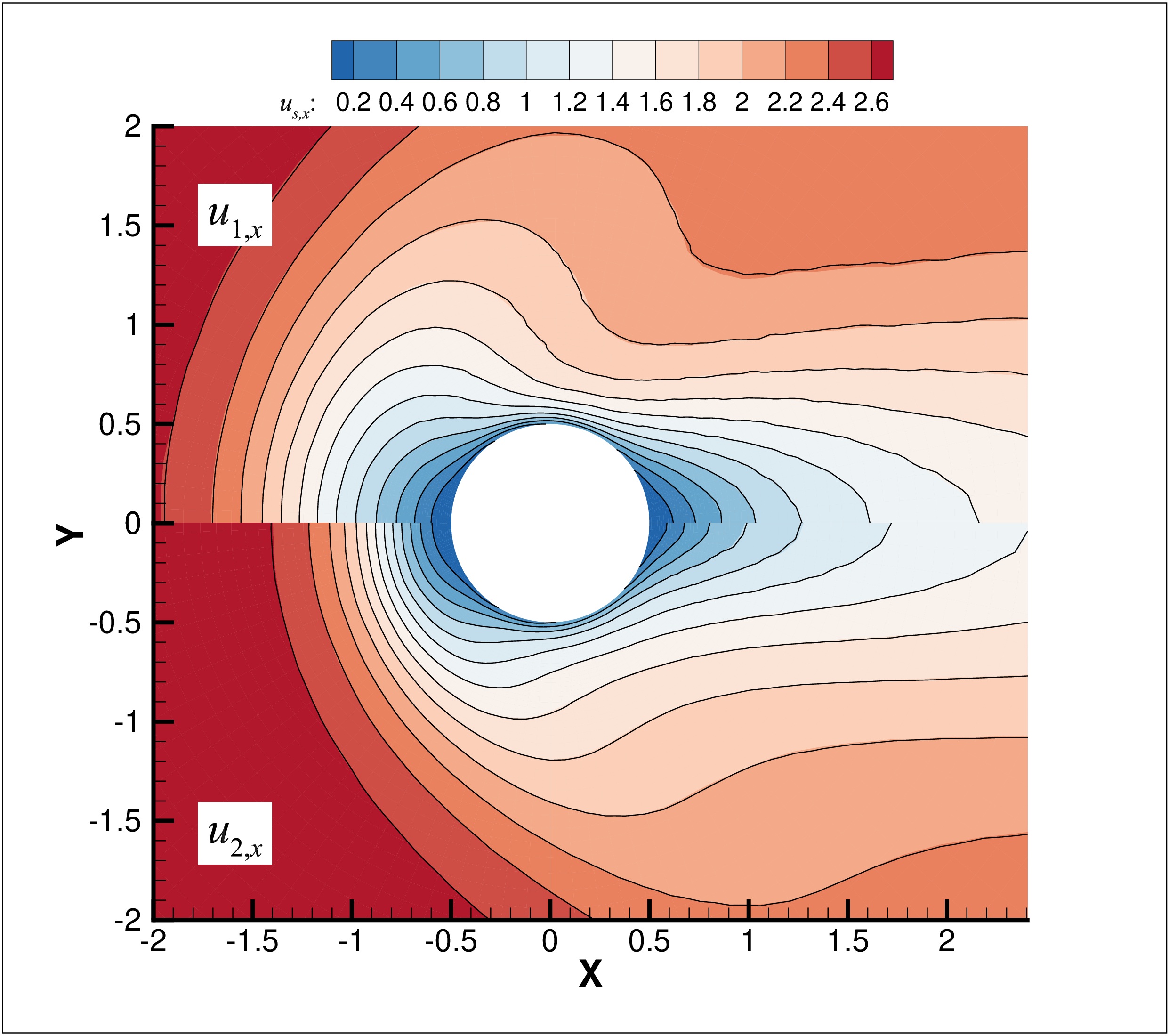}
\hspace{-11mm}
\includegraphics[width=0.38\textwidth,trim=10pt 10pt 10pt 10pt,clip]{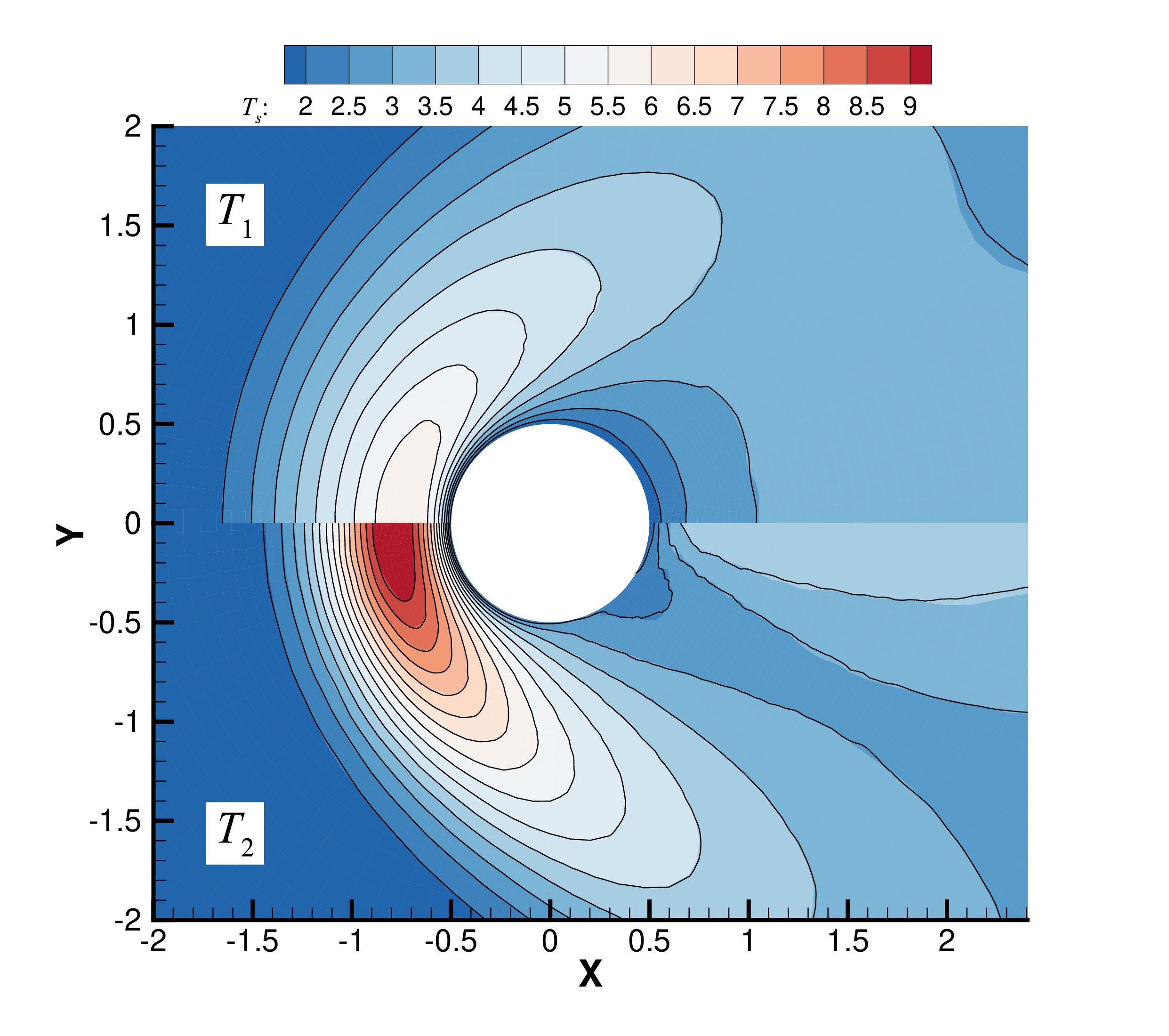}
\\
\vspace{-1.5mm}
\hspace{-11mm}
\includegraphics[width=0.38\textwidth,trim=10pt 10pt 10pt 10pt,clip]{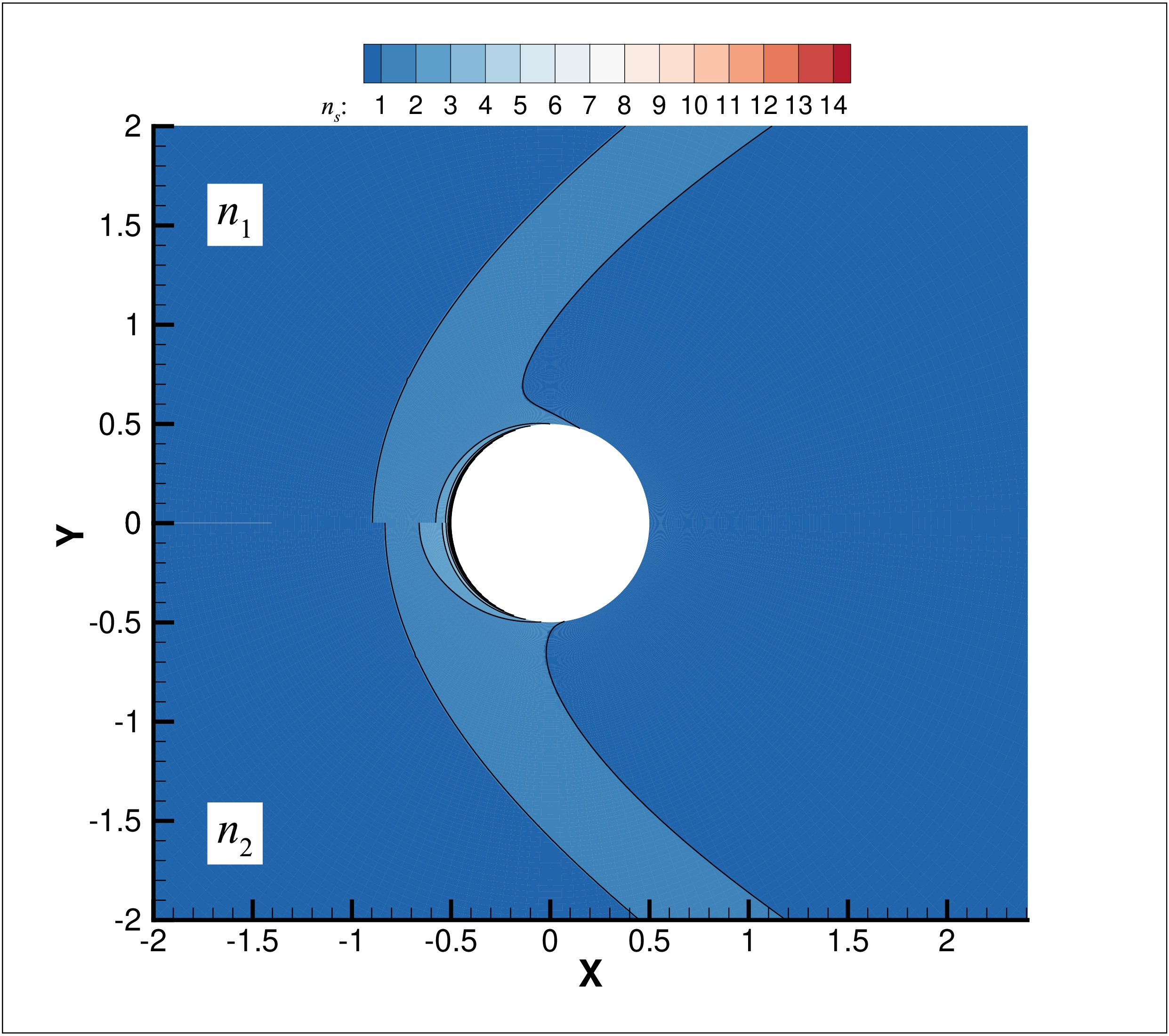}
\hspace{-11mm}
\includegraphics[width=0.38\textwidth,trim=10pt 10pt 10pt 10pt,clip]{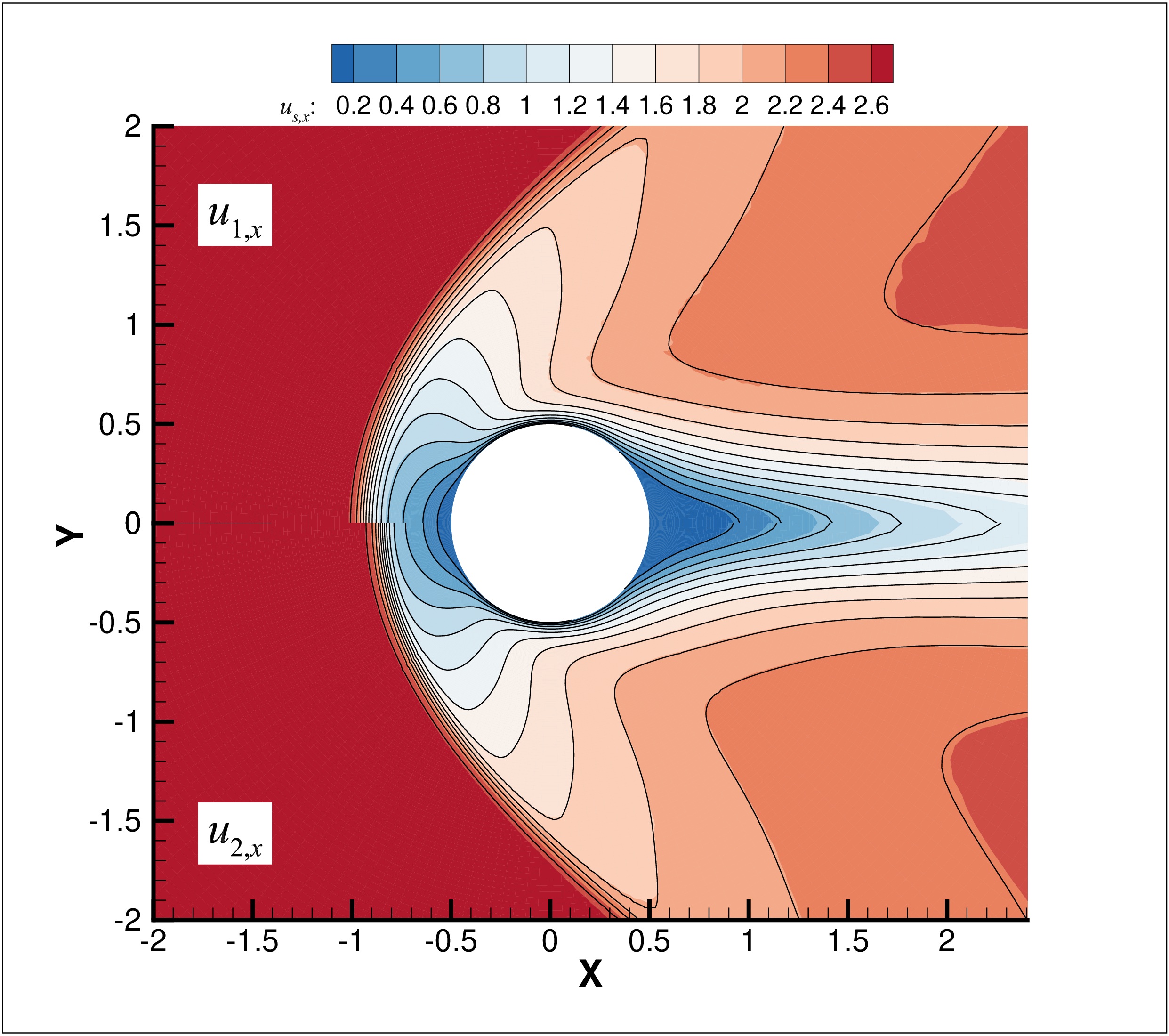}
\hspace{-11mm}
\includegraphics[width=0.38\textwidth,trim=10pt 10pt 10pt 10pt,clip]{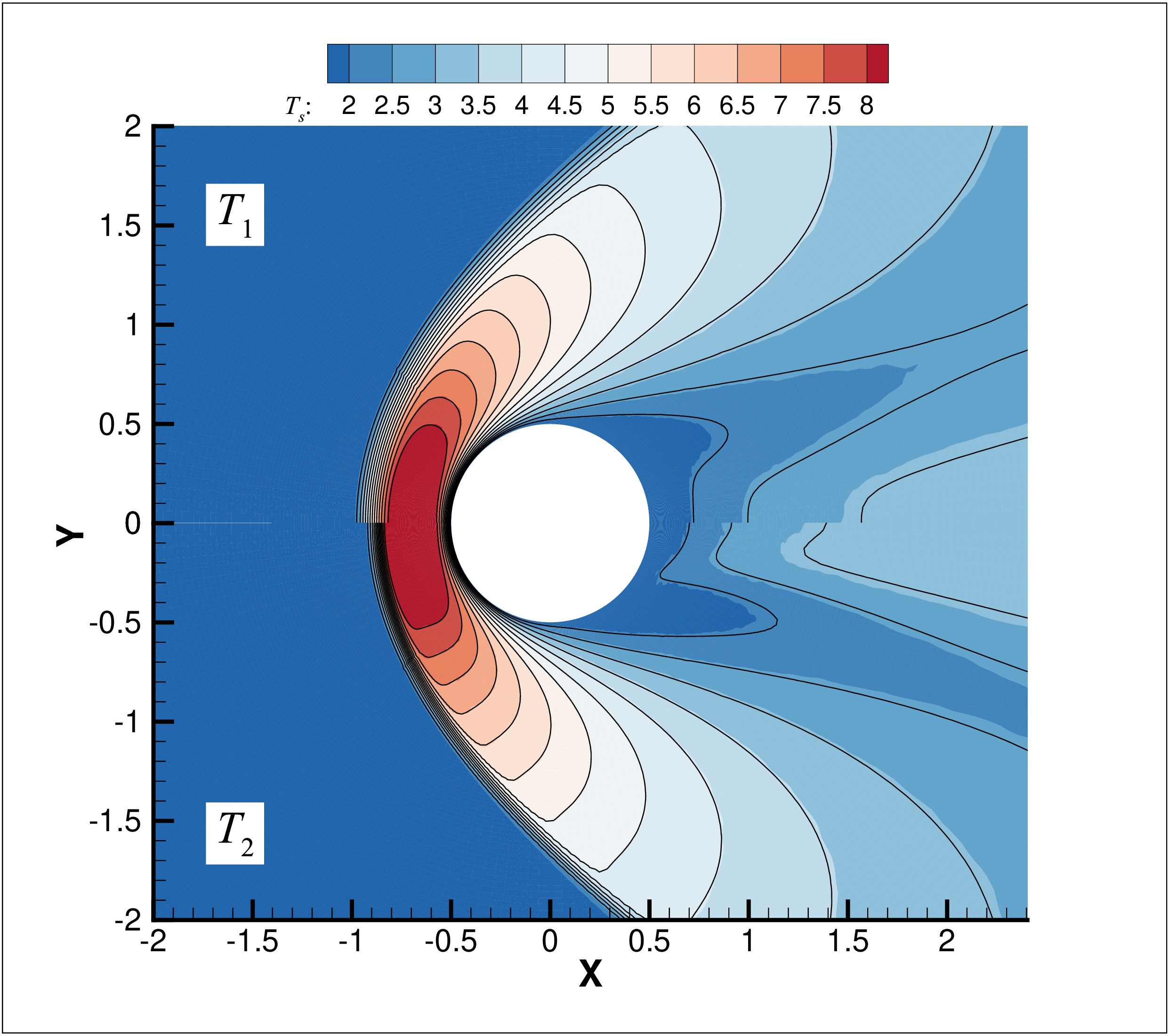}
\caption{Comparisons of macroscopic properties predicted by DIG (lines) and DSMC (contours) for the type 1 mixture with a mass ratio $m_r=10$ and an incoming Mach number of 5. Results in the top and bottom rows correspond to global Knudsen numbers of 0.1 and 0.01, respectively.}
\label{fig:Contour_Mr10_Maxwell}
\end{figure}

Figure~\ref{fig:meshcompare_Mr10_Maxwell} compares the computational meshes used in the DSMC and DIG for a type-1 gas mixture with an inflow Mach number of 5. In the open-source DSMC code SPARTA, Cartesian grids are adaptively refined to be less than one-third of the mean free path. As a result, DSMC requires approximately 250,000 cells for $\text{Kn}=0.1$ and 1.9 million cells for $\text{Kn}=0.01$. In contrast, the spatial cell sizes in DIG are significantly larger than the mean free path, particularly in the near-continuum regime. For example, at $\text{Kn}=0.01$, the cell size near the cylinder surface in DIG can be nearly 15 times larger than that required by standard DSMC. Consequently, only $1.28\times 10^4$ cells for $\text{Kn}=0.1$ and $4.0\times 10^4$ cells for $\text{Kn}=0.01$ are used in DIG.

\begin{figure}[t]
\centering
\vspace{-1.5mm}
\hspace{-11mm}
\includegraphics[width=0.38\textwidth,trim=10pt 10pt 10pt 10pt,clip]{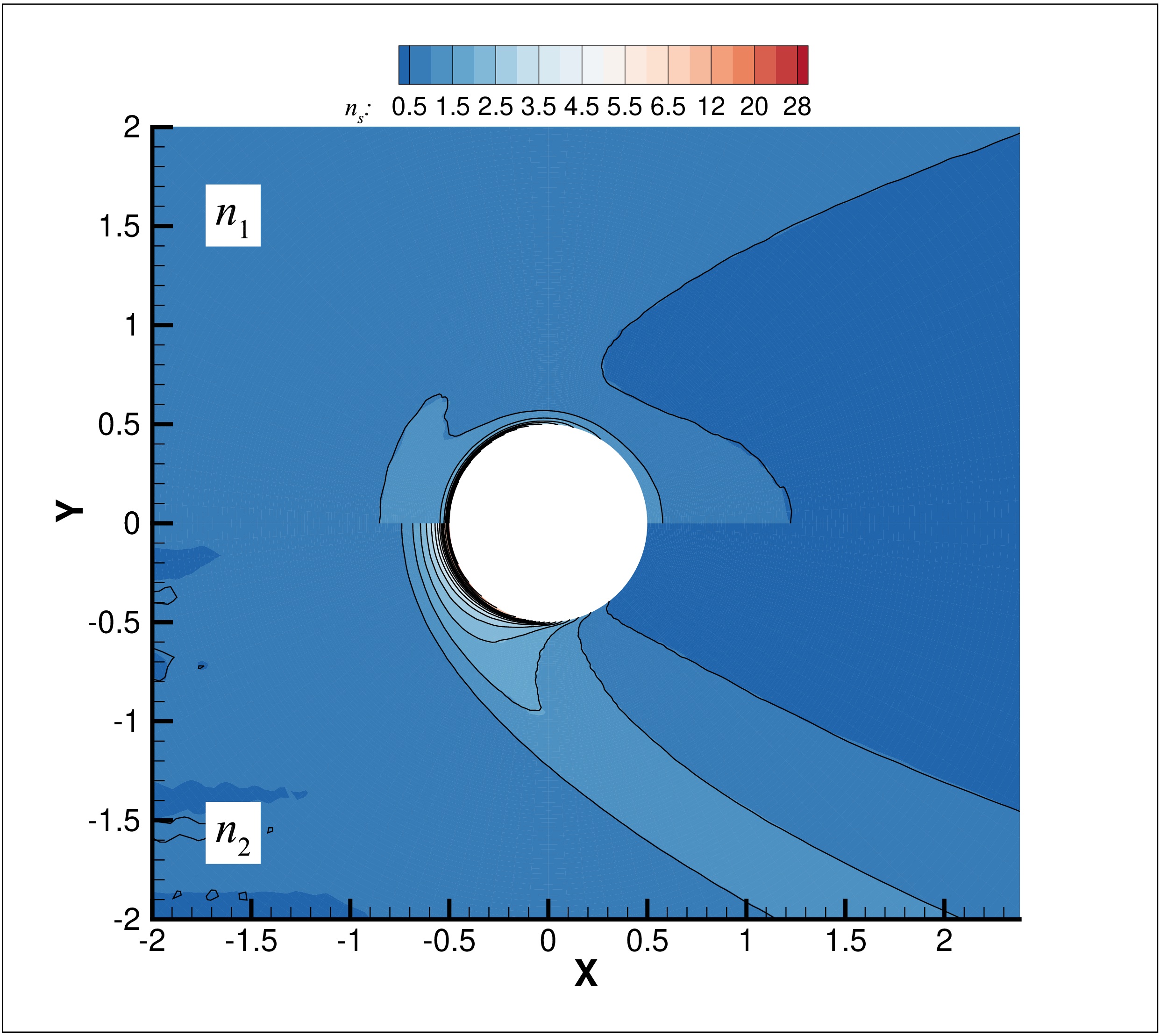}
\hspace{-11mm}
\includegraphics[width=0.38\textwidth,trim=10pt 10pt 10pt 10pt,clip]{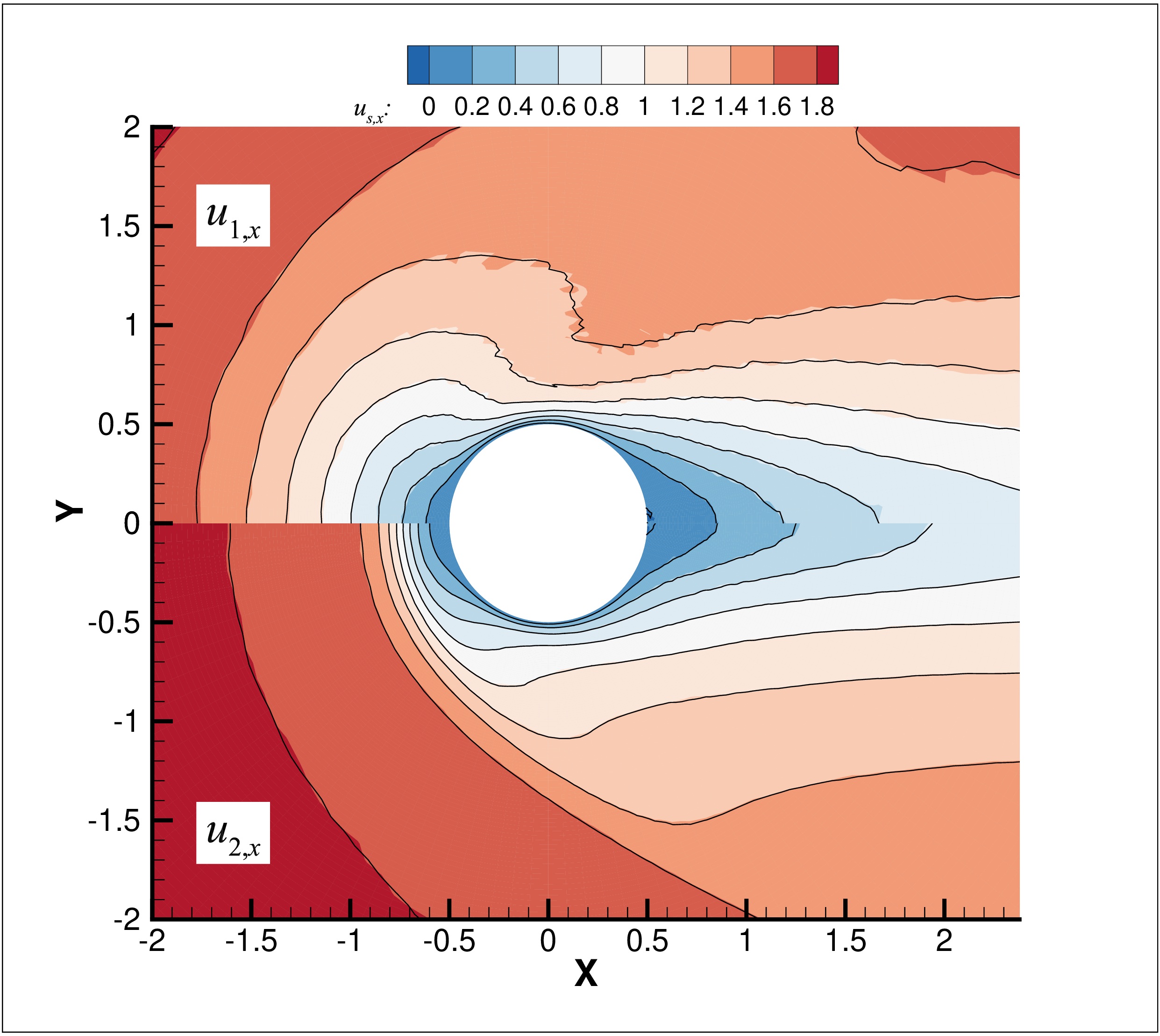}
\hspace{-11mm}
\includegraphics[width=0.38\textwidth,trim=10pt 10pt 10pt 10pt,clip]{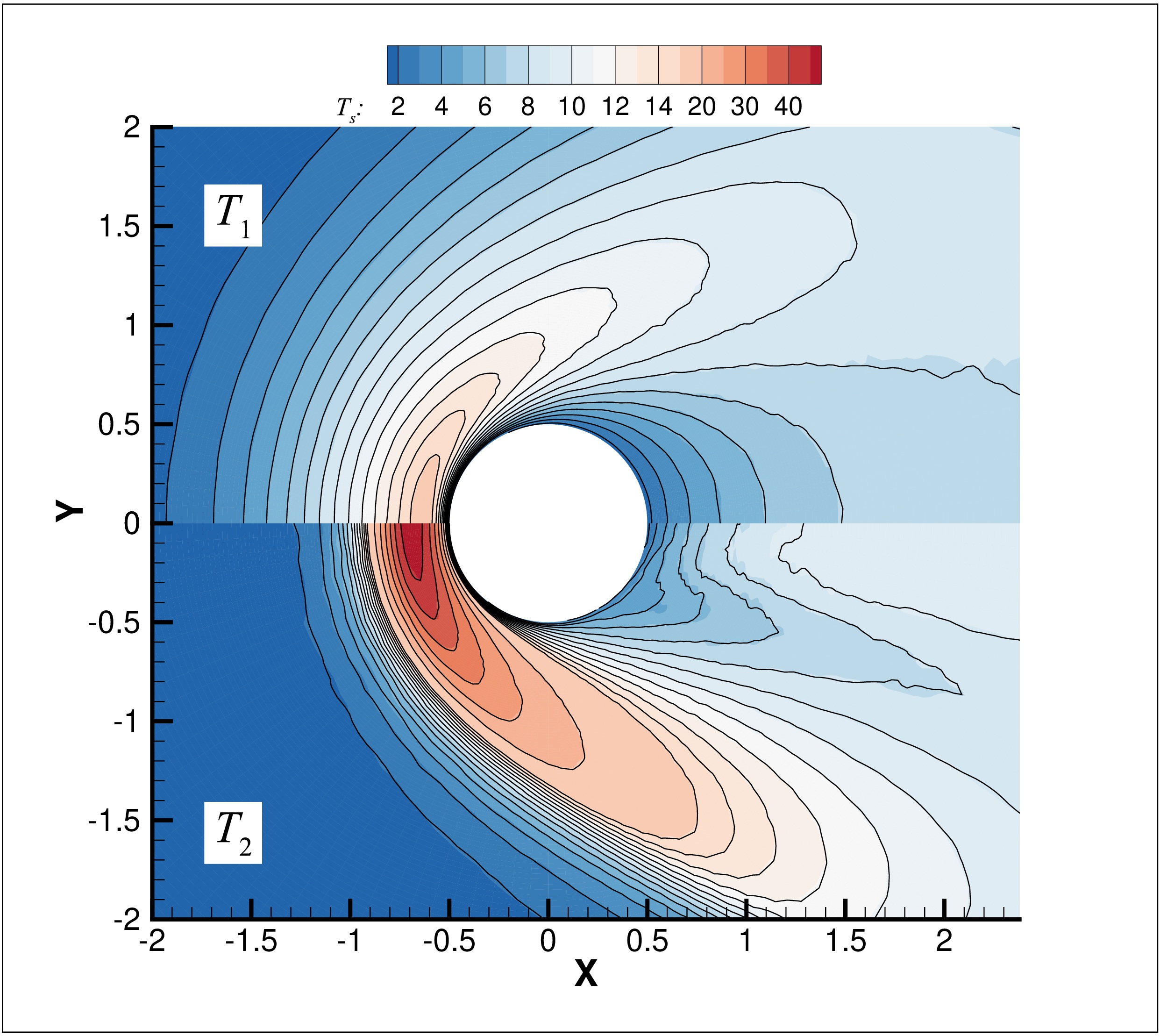}
\\
\vspace{-1.5mm}
\hspace{-11mm}
\includegraphics[width=0.38\textwidth,trim=10pt 10pt 10pt 10pt,clip]{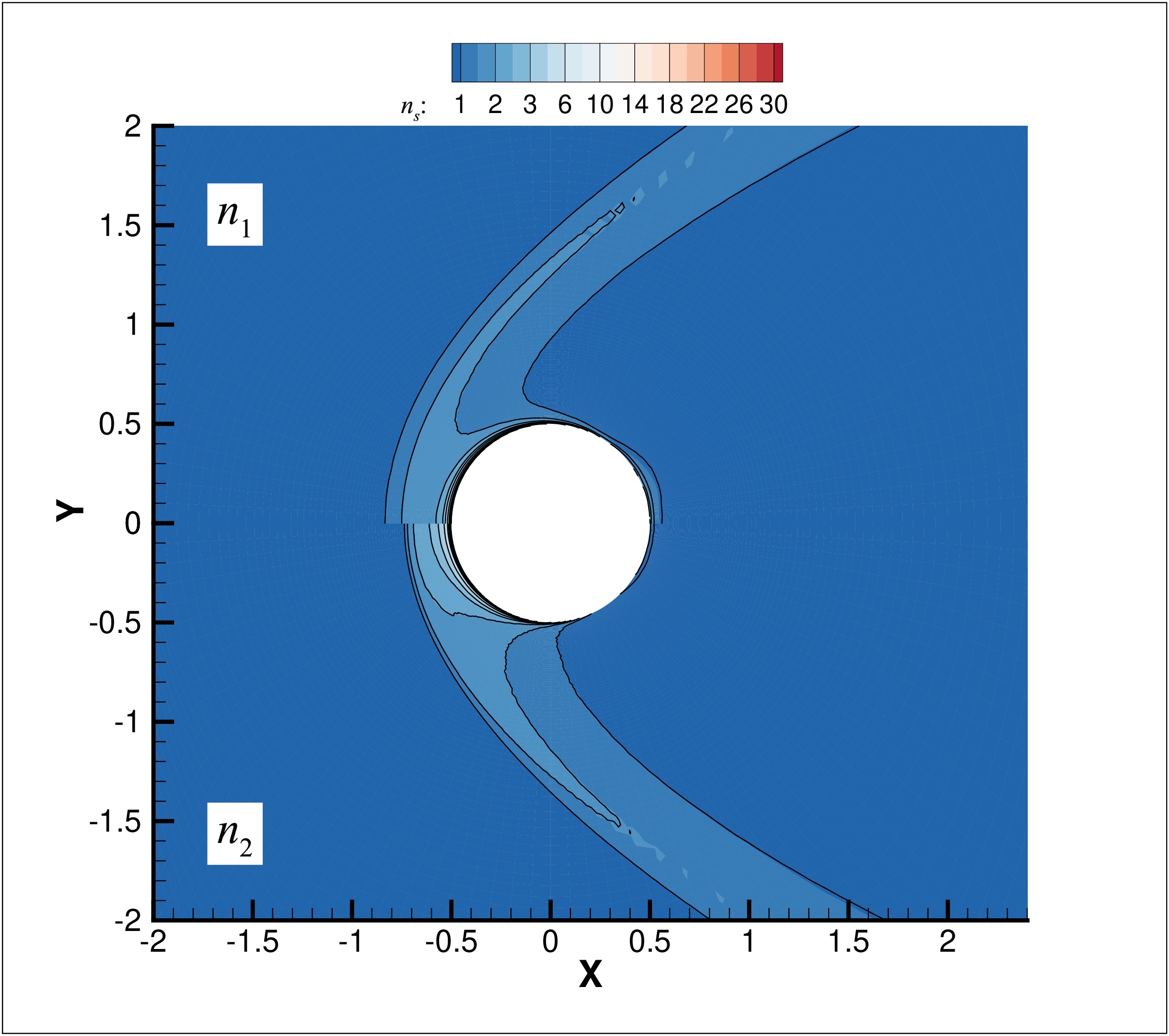}
\hspace{-11mm}
\includegraphics[width=0.38\textwidth,trim=10pt 10pt 10pt 10pt,clip]{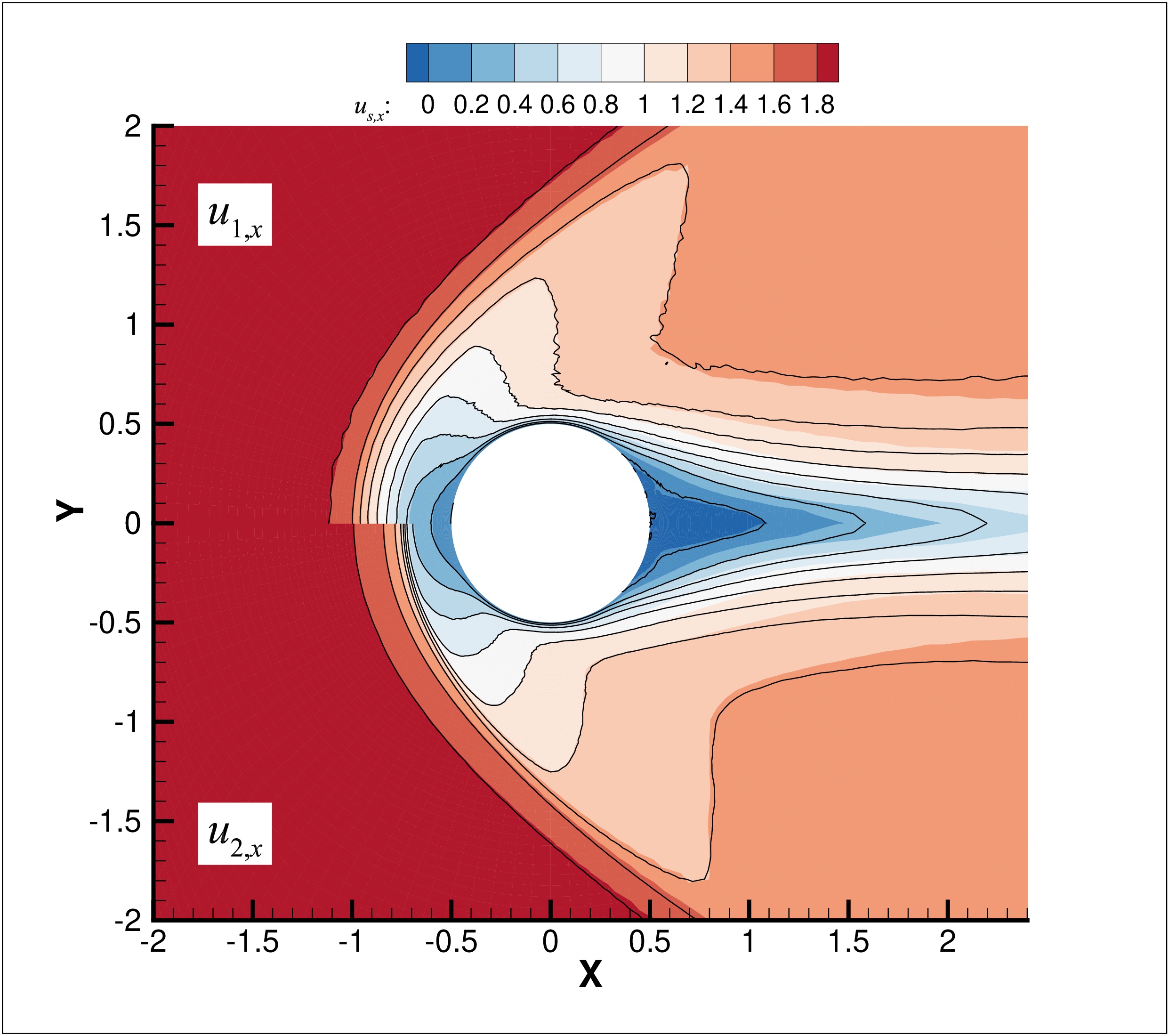}
\hspace{-11mm}
\includegraphics[width=0.38\textwidth,trim=10pt 10pt 10pt 10pt,clip]{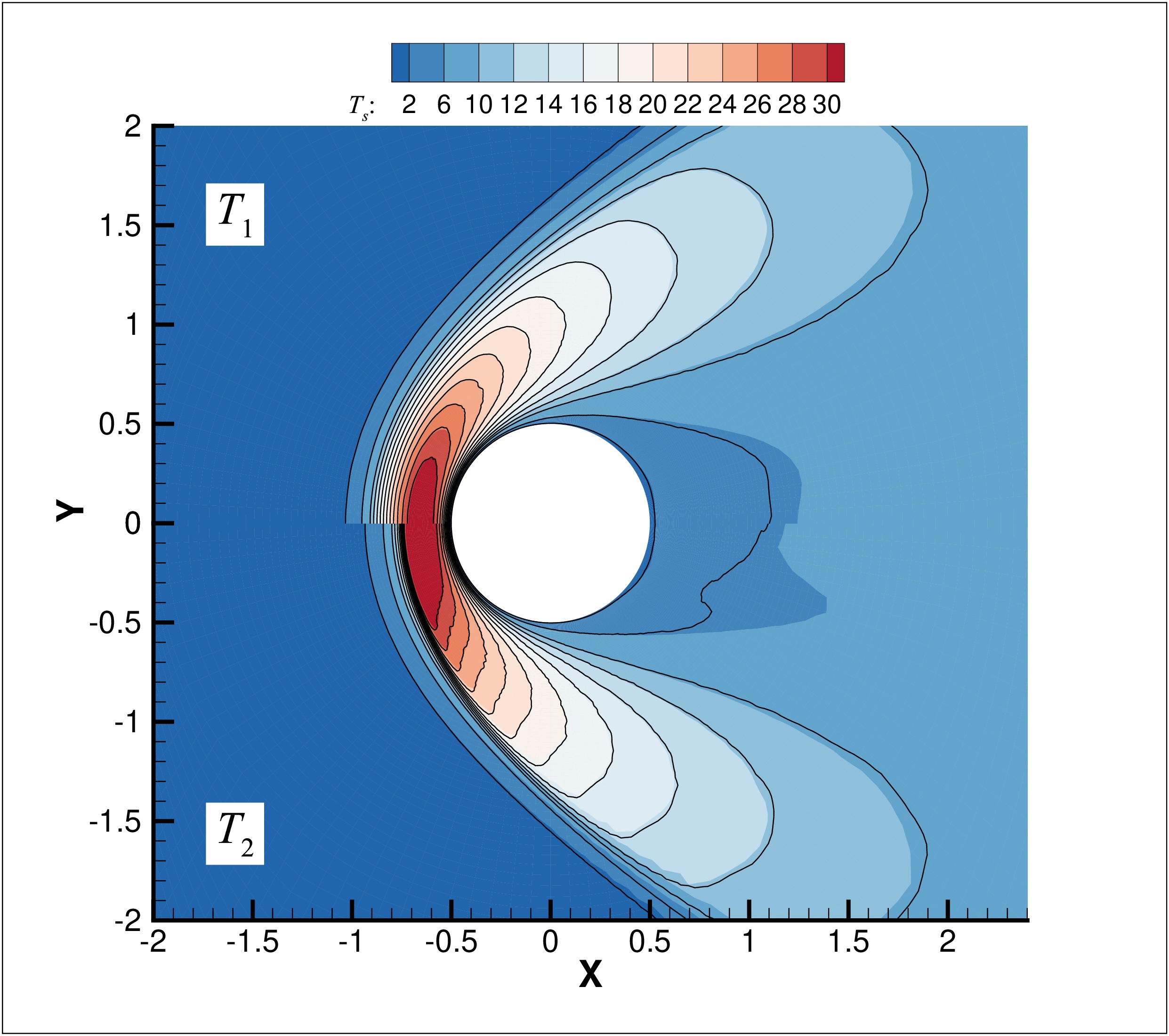}
\caption{Comparisons of macroscopic properties predicted by DIG (lines) and DSMC (contours) for the type 2 mixture with a mass ratio $m_r=100$ and an incoming Mach number of 10. Results in the top and bottom rows correspond to global Knudsen numbers of 0.1 and 0.01, respectively.}
\label{fig:Contour_Mr100_Maxwell}
\end{figure}

Figure~\ref{fig:Contour_Mr10_Maxwell} compares the macroscopic fields predicted by DIG and DSMC for the type-1 gas mixture at global Knudsen numbers of $0.1$ and $0.01$. The results show pronounced compression of the flow upstream of the cylinder and expansion downstream. As the Knudsen number decreases, the shock layer of each species becomes progressively thinner and shifts closer to the cylinder surface. At $\text{Kn}=0.1$, the peak temperature in the shock layer of the heavier species is nearly twice that of the lighter species, since the effective Mach number of the lighter gas is smaller and thus the gas is less compressed. When $\text{Kn}$ is reduced to $0.01$, the peak temperatures of both species become nearly identical due to frequent inter-species collisions that drive the system toward thermal equilibrium. Nevertheless, visible discrepancies in macroscopic properties remain, primarily due to the different transport characteristics of the two species. As expected, the grid size in DIG is not constrained by the mean free path, as in DSMC, enabling the use of significantly fewer computational cells while still achieving results in close agreement with DSMC.

Figure~\ref{fig:Contour_Mr100_Maxwell} further compares the macroscopic property contours predicted by the DIG and DSMC when the mass ratio is increased to 100. The incoming Mach number is increased to 10, so as to accentuate the inter‑species differences in the macroscopic fields. 
Similar to the type 1 mixture case, an increase in the Knudsen number leads to a thicker shock layer. At $\text{Kn}=0.1$, the peak temperature of the heavy species is nearly four times that of the light species, reflecting the pronounced non‑equilibrium between speciess. When the Knudsen number decreases to $0.01$, the peak temperatures of the two species become nearly identical, although minor differences in other macroscopic properties remain due to non-negligible rarefaction effects. The larger mass ratio increases the disparity in mean free paths between species, which, in turn, necessitates a finer mesh to obtain the reference DSMC results when using the adaptive grid refinement based on the local mean free path. However, we employ the same grid as previous type-1 gas mixture in DIG; despite the substantial difference in the number of computational cells, the macroscopic results obtained from the DSMC and DIG exhibit excellent agreement.

\begin{figure}[!t]
\centering
\includegraphics[width=0.4\textwidth,trim=20pt 20pt 50pt 50pt,clip]{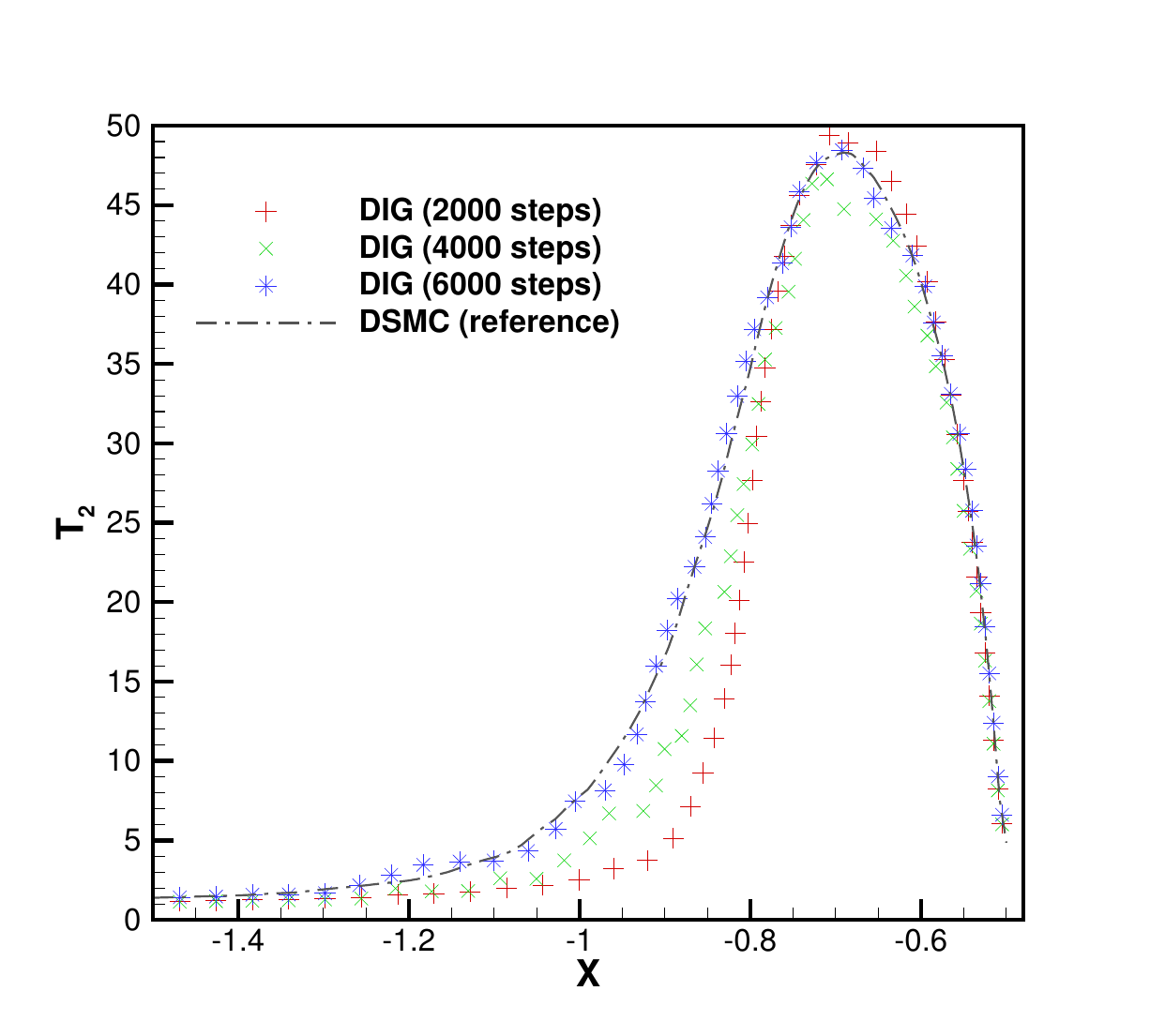}
\hspace{0.5cm}
\includegraphics[width=0.4\textwidth,trim=20pt 20pt 50pt 50pt,clip]{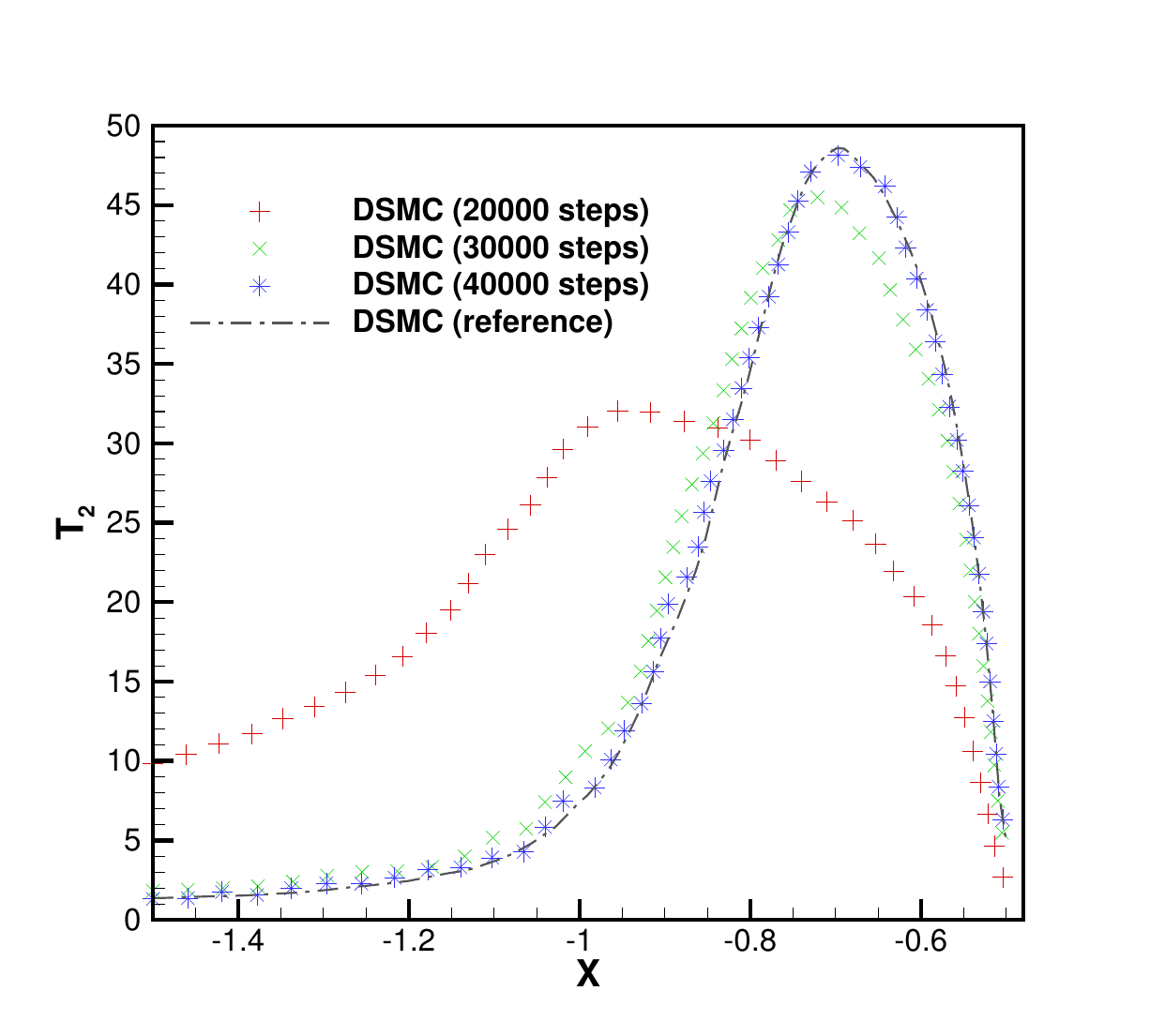}\\
\includegraphics[width=0.4\textwidth,trim=20pt 20pt 50pt 50pt,clip]{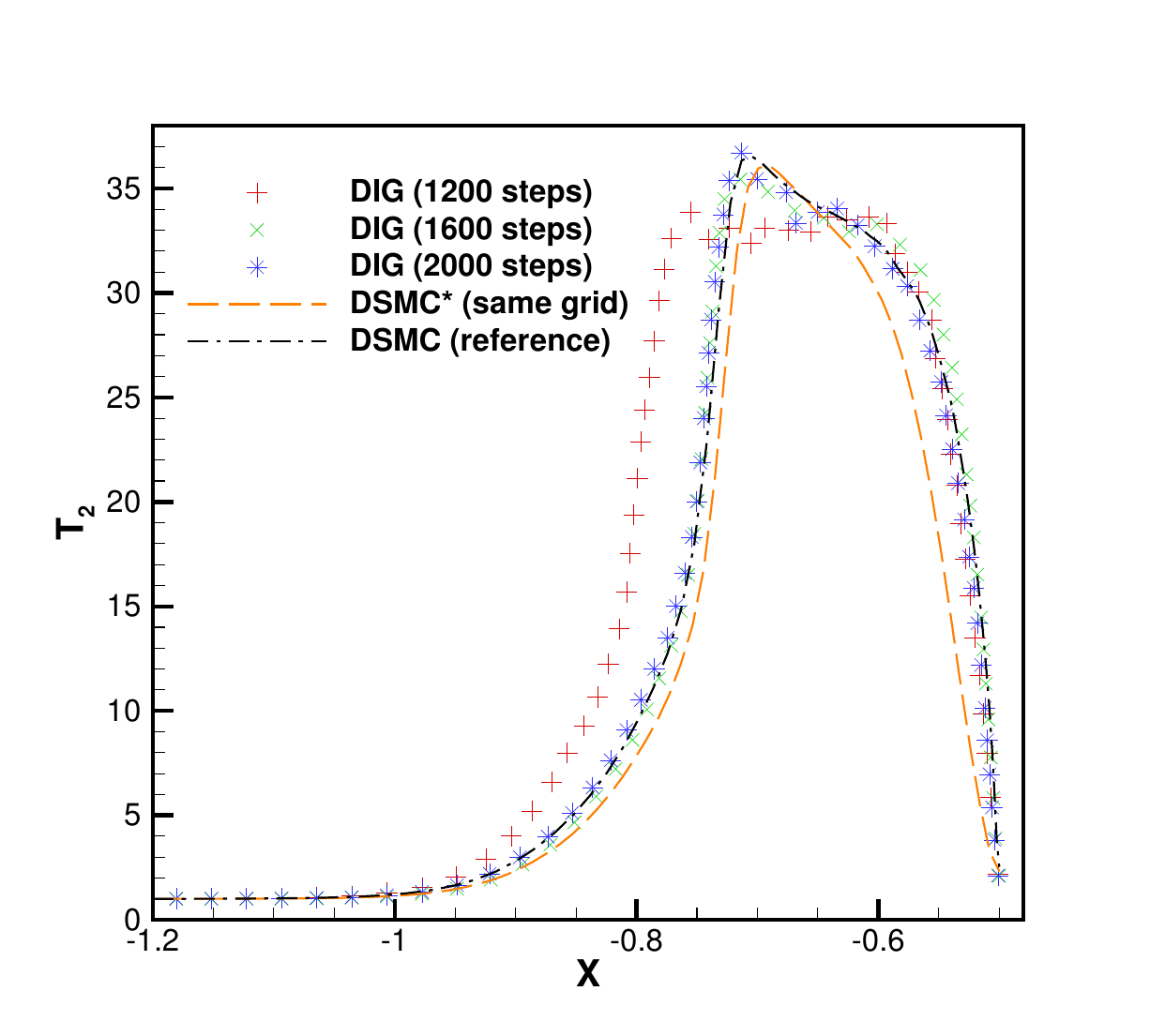}
\hspace{0.5cm}
\includegraphics[width=0.4\textwidth,trim=20pt 20pt 50pt 50pt,clip]{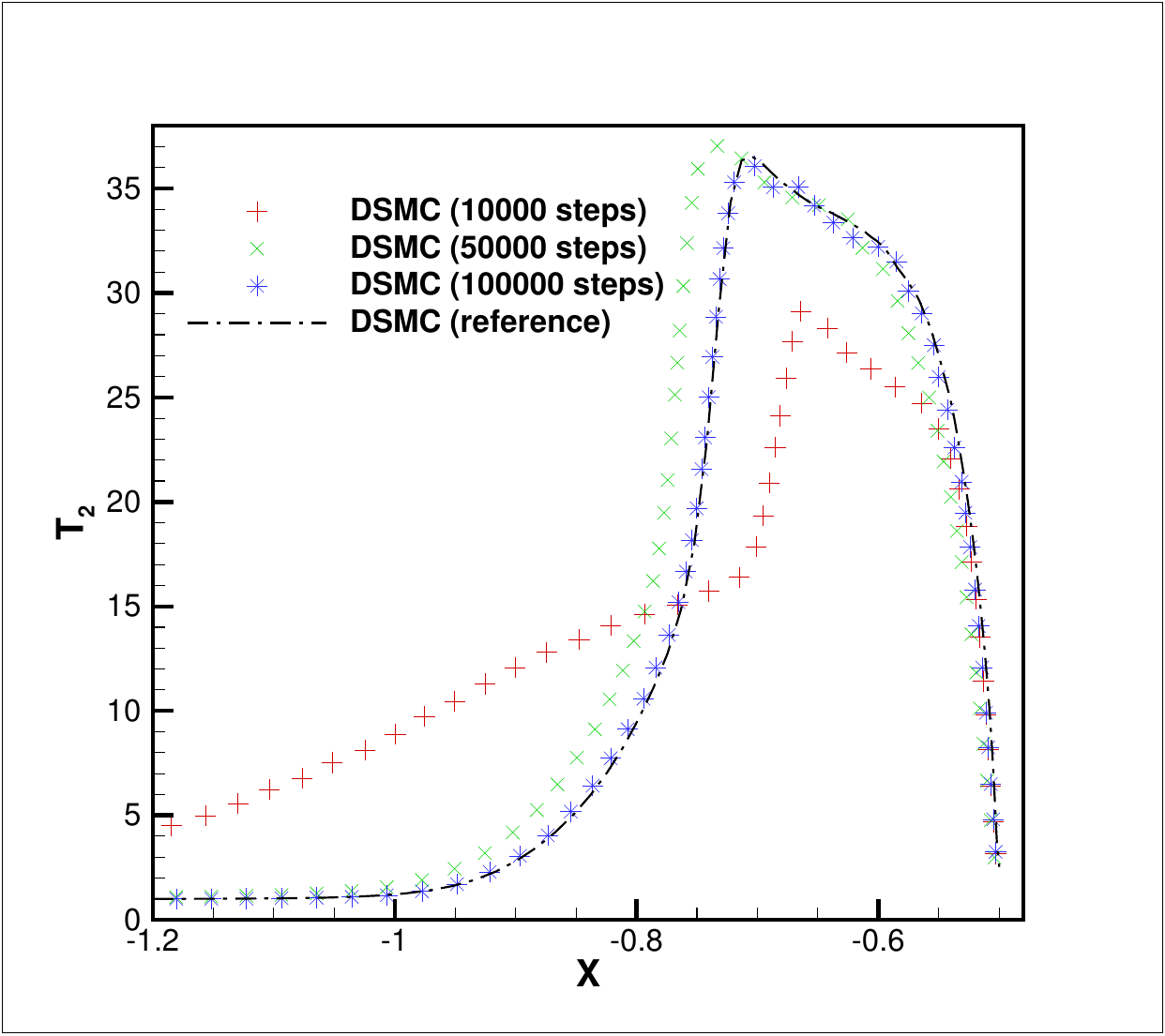}
\caption{Maxwell gas mixture with $m_r=100$: the evolution of the temperature for heavy species before the flow field reaches to the steady state when $\text{Kn}=0.1$ (top row) and $\text{Kn}=0.01$ (bottom row). DSMC results are obtained in SPARTA (the mesh size is adaptive refined to be smaller than the local mean free path), whereas the DSMC* results are computed on the same mesh as employed in the DIG (the mesh size are much larger than the mean free path). }
\label{fig:evolution_mr100Ma10_Maxwellian}
\end{figure}

Figure~\ref{fig:evolution_mr100Ma10_Maxwellian} compares the temporal evolution of the heavy species temperature in the type-2 gas mixture between DIG and DSMC. Since the light species typically equilibrates faster, the heavy species is selected for detailed comparison. The same time step is used in both approaches, and all simulations are initialized from a Maxwellian distribution corresponding to the freestream macroscopic properties. As shown in the figure, when $\text{Kn}=0.1$, DSMC requires nearly 40,000 steps for the heavy species to reach equilibrium, whereas DIG requires only about 6,000 steps, resulting in nearly an order-of-magnitude reduction in evolution steps. This is due to the iterative solution of the macroscopic synthetic equations, which guides the DSMC simulation toward the steady state by skipping many unnecessary intermediate evolutions.
When $\text{Kn}=0.01$, the acceleration to steady state achieved by DIG becomes even more pronounced, with the required number of steps reduced by nearly a factor of 50 compared with standard DSMC. Furthermore, solving the macroscopic synthetic equations endows DIG with the asymptotic-preserving property. When $\text{Kn}=0.01$, if the intermittent macroscopic equation guidance is switched off after obtaining the steady-state solution by the DIG, the solution reverts to that of DSMC on the coarser mesh (see the DSMC* results), resulting in noticeable deviations from the standard fine-mesh DSMC results. This occurs because the spatial cell size is much larger than the mean free path, so that numerical dissipation sets in.

\begin{table}[!t]
 \centering
 \caption{\label{tab:tab1}Computational overheads of the DSMC and DIG for hypersonic Maxwell gas mixtures past a cylinder. The DSMC simulations are performed using the SPARTA code with 400 cores and adaptive mesh refinement, whereas our in-house DIG uses 280 cores. For each Knudsen number, both methods employ the same time step and are initialized from uniform freestream conditions. The simulation time is given in core$\times$hours.  Note that without DIG, our in-house  DSMC code is slower than SPARTA by one order of magnitude. }

\begin{threeparttable} 
  \begin{tabular}{c c c c c c c c c}\toprule
 {mass} & \multirow{2}{*}{Ma}  & \multirow{2}{*}{Kn} & \multirow{2}{*}{method} & \multirow{2}{*}{$N_{\text{cell}}$}  & \multicolumn{2}{c}{Transition state}  &  \multicolumn{2}{c}{Steady state} \\ \cmidrule(r){6-9}
  ratio & ~ & ~ & &  & steps&time&steps   &    time\\ \hline
 \multirow{4}{*}{10} &  \multirow{4}{*}{5} & \multirow{2}{*}{0.1} &DSMC & 251251  &    20000   & 112 &  10000 &      53    \\
~   & ~  & ~ &DIG & $100\times 128$    &  5000   & 39 &  3000 &      23     \\ \cmidrule(r){3-9}
~   & ~  & \multirow{2}{*}{0.01} &DSMC & 1919737   &   60000  & 833 &  10000 &      128    \\ 
~   & ~  & ~ &DIG & $200\times 200$   &  2000   & 27 &  3000 &      41    \\ \hline
\multirow{4}{*}{100} &  \multirow{4}{*}{10} & \multirow{2}{*}{0.1} &DSMC & 1372999  &    40000   & 231 &  10000 &      48     \\
~   & ~  & ~ &DIG & $100\times 128$    &  6000   & 31 &  3000 &      15     \\ \cmidrule(r){3-9}
~   & ~  & \multirow{2}{*}{0.01} &DSMC & 5935153   &  100000   & 1467 &  10000 &    146      \\ 
~   & ~  & ~ &DIG & $200\times 200$   &  2000   & 25 &  3000 &      38    \\
\bottomrule
\end{tabular}
 \end{threeparttable}
\end{table}

The CPU times for DSMC and DIG are compared in Table~\ref{tab:tab1}. The parallel efficiency of our in-house DSMC code is roughly an order of magnitude lower than that of SPARTA, primarily due to its Cartesian grid structure and the more advanced parallelization strategy implemented in SPARTA. Nevertheless, for a mass ratio of 10, the total computational time (including both transitional and steady-state phases) of DIG is reduced by approximately a factor of 3 for $\text{Kn} = 0.1$ and a factor of 15 for $\text{Kn} = 0.01$. This advantage becomes even more pronounced at a mass ratio of 100, where the DIG achieves a CPU time reduction of more than a factor of 25 for $\text{Kn} = 0.01$ compared with SPARTA.


\subsubsection{Hard-sphere gas}

It should be noted that the corrections to the exchange terms, $\Delta \bm{Q}$ in Eq.~\eqref{eq:sourcetermexpression1}, vanish for mixtures of Maxwell molecules. For other gas models, however, these correction terms must be accurately extracted from DSMC in order to correctly capture inter-species momentum and energy exchange in the macroscopic synthetic equation. In this work, we focus on hard-sphere gases, for which the relative contribution of $\Delta \bm{Q}$ is largest in noble gases.  We investigate hypersonic flow over a cylinder at $\text{Ma} = 10$ for gas mixture type-3 for an example.

\begin{figure}[!t]
\centering
\includegraphics[width=0.4\textwidth,trim=20pt 20pt 50pt 10pt,clip]{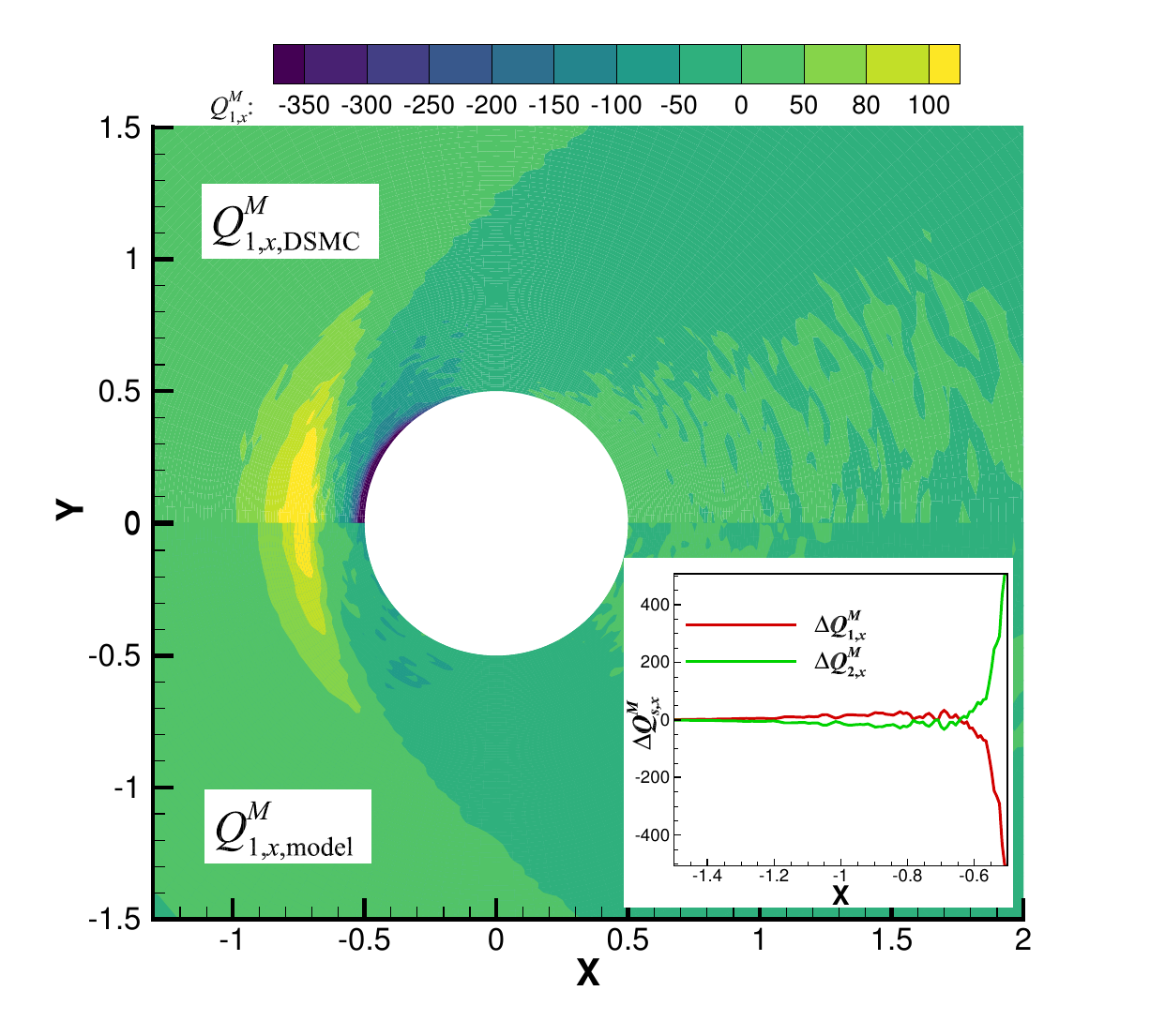}
\includegraphics[width=0.4\textwidth,trim=20pt 20pt 50pt 10pt,clip]{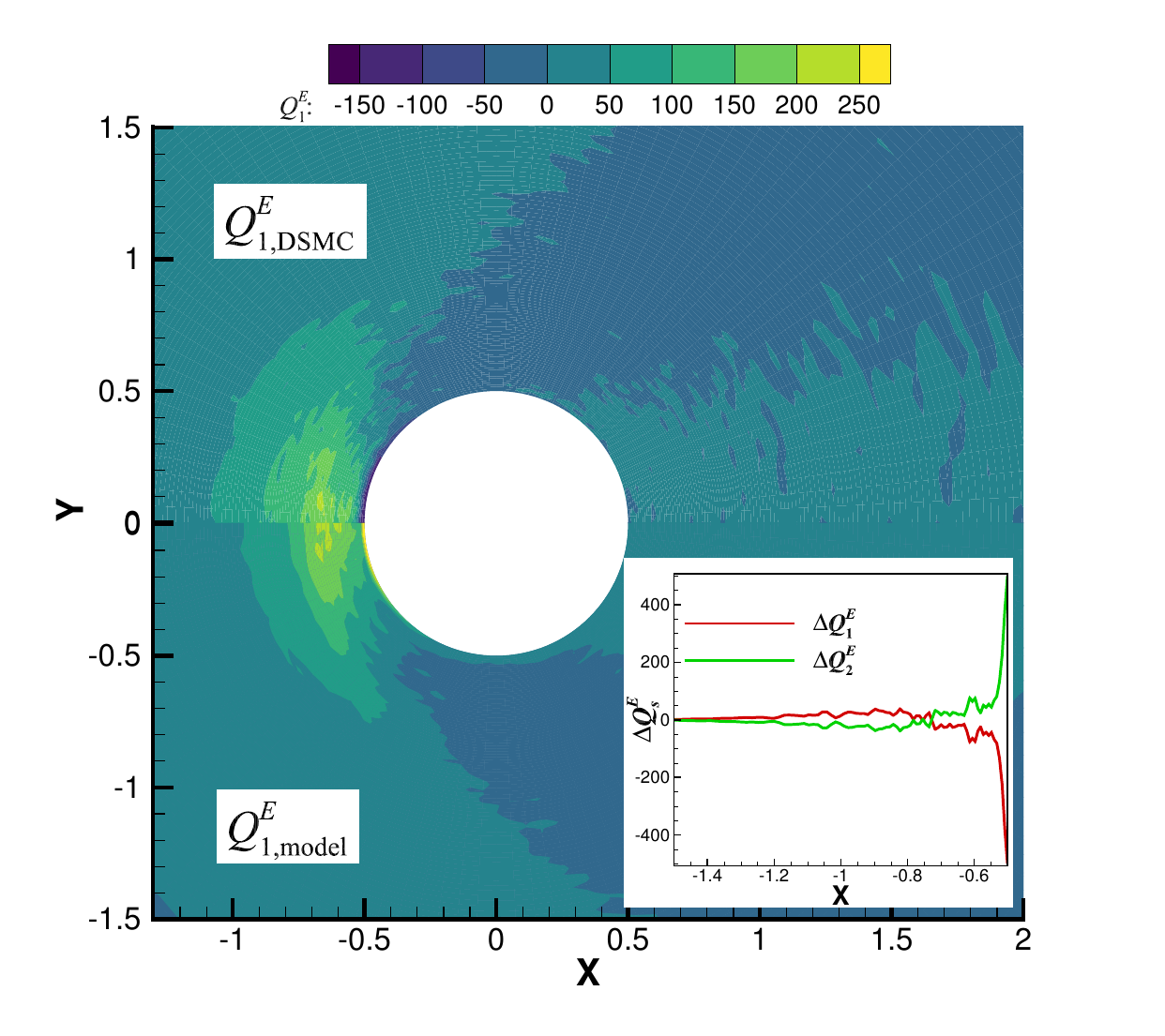}\\
\includegraphics[width=0.4\textwidth,trim=20pt 20pt 50pt 10pt,clip]{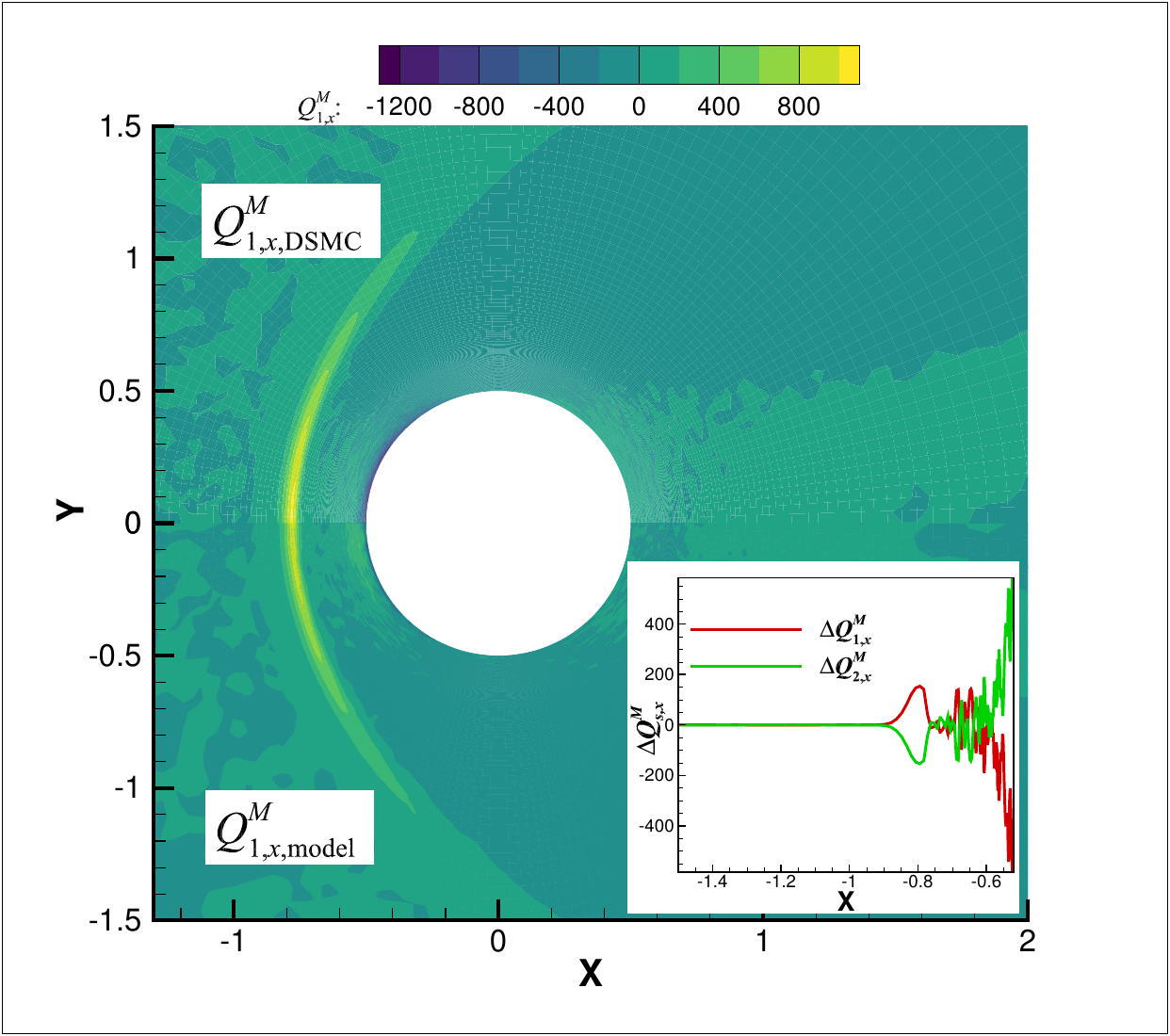}
\includegraphics[width=0.4\textwidth,trim=20pt 20pt 50pt 10pt,clip]{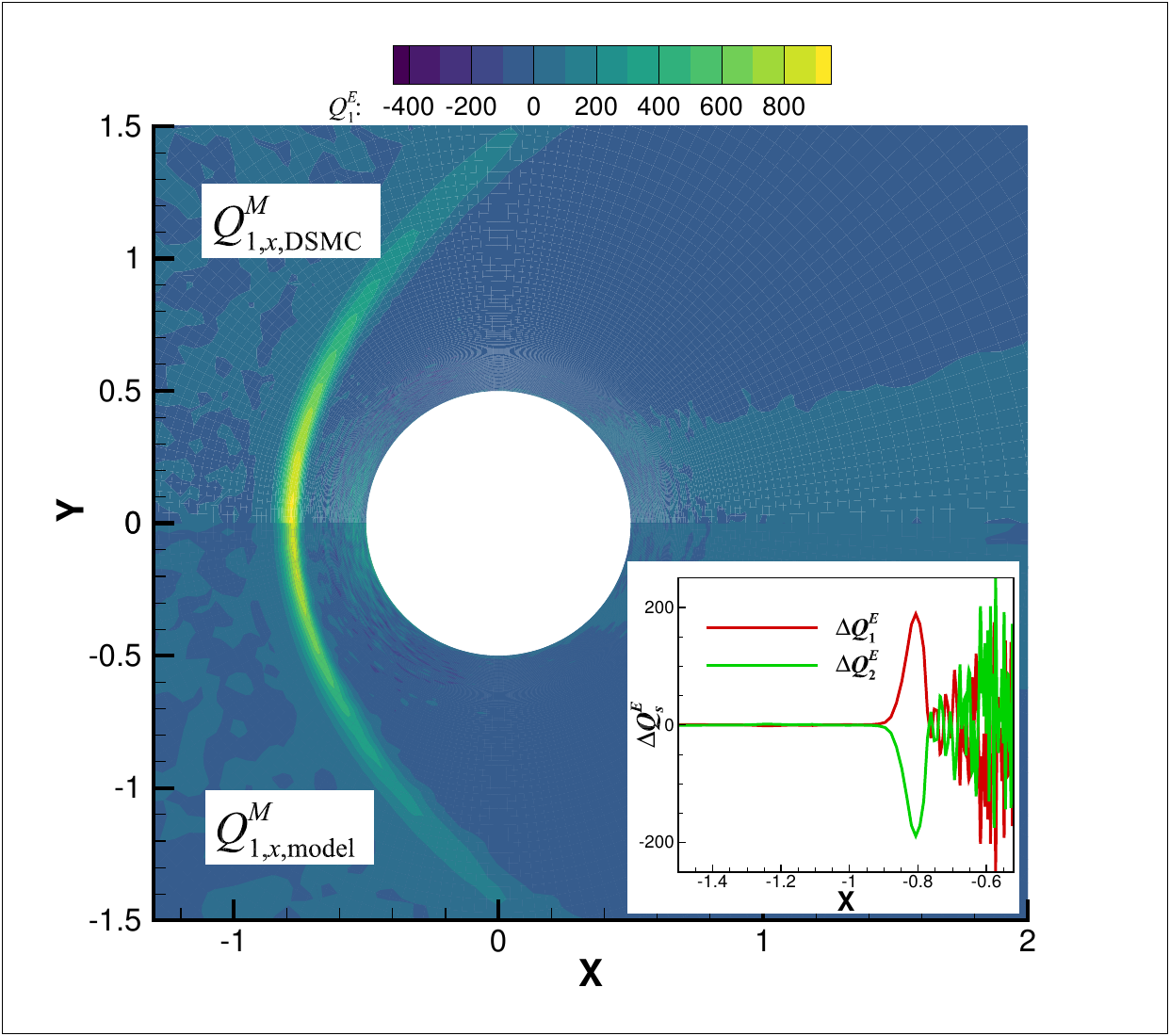}
\caption{The momentum and energy exchange terms for hard-sphere gas mixtures with $m_r = 100$ and an incoming Mach number of 10. Results are shown for Knudsen numbers $\text{Kn} = 0.1$ (top row) and $\text{Kn} = 0.01$ (bottom row). The inset illustrates the correction terms along the stagnation line. }
\label{fig:correctionterms}
\end{figure}

Figure~\ref{fig:correctionterms} shows the inter-species momentum and energy exchange terms~\eqref{eq:sourcetermexpression1} for $\text{Kn} = 0.1$ and $0.01$. Only light species are considered, since in binary gas mixtures the contributions from heavy species have the opposite sign. Note that the summations $\sum_s \Delta \bm{Q}_s^M$ and $\sum_s \Delta \bm{Q}_s^E$ also vanish as a consequence of momentum and energy conservation.
These exchange terms are most significant in the shock region and near the cylinder’s frontal surface. As indicated in Eq.~\eqref{eq:modelsourceterm}, because the mean collision time appears in the denominator, the magnitudes of the exchange terms decrease as the Knudsen number increases. The corresponding correction terms—defined as the difference between the DSMC results and the simplified kinetic models—are shown along the stagnation line. These corrections are large near the stagnation point and at the center of the bow shock, where the gas rarefaction effects are strong. In particular, for the momentum exchange term at the stagnation point, the prediction from the kinetic model vanishes because the flow velocity is zero, see Eq.~\eqref{eq:modelsourceterm}. Consequently, for $\text{Kn} = 0.1$, the correction terms are of the same order as the corresponding exchange terms. As the Knudsen number decreases, rarefaction effects at the cylinder surface diminish, and the relative magnitude of the correction terms decreases accordingly.

\begin{figure}[!t]
\hspace{-11mm}
\includegraphics[width=0.38\textwidth,trim=10pt 10pt 10pt 10pt,clip]{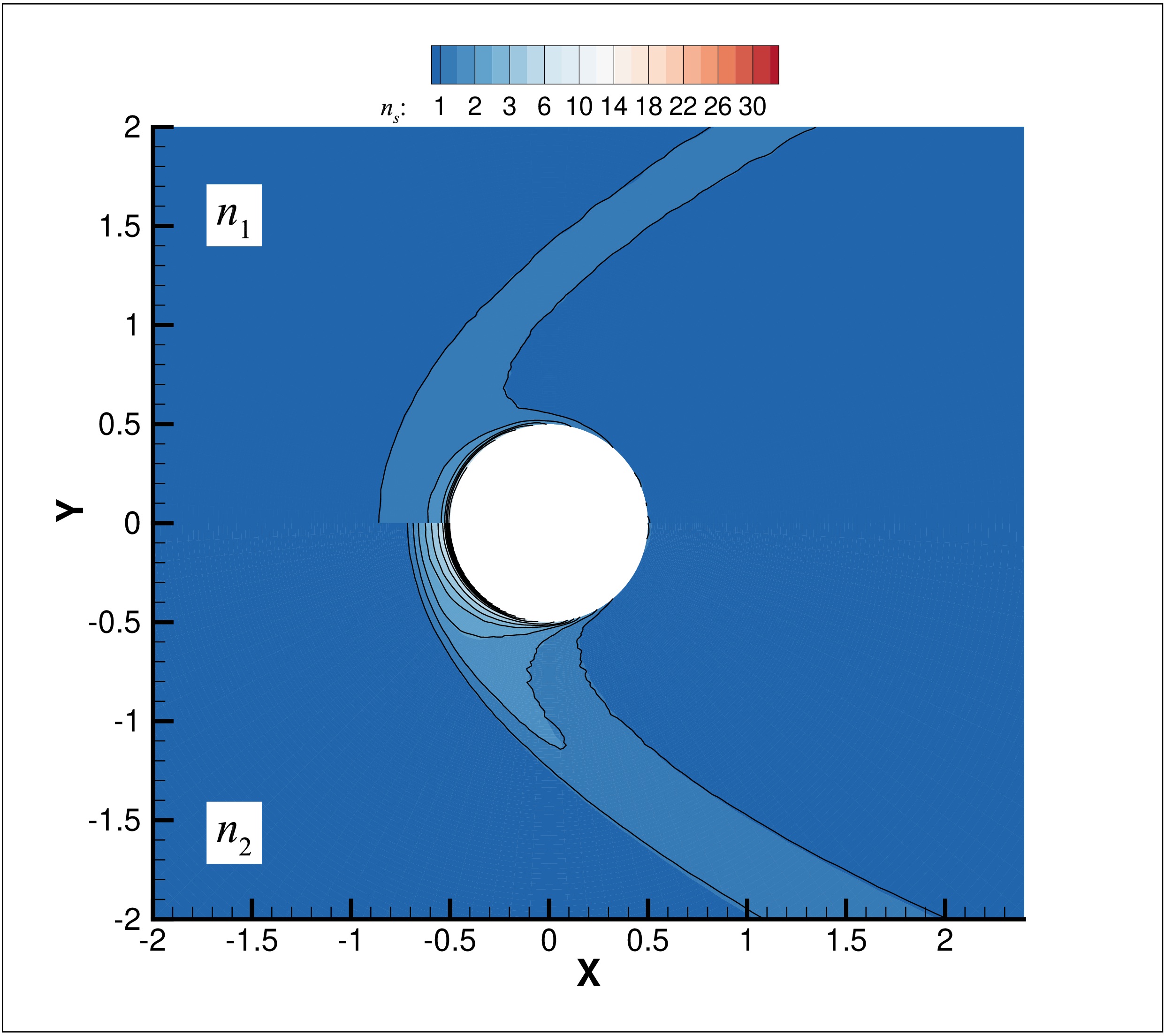}
\hspace{-11mm}
\includegraphics[width=0.38\textwidth,trim=10pt 10pt 10pt 10pt,clip]{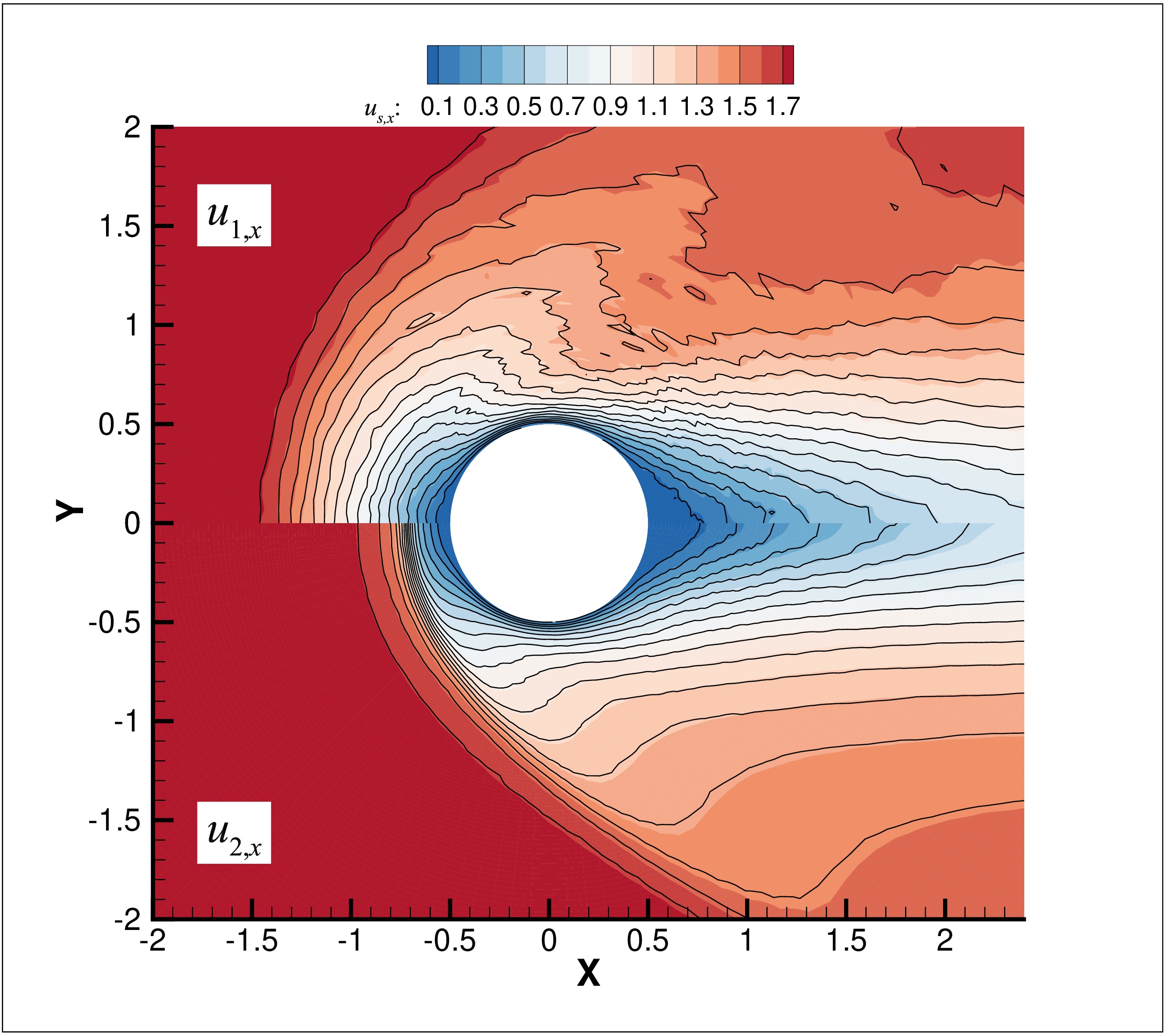}
\hspace{-11mm}
\includegraphics[width=0.38\textwidth,trim=10pt 10pt 10pt 10pt,clip]{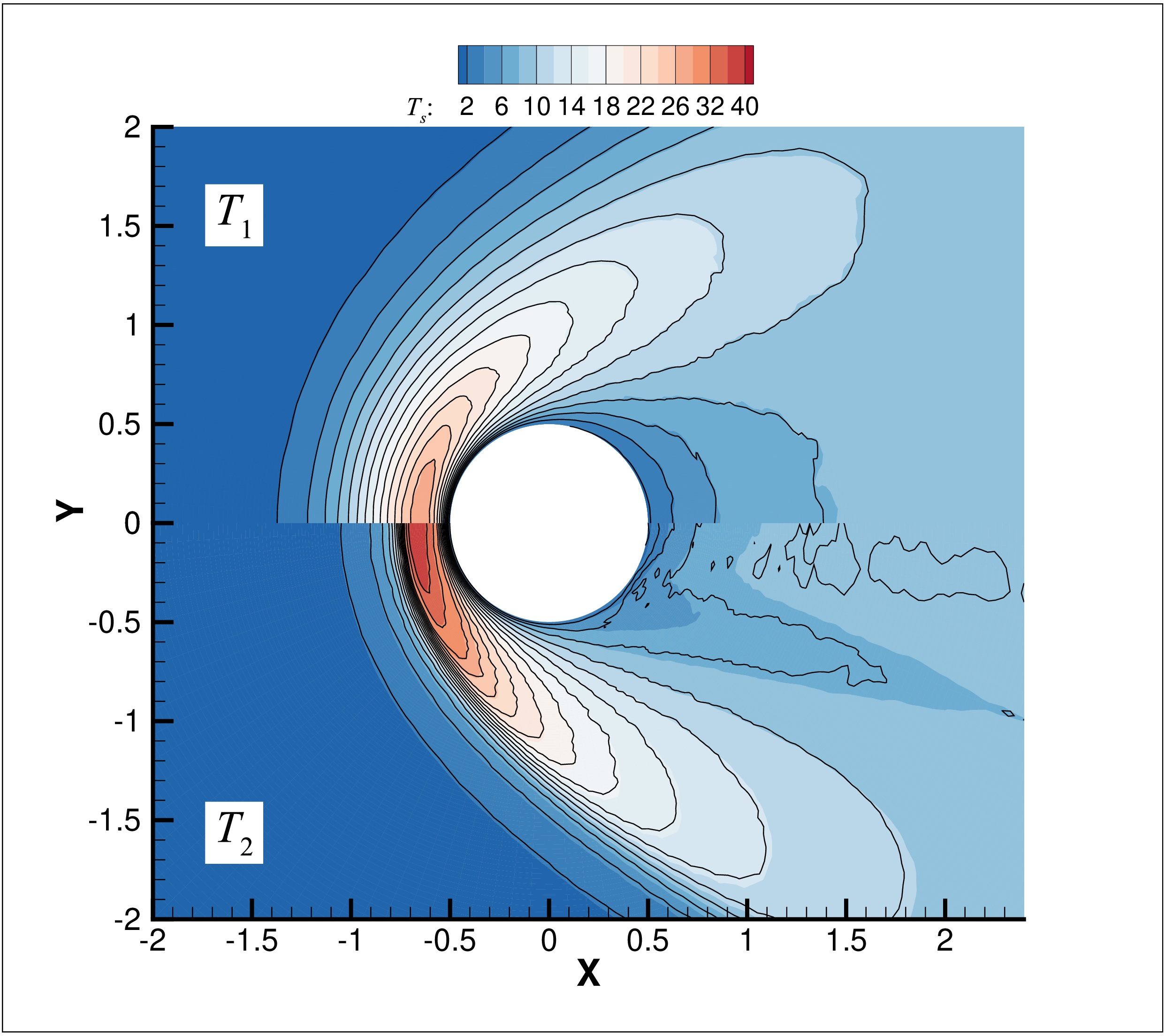}\\
\vspace{-1.5mm}
\hspace{-11mm}
\includegraphics[width=0.38\textwidth,trim=10pt 10pt 10pt 10pt,clip]{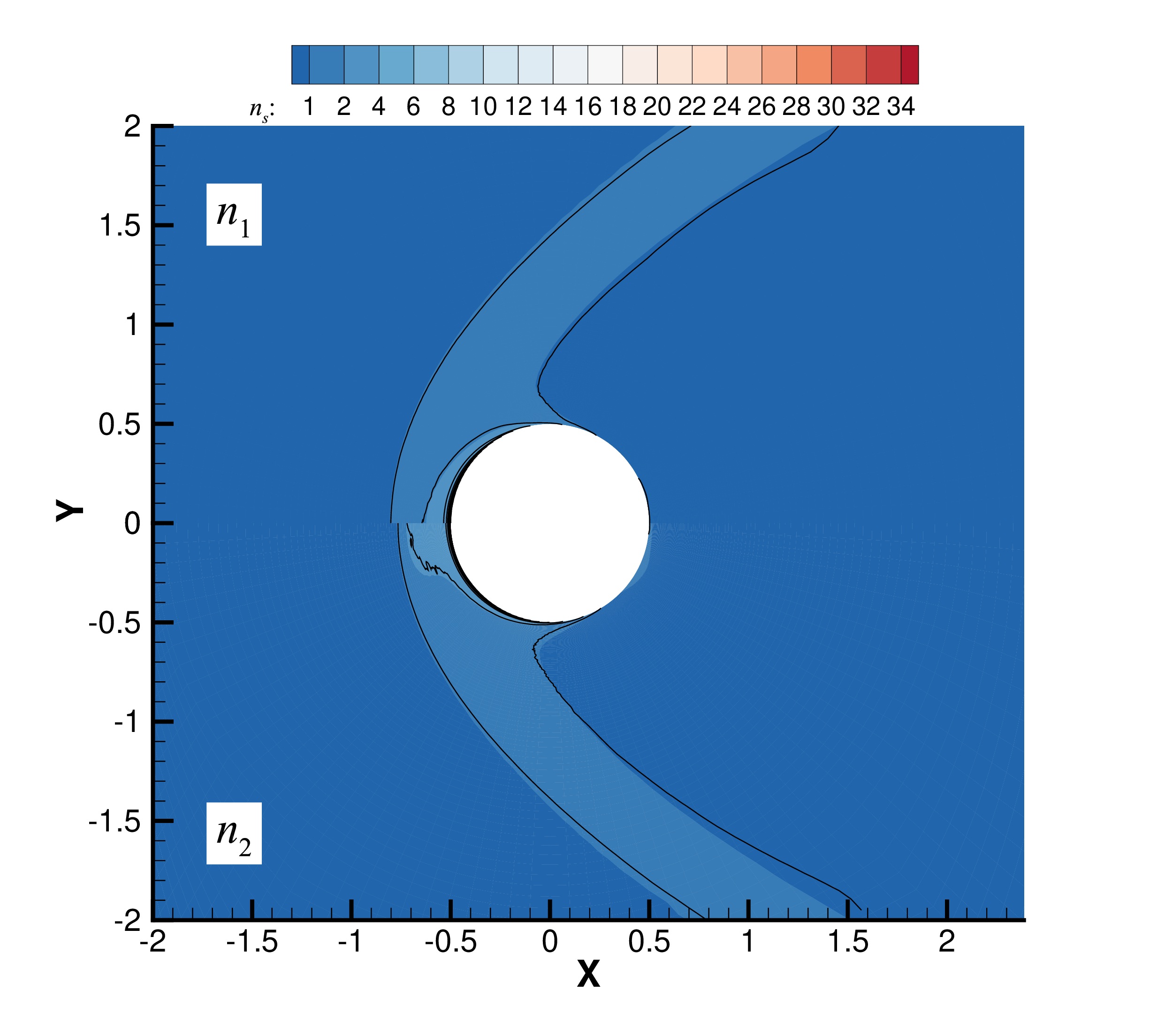}
\hspace{-11mm}
\includegraphics[width=0.38\textwidth,trim=10pt 10pt 10pt 10pt,clip]{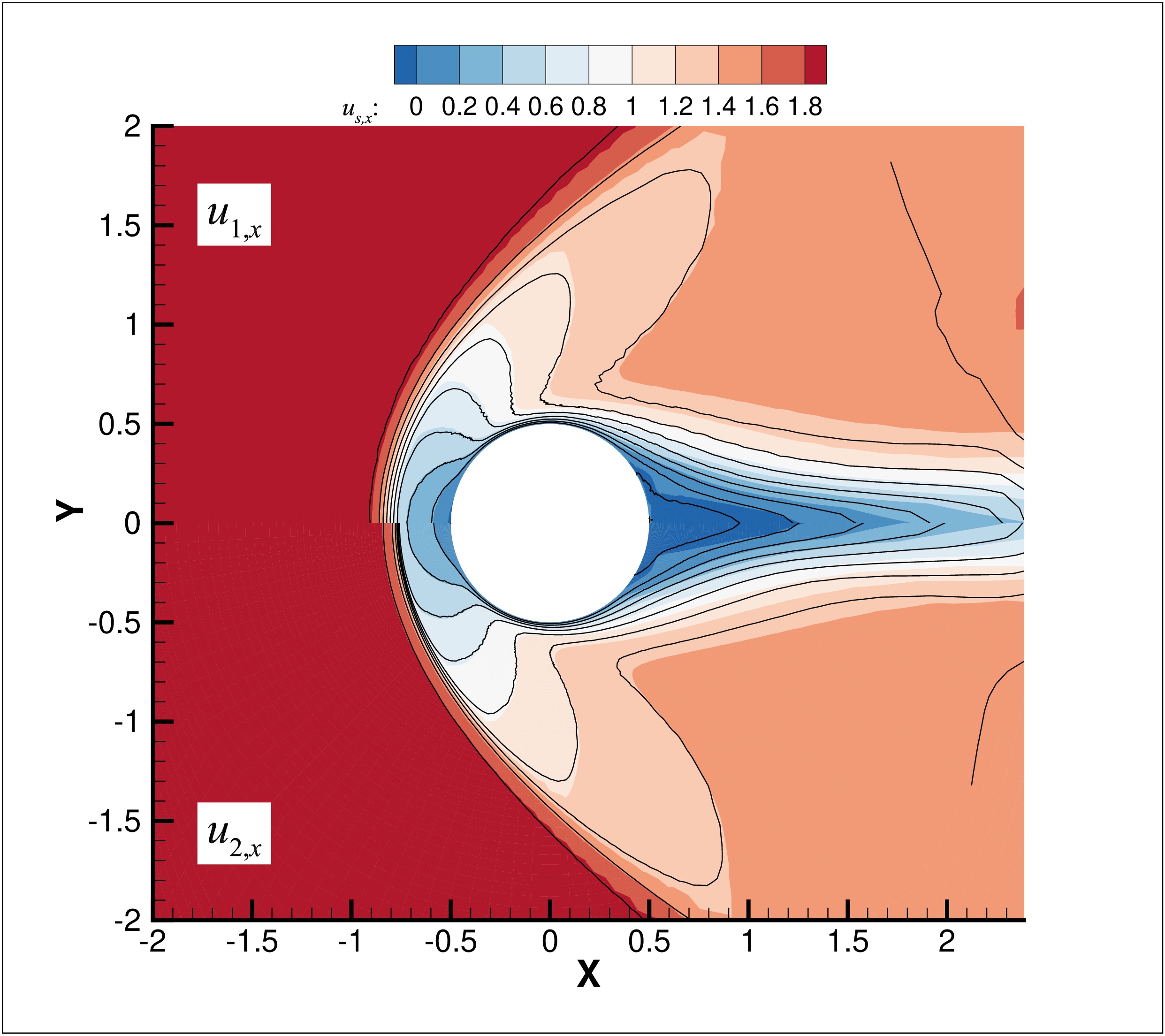}
\hspace{-11mm}
\includegraphics[width=0.38\textwidth,trim=10pt 10pt 10pt 10pt,clip]{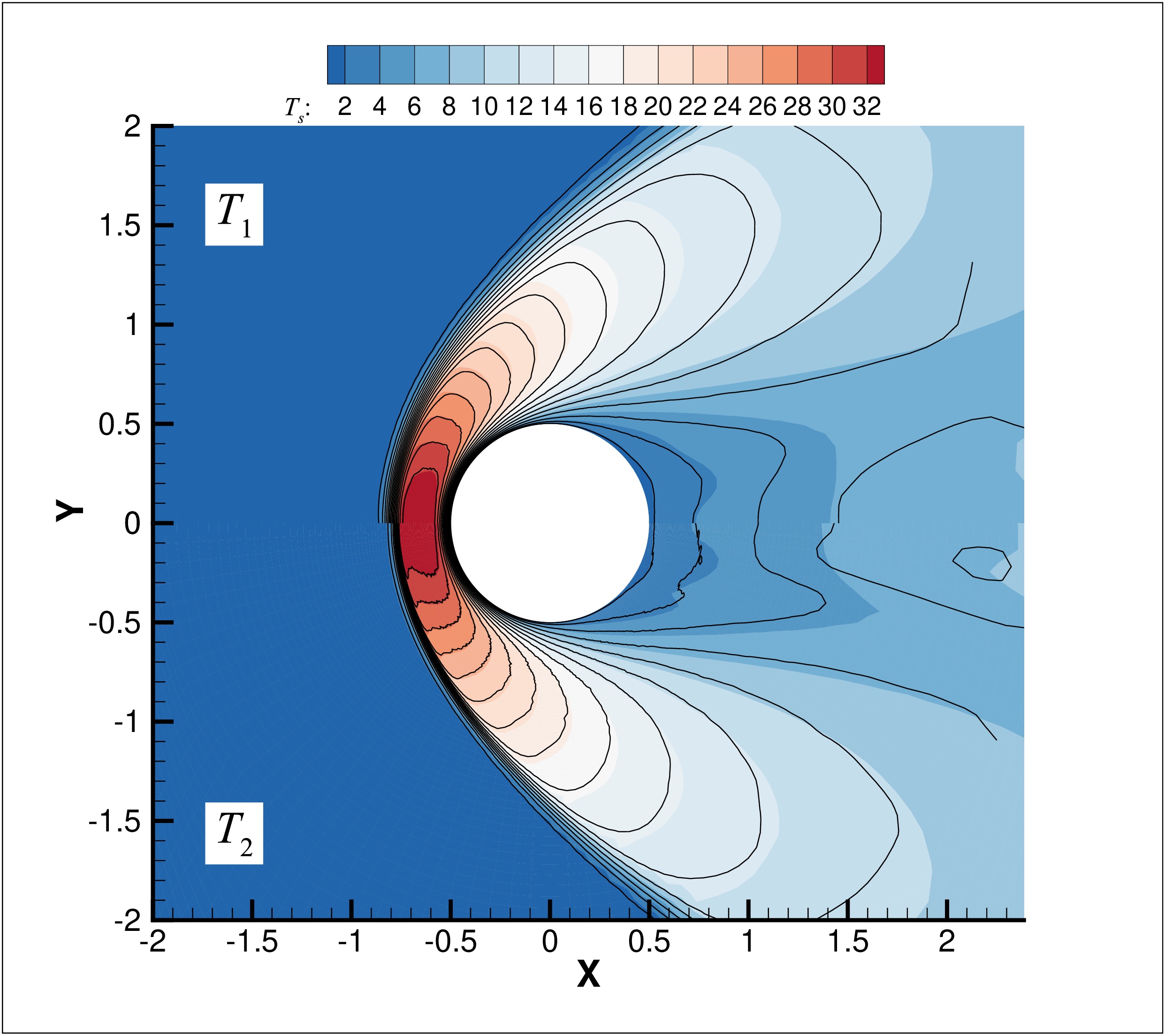}
\caption{Comparisons of macroscopic properties predicted by DIG (lines) and DSMC (contours) for the type 3 mixture with $m_r=10$ and an incoming Mach number of 10. Results in the top and bottom rows correspond to global Knudsen numbers of 0.1 and 0.01, respectively.}
\label{fig:Contour_Mr100_HS}
\end{figure}

Figure~\ref{fig:Contour_Mr100_HS} compares the macroscopic fields predicted by DIG and DSMC for the type-3 gas mixture at $\text{Kn}=0.1$ and $0.01$ with an incoming Mach number of 10. Consistent with the trends observed in previous sections, when $\text{Kn}=0.1$, the peak temperature of the heavier species in the shock region is nearly 1.5 times that of the lighter species. As $\text{Kn}$ decreases to 0.01, the peak temperatures of both species become nearly identical. Nevertheless, substantial inter-species differences in other macroscopic properties remain, primarily due to the large mass ratio and the persistence of non-negligible rarefaction effects. Although SPARTA employs a finer mesh than the DIG method, both approaches yield macroscopic fields in close agreement.

\begin{figure}[!t]
\centering
\includegraphics[width=0.4\textwidth,trim=20pt 20pt 50pt 50pt,clip]{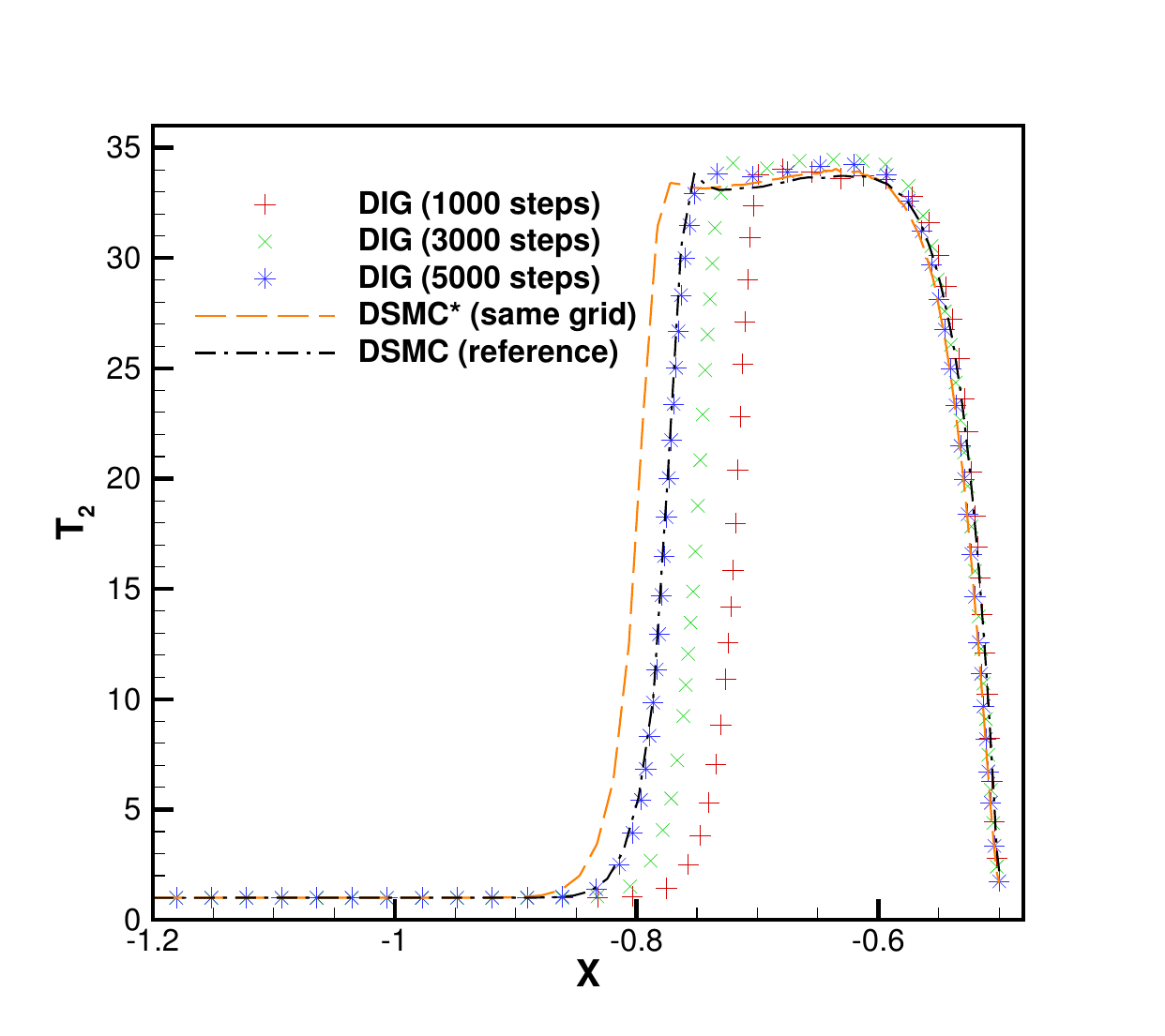}
\hspace{0.5cm}
\includegraphics[width=0.4\textwidth,trim=20pt 20pt 50pt 50pt,clip]{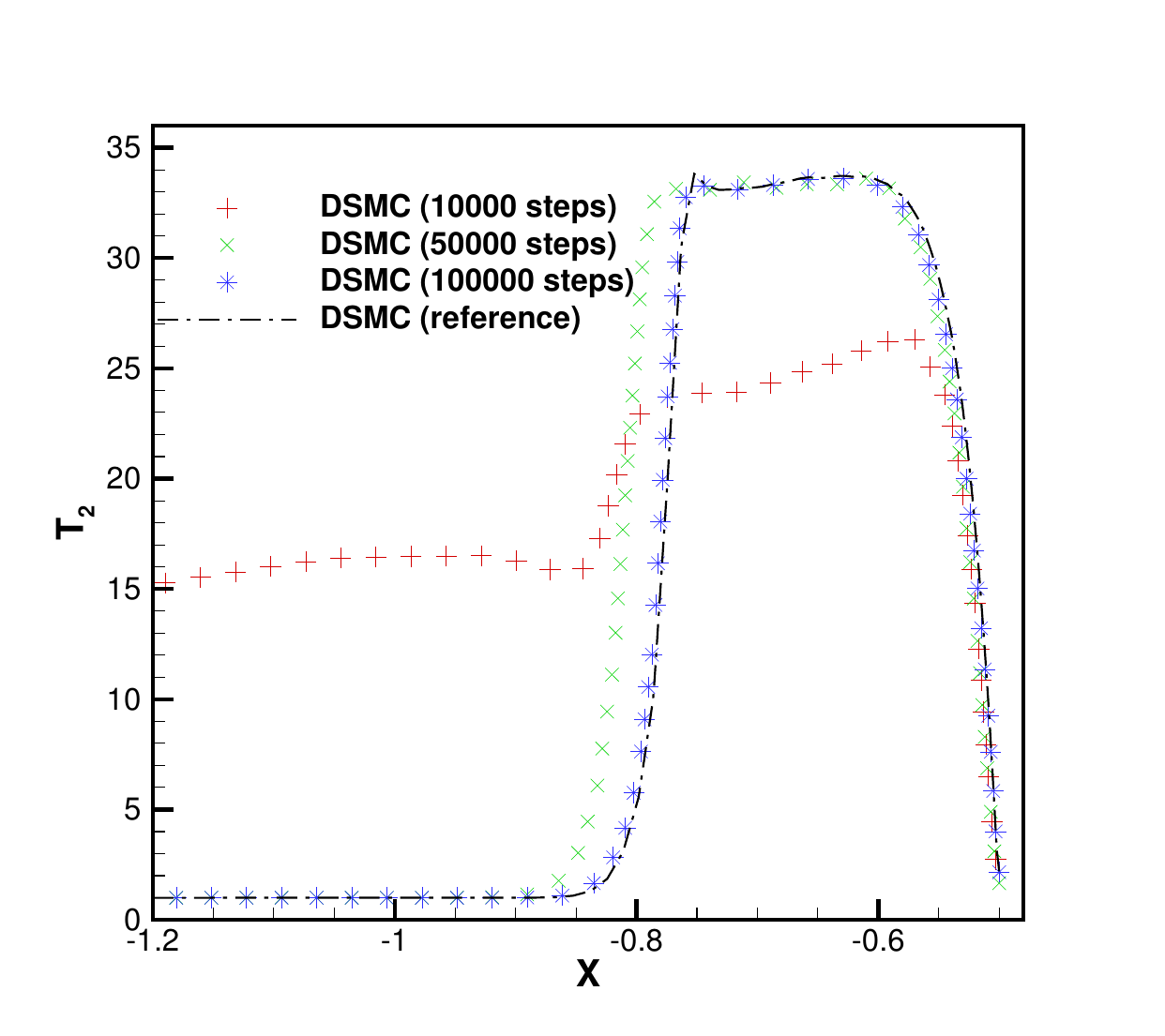}
\caption{Hard-sphere gas mixture with $m_r=100$: the evolution of the temperature for heavy species before the flow field reaches the steady state when $\text{Kn}=0.01$. DSMC results are obtained in SPARTA with adaptive mesh, whereas the DSMC* results are computed on the same mesh as employed in the DIG (the mesh size are much larger than the mean free path).  }
\label{fig:evolution_mr100Ma10_HS}
\end{figure}

Figure~\ref{fig:evolution_mr100Ma10_HS} compares the temperature evolution of the heavy-species in the type-3 mixture predicted by DSMC and DIG when $\text{Kn}=0.01$. Owing to the implicit treatment of the macroscopic synthetic equations in DIG, the particle distribution reaches steady state in substantially fewer steps than in DSMC, even though both solvers employ the same DSMC time step. 
Note that, unlike Maxwell gas mixtures, the exchange terms in the macroscopic synthetic equations for hard-sphere gases require a sufficient number of samples to suppress statistical fluctuations. As a result, the number of evolution steps needed to achieve steady state is larger in the latter case. Nevertheless, DSMC requires approximate 100,000 steps to reach the steady state, whereas DIG requires only 5000 steps, representing a reduction in evolution steps by nearly a factor of 20. Together with the asymptotic-preserving property, which enables the DIG to use significantly fewer spatial grids, the overall computational time is about 10 times shorter for $\text{Kn}=0.1$ and 60 times shorter for $\text{Kn}=0.01$, compared to SPARTA, see Table~\ref{tab:tab2}.

If the macroscopic synthetic equations is switched off after reaching steady state, the DIG results will evolve to that obtained by DSMC on the coarse mesh, resulting in substantial discrepancy compared to fine-mesh DSMC results, see the DSMC* result in Fig.~\ref{fig:evolution_mr100Ma10_HS}. 

\begin{table}[t!]
\centering
\caption{\label{tab:tab2} Same as Table~\ref{tab:tab1}, except that the hard-sphere gas mixture is considered, when the incoming Mach number is 10. }
\begin{tabular}{@{}c c c c c c c@{}}
\toprule
\multirow{2}{*}{ $ \text{Kn} $} & \multirow{2}{*}{Method} & \multirow{2}{*}{$N_{\text{cell}}$} & \multicolumn{2}{c}{Transition state} & \multicolumn{2}{c}{Steady state} \\
\cmidrule(lr){4-7}
 &  &                         & Steps & Time & Steps & Time \\
\midrule
0.1  & DSMC & 887893          & 40000 & 449  & 10000 & 93  \\
     & DIG  & $100\times128$  & 8000  & 45   & 3000  & 15  \\
\midrule
0.01 & DSMC & 6988054         & 100000 & 5021 & 10000 & 489 \\
     & DIG  & $200\times200$  & 5000  & 56   & 3000  & 31  \\
\bottomrule
\end{tabular}
\end{table}

\section{Conclusions and outlook}\label{sec:conclusion}

In summary, we have developed the DIG method to efficiently simulate multiscale flows of monatomic gas mixtures. The approach retains the fundamental framework of the conventional DSMC method while intermittently coupling it with a macroscopic solver, enabling rapid and accurate steady-state solutions even on coarse spatial grids. The DSMC simulation supplies high-order constitutive relations, along with momentum and energy exchange terms, to the macroscopic synthetic equations to account for rarefaction effects. The macroscopic synthetic equations, when solved toward the steady state, in turn guide the evolution of DSMC particles by correcting their velocities and numbers. Together, these two components ensure both rapid convergence and the asymptotic-preserving property for multiscale gas-mixture problems.

Numerical simulations of hypersonic Maxwell and hard-sphere gas-mixture flows past a cylinder demonstrate the superior efficiency and accuracy of the DIG method compared with SPARTA-DSMC, particularly in near-continuum regimes with large mass ratios. For instance, at $\text{Kn} = 0.01$ with a mass ratio of 100, the DIG method achieves a nearly 30-fold reduction in total computational time while maintaining close agreement with the fine-mesh SPARTA-DSMC results.

In contrast to the NS–DSMC hybrid method~\cite{schwartzentruber-2006,schwartzentruber-2007}, the proposed DIG framework offers three distinct advantages. First, it eliminates the need for an exact a priori delineation of rarefied and continuum regions across the entire flow field. Second, by solving the macroscopic synthetic equations separately for each species, it accurately captures inter-species non-equilibrium effects. Third, compared with NS–DSMC and other explicit stochastic methods~\cite{Binzhang-2024,FEI2025114196}, although DIG can employ a computational grid of similar size, it avoids many unnecessary intermediate evolution steps because the macroscopic synthetic equations are solved implicitly and drive the DSMC simulation directly toward the steady state.

Our in-house code, implemented within a finite-volume framework on a non-uniform structured grid, is approximately ten times slower than SPARTA DSMC when using a comparable number of computational cells. Nevertheless, for Knudsen numbers below 0.1, our in-house DIG code remains faster than SPARTA DSMC, and the performance advantage of DIG becomes even more significant as the Knudsen number decreases. Like the SPARTA DSMC, future work of DIG could focus on implementing the synthetic equation on adaptive grids as well. Numerous open-source codes, such as ECOGEN (\url{https://github.com/code-mphi/ECOGEN}) and AMREX (\url{https://github.com/AMReX-Codes/amrex}), are available to facilitate this, potentially further boosting DIG simulation performance by approximately an order of magnitude.

The present work not only offers an efficient and accurate method for solving the Boltzmann equation for monatomic gas mixtures with disparate masses, but also represents a critical step toward improving the DSMC method for hypersonic flows involving chemical reactions. The underlying physics remains essentially the same as in binary gas mixtures, provided that chemical reactions—which primarily involve mass and energy exchange among species—are treated in a manner analogous to the treatment of momentum and energy exchange in Eq.~\eqref{eq:sourcetermexpression1}.

\appendix
\section{Momentum and energy exchanges in the simplified kinetic model}\label{Appendix_model}

Inspired by the exact formulas for the Maxwell gas, the momentum and energy exchange terms derived from the simplified Boltzmann equation with multi-relaxation collision operators is expressed as~\cite{li2024jfm},
\begin{equation}
\begin{aligned}
\bm{Q}_{s,\text{model}}^{M} =& \sum_{r \ne s} \frac{\rho_s \left( \bm{u}_{sr} - \bm{u}_s \right)}{\tau_{sr}}, 
\\
Q_{s,\text{model}}^{E} =& \sum_{r \ne s} \frac{E_{sr} - E_s}{\tau_{sr}},
\end{aligned}
\label{eq:modelsourceterm}
\end{equation}
where $\tau_{ss}$ and $\tau_{sr}$ are corresponding dimensionless relaxation time for different species, and can be written in accordance with the Knudsen number as:
\begin{equation}
    \tau_{ss}=\mathrm{Kn}_s\sqrt{\frac{2m_s}{\pi}}\frac{T_s^{\omega_s-1}}{n_s},\quad\tau_{sr}=\tau_{ss}\phi_{sr}^{-1}\frac{n_s}{n_r},
    \label{eq:tau_relaxation}
\end{equation}
where $\phi_{sr}$ denotes the ratio of the relaxation times for intra-species and inter-species collisions, and can be determined by matching the mixture shear viscosity~\cite{li2024jfm}.

\begin{table}[t]
\centering
\caption{The parameters for different types of gas mixture~\cite{li2024jfm}. Note that the corresponding parameters $\phi_{sr}$ and $\varphi_{sr}$ are determined by matching the mixture viscosity and thermal conductivity from DSMC with the variable-soft-sphere collision model, respectively. The ratio of the species viscosity in a mixture satisfies $\mu_2/\mu_1=\sqrt{m_2/m_1}(d_1/d_2)^2$, while the Knudsen number ratio is $\text{Kn}_2/\text{Kn}_1=(d_1/d_2)^2$.}
\label{tab:mixture}
\begin{tabular}{cccccccccc}
\toprule
Mixture & Gas type & $m_2/m_1$ & $d_2/d_1$ & $\omega_{12}$ & $\alpha_{12}$ & $\phi_{12}$ & $\phi_{21}$ & $\varphi_{12}$ & $\varphi_{21}$ \\
\midrule
1 & Maxwell & 10 & 1 & 1.0 & 2.14 & 1.214 & 0.515 & 1.035 & 1.779 \\
2 & Maxwell & 100 & 3 & 1.0 & 2.14 & 5.505 & 0.078 & 0.988 & 2.093 \\
3 & Hard-sphere & 100 & 2 & 0.5 & 1.0 & 2.955 & 0.127 & 1.425 & 1.261 \\
\bottomrule
\end{tabular}
\end{table}


The inter-species kinetic energy $E_{sr}$ can be represented in terms of the auxiliary flow properties, including the auxiliary flow velocity $\hat{\bm{u}}_{sr}$ and auxiliary temperature $\hat{T}_{sr}$, and is given by 
\begin{equation}
    E_{sr} = \frac{3}{2} n_s \hat{T}_{sr} + \frac{1}{2} \rho_s \hat{\bm{u}}_{sr}^2,
\end{equation} 
where $\hat{\bm{u}}_{sr}$ and $\hat{T}_{sr}$ are constructed to recover the diffusion and energy relaxation processes between species:
\begin{equation}
\begin{aligned}
\hat{\boldsymbol{u}}_{sr}&=\boldsymbol{u}_{s}-\frac{\rho_{r}\tau_{sr}}{\rho_{s}\tau_{s}+\rho_{r}\tau_{sr}}\boldsymbol{X}_{sr},\\
\hat{T}_{sr}&=T_{s}-\frac{n_{r}\tau_{sr}}{n_{s}\tau_{rs}+n_{r}\tau_{sr}}Y_{sr}-\frac{\rho_{s}\rho_{r}\tau_{sr}\tau_{rs}\boldsymbol{X}_{sr}\cdot[\boldsymbol{X}_{sr}-2(\boldsymbol{u}_{s}-\boldsymbol{u}_{r})]}{3(n_{s}\tau_{rs}+n_{r}\tau_{sr})(\rho_{s}\tau_{rs}+\rho_{r}\tau_{sr})},
\end{aligned}
\label{eq:auxiliary_variables}
\end{equation}
with $\boldsymbol{X}_{s r}  =a_{s r}\left(\boldsymbol{u}_{s}-\boldsymbol{u}_{r}\right)+b_{s r}\left(\nabla \ln T_{s}+\nabla \ln T_{r}\right)$ and $Y_{s r}  =c_{s r}\left(T_{s}-T_{r}\right)+d_{s r}\left(\boldsymbol{u}_{s}-\boldsymbol{u}_{r}\right)^{2}$,
where $ a_{sr}=a_{rs}, \, b_{sr}=-b_{rs}, \, c_{sr}=c_{rs}, \, d_{sr}=-d_{rs} $ are adjustable parameters that characterize the rate at which equilibrium between different gas species is established via inter-species collisions.  These parameters can be determined based on the transport properties of the gas mixture as
\begin{equation}
\begin{aligned}
a_{sr} &= \frac{(\rho_{s}\tau_{rs} + \rho_{r}\tau_{sr})T}{m_{s} m_{r}(n_{s} + n_{r})D_{sr}}, \\
b_{sr} &= \frac{(n_{s} + n_{r})T(\rho_{s}\tau_{rs} + \rho_{r}\tau_{sr})k_{T,sr}}{2\rho_{s}\rho_{r}}, \\
c_{sr} &= \frac{2a_{sr}(n_{s}\tau_{rs} + n_{r}\tau_{sr})m_{s}m_{r}}{(m_{s} + m_{r})(\rho_{s}\tau_{rs} + \rho_{r}\tau_{sr})}, \\
d_{sr} &= \frac{a_{sr}m_{s}m_{r}}{3(\rho_{s}\tau_{rs} + \rho_{r}\tau_{sr})} \left[ \frac{a_{sr}\left(n_{r}|\rho_{r}\tau_{sr}^{2} - n_{s}\rho_{s}\tau_{rs}^{2}\right)}{\rho_{s}\tau_{rs} + \rho_{r}\tau_{sr}} - \frac{2\left(\rho_{r}\tau_{sr} - \rho_{s}\tau_{rs}\right)}{m_{s} + m_{r}} \right],
\end{aligned}
\end{equation}
where $D_{sr}$ is the binary diffusion coefficient, and $k_{T,sr}$ is the thermal diffusion ratio. Relevant parameters are summarized in Table~\ref{tab:mixture}.

\section{Finite volume method for macroscopic synthetic equations}\label{Finite_volume_synthetic}

The macroscopic synthetic equations for gas mixture~\eqref{eq:macroscopicequation}  can be considered to be the conventional NS equations with the HoTs~\eqref{eq:highorderterms_DSMC} and the correction terms~\eqref{eq:sourcetermexpression1}. Therefore, they can be solved efficiently by conventional CFD techniques. In this paper, the cell-centered finite volume method is used, and the synthetic equations can be discretized as
\begin{equation}
\frac{\partial \bm{W}_i}{\partial t} + \frac{1}{V_i} \sum_{j \in N(i)} (\bm{F}_{ij}+\bm{F}_{ij}^{HoT}) \bm{S}_{ij} = \bm{Q}_i+\Delta \bm{Q}^*_i + \bm{Q}^F_i.
\label{eq:discreteform}
\end{equation}
Here, $\boldsymbol{W} =[\rho_{s}, \rho_{s} \bm{u}_{s}, E_{s}]^\top$ is the vector of macroscopic variables for the discrete cell $i$, and $N(i)$ represents the set of neighboring cells adjacent to cell $i$, with $j$ indicating a specific neighboring cell. The interface shared by cells $i$ and $j$ is denoted by the subscript $ij$. The volume of cell $i$ is $V_i$, while $S_{ij}$ denotes the area of the interface $ij$, with its outward normal vector directed from cell $i$ to cell $j$. $F_{ij}$ is the interface flux $\bm{F}_{ij}=\bm{F}_{c,ij}+\bm{F}_{{v},ij}$ including both convection and viscosity flux. Moreover, the viscous flux term depends on the HoTs  in shear stress and heat flux $\bm{F}_{v,ij}^{HoT}=\bm{F}_{v}(\bm{\sigma}_s^{\text{DSMC}},\bm{q}_s^{\text{DSMC}})-\bm{F}_{v}(\bm{\sigma}_s^{\text{NS}*},\bm{q}_s^{\text{NS}*})$. The source term $\bm{Q}$ is determined based on the macroscopic variables $\bm{W}$, and $\Delta\bm{Q}^*$ corresponds to the source correction terms~\eqref{eq:sourcetermexpression1}. In  two-dimensional cases, the conservative variables in Eq.~\eqref{eq:discreteform} can be written as,
\begin{equation}
\begin{aligned}
\boldsymbol{F}_{c} &= \begin{bmatrix}
\rho_{s} u_{n} \\
\rho_{s} u_{s,x} u_{n} + n_{x} p_{s} \\
\rho_{s} u_{s,y} u_{n} + n_{y} p_{s} \\
u_{s,n}(E_{s} + p_{s})
\end{bmatrix},
\quad 
\boldsymbol{F}_{v} = \begin{bmatrix}
0 \\
n_{x} \sigma_{s,xx} + n_{y} \sigma_{s,xy} \\
n_{x} \sigma_{s,yx} + n_{y} \sigma_{s,yy} \\
n_{x} \Theta_{s,x} + n_{y} \Theta_{s,y}
\end{bmatrix},  \\ 
\boldsymbol{Q} &= \left[0,\frac{\rho_{s}(u_{sr,x}-u_{s,x})}{\tau_{sr}},\frac{\rho_{s}(u_{sr,y}-u_{s,y})}{\tau_{sr}},\frac{E_{sr}-E_{s}}{\tau_{sr}}\right]^\top, \\ \Delta\boldsymbol{Q}^* &= \left[0,\Delta Q_{s,x}^{M*},\Delta Q_{s,y}^{M*},\Delta Q_{s}^{E*}\right]^\top,\quad \bm{Q}^F=[0,\rho_sa_x,\rho_sa_y,\rho_s(a_xu_{s,x}+a_yu_{s,y})]
\end{aligned}
\label{eq:conservation_system}
\end{equation}
where $\Theta_{s,x} = u_{s,x}\sigma_{s,xx} + u_{s,y}\sigma_{s,xy} + q_{s,x}$, $\Theta_{s,y} = u_{s,x}\sigma_{s,yx} + u_{s,y}\sigma_{s,yy} + q_{s,y}$,  and $u_{s,n}=u_{s,x}n_x+u_{s,y}n_y$ is defined as the scalar product of velocity vector and the unit normal vector. 

Equation~\eqref{eq:discreteform} is further discretized using the implicit backward Euler scheme:
\begin{equation}
\frac{\bm{W}_{i}^{k+1} - \bm{W}_{i}^{k}}{\Delta t_{i}} + \frac{1}{V_{i}} \sum_{j \in N(i)} (\bm{F}_{ij}^{k+1}+\bm{F}_{v,ij}^{HoT}) \bm{S}_{ij} = \bm{Q}_{i}^{k+1}+\Delta \bm{Q}_i^*+\bm{Q}_i^F,
\label{eq:discrete_conservation}
\end{equation}
where $\Delta t=t^{k+1}-t^k$ is the numerical time step given in the implicit process. By introducing the incremental variables $\Delta\bm{W}^{k}_i=\bm{W}^{k+1}_i-\bm{W}^k_i$ and $\bm{F}_{ij}^{k+1}=\bm{F}_{ij}^k+\Delta\bm{F}_{ij}^k$, the delta-form governing equation of Eq.~\eqref{eq:discrete_conservation} for implicit iterative algorithm can be written as,
\begin{equation}
\scalebox{0.8}{ $
\displaystyle 
\left[\frac{1}{\Delta t_{i}} - \left(\frac{\partial (\bm{Q}_{i}+\bm{Q}_i^F)}{\partial \bm{W}_{i}}\right)\right]
\Delta \bm{W}_{i}^{k} + 
\frac{1}{V_{i}}\sum_{j\in N(i)}\Delta \bm{F}_{ij}^{k}\bm{S}_{ij} = 
\underbrace{
-\frac{1}{V_{i}}\sum_{j\in N(i)}(\bm{F}_{ij}^{k}+\bm{F}_{v,ij}^{HoT})\bm{S}_{ij} + \bm{Q}_{i}^{k}+\Delta\bm{Q}_i^*
+\bm{Q}^F_i}_{\bm{R}_{i}^{k}},
$}
\label{eq:deltaform_discretization}
\end{equation}
where $\bm{R}_{i}^{k}$ is the macroscopic residuals in the $k$-th step. 

In general, the macroscopic implicit fluxes in left-hand-side of Eq.~\eqref{eq:deltaform_discretization} is approximated by the first-order flux in the Euler equation:
\begin{equation}
\Delta\bm{F}_{ij}^k = \frac{1}{2}\left[\Delta\bm{F}_i^k+\Delta\bm{F}_j^k+\Gamma_{ij}\left(\Delta \bm{W}_i^k-\Delta \bm{W}_j^k\right)\right],\,\, \bm{F}_{ij}=\bm{F}(\bm{W}_L,\bm{W}_R,S_{ij}),
\label{eq:Eulerfluxes}
\end{equation}
where $\Gamma_{ij}=|u_{s,n}|+c_{s}+2\mu_s/\rho_s|\bm{n}_{ij}\cdot(\bm{x}_{j}-\bm{x}_i)|$ is the approximate spectral radius for each species, and $c_{s}$ is the speed of sound for $s$ species. The reconstructed macroscopic variables of the left and right sides of the interface can be obtained as $\bm{W}_{L/R}=\bm{W}_{i/j}+\phi\nabla(\bm{W}_{i/j}\cdot\bm{x})$, where $\phi$ is calculated using the Venkatakrishnan limiter. In this paper, we apply the Rusanov scheme~\cite{sod-1978} for reconstruction to enhance the numerical stability. Moreover, since the control volume satisfies $\sum_{j\in N(i)}\bm{F}_i\bm{S}_{ij}=0$, the flux can be directly represented by the convection flux, and thus the incremental flux from $j$-th cell can be expressed as $\Delta\bm{F}_j^k=\bm{F}_c(\bm{W}_j^k+\Delta \bm{W}_j^k)-\bm{F}_c(\bm{W}_j^k)$. Thus, the general implicit governing equations for macroscopic properties in Eq.~\eqref{eq:deltaform_discretization} can be expressed as:
\begin{equation}
\scalebox{0.8}{ $
\displaystyle 
\left[\frac{1}{\Delta t_{i}} + \frac{1}{2 V_{i}}\sum_{j \in N(i)} \Gamma_{i j}S_{ij} - \left(\frac{\partial \bm{Q}_i}{\partial \bm{W}_i}\right)^{k}-\left(\frac{\partial \bm{Q}_i^F}{\partial \bm{W}_i}\right)^{k}\right] \Delta \bm{W}_{i}^{k} + \frac{1}{2 V_{i}} \sum_{j \in N(i)} \left( \Delta \bm{F}_{j}^{k} - \Gamma_{ij} \Delta \bm{W}_{j}^{k} \right) \bm{S}_{ij} = \bm{R}_{i}^{k},$ }
\end{equation}
which can be efficiently solved using the classical Lower-Upper Symmetric Gauss-Seidel iteration technique. Furthermore, the Jacobian $\frac{\partial \bm{Q}_i}{\partial \bm{W}_i}$ and $\frac{\partial \bm{Q}_i^F}{\partial \bm{W}_i}$ can be explicitly formulated in terms of macroscopic properties. Note that the correction term $\Delta\bm{Q}_i^*$ remains constant throughout the iteration of the macroscopic equations. Consequently, its Jacobian $\partial \bm{\Delta Q}^*_i / \partial \bm{W}_i$ is zero. 

\bibliographystyle{elsarticle-num}
\bibliography{ref}
\end{document}